# Metal Halide Perovskite Nanocrystals: Synthesis, Post-Synthesis Modifications and Their Optical Properties

Javad Shamsi[†], Alexander S. Urban[ϕ], Muhammad Imran[†], Luca De Trizio*[†], Liberato Manna*[†]

[†] Nanochemistry Department, Istituto Italiano di Tecnologia, Via Morego 30, 16163 Genova, Italy

[ϕ]Nanospectroscopy Group, Department of Physics and Center for Nanoscience (CeNS), Ludwig-Maximilians-Universität (LMU), Amalienstaße 54, 80799 Munich, Germany

**ABSTRACT:** Metal halide perovskites represent a flourishing area of research, driven by both their potential application in photovoltaics and optoelectronics, and for the fundamental science underpinning their unique optoelectronic properties. The advent of colloidal methods for the synthesis of halide perovskite nanocrystals has brought to the attention interesting aspects of this new type of materials, above all their defect-tolerance. This review aims to provide an updated survey of this fast-moving field, with a main focus on their colloidal synthesis. We examine the chemistry and the capability of different colloidal synthetic routes to control the shape, size and optical properties of the resulting nanocrystals. We also provide an up to date overview of their post-synthesis transformations, and summarize the various solution processes aimed at fabricating halide perovskite-based nanocomposites. We then review the fundamental optical properties of halide perovskite nanocrystals, by focusing on their linear optical properties, on the effects of quantum confinement and, then, on the current knowledge of their exciton binding energies. We also discuss the emergence of non-linear phenomena such as multiphoton absorption, biexcitons and carrier multiplication. At last, we provide an outlook in the field, with the most cogent open questions and possible future directions.

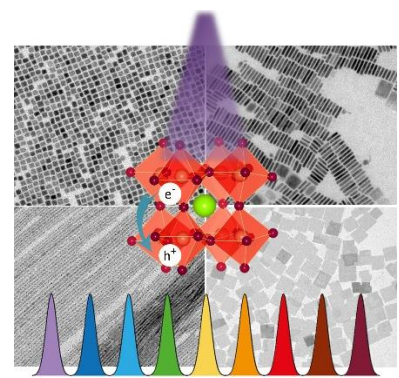

## Contents

## Introduction

Metal halide perovskites (MHPs) have aroused strong interest in scientific and engineering communities due to their intriguing optical and electronic properties. They were first reported in 1893[1] and experienced a brief climax, with a focus on light emitting devices and transistors, in the early 1990s. However, it took until 2012 for the real potential of these materials to be discovered. MHPs were used initially as sensitizing materials in dye-sensitized solar cells. However, it was rapidly determined that they not only boost the absorption cross-section of the resulting device, but also exhibit impressive charge transport properties.[2-5] These findings spawned a massive interest in halide perovskites and, in a narrow time span, the efficiency of single cell perovskite-based photovoltaic devices surpassed 23%.[6-9]

Interestingly, despite seemingly counter-intuitive, perovskites were proved not only to be good for separating charges and creating electricity, but also for bringing charges together to create light.[10-12] Besides their relatively low non-radiative recombination rates, their high color purity makes them interesting candidates for light-emitting diodes (LEDs) and lasers.[13] Unfortunately, the bulk perovskite structures seemed limited in their photoluminescence quantum yield (PLQY), which is around 20%, mainly due to two key limiting factors: i) the presence of mobile ionic defects, which are characterized by a low formation energy, and ii) the small exciton binding energy in MHPs, which results in low electron–hole capture rates for radiative recombination. Moreover, in MHP films prepared from precursor solutions, dominant intrinsic defects are not as benign as initially thought.[14] This was demonstrated by the grain-to-grain variation in the PL intensity with grain boundaries normally being dimmer and exhibiting faster nonradiative decay.[15]

Consequently, as with conventional semiconducting materials, researchers turned their attention into perovskite nanocrystals (NCs), with the intention not only of boosting the PLQY, but also of accessing the quantum-confinement size regime, which could be used as an additional method for tuning the emission of such materials. The first perovskite NCs were synthesized in 2014,[16] and since then research on these compounds has virtually exploded. In the preparation of MHPs NCs, organic capping ligands enable the growth of crystals in the nanometer size range and actively passivate surface defects, similar to the synthesis of NCs of a more traditional materials. It is also possible to finely tune the size and the shape of the NCs, so that one can prepare either bulk-like NCs (*i.e.* particles that are large enough to exhibit optical properties as in bulk crystals or films) or nanostructures like nanoplatelets (NPLs), nanowires (NWs) and quantum dots (QDs). The sizes of these nanostructures can be controlled down to a single perovskite layer, and, consequently, significantly below the exciton Bohr radius (hence in the strong quantum confinement regime).[17-20] The composition, structure, and size of the NCs can be tuned not only during the synthesis, but also via post-synthesis transformations, for example through ion exchange and exfoliation.[21-24]

The peculiar nature of the band structure of MHPs is such that defect states tend to be either localized within the valence and conduction bands or to be essentially "inert", resulting in a high PL efficiency of perovskite NCs, a property that is often dubbed "defect-tolerance". This does not mean that there are no defects which induce non-radiative recombination in perovskite NCs, but rather that the need for extensive passivation, as for example in metal chalcogenide NCs, is less demanding. In a time span of a few years, MHP NCs have been optimized towards emission wavelengths with tunability throughout the entire visible spectral range (and beyond) and quantum yields that have approached 100%.[25-26] Nearly five years have passed since the pioneering colloidal syntheses of MHP NCs,[16, 27] and since then countless innovations in their synthesis and processing have delivered materials that have been tested in solar cells, both as active materials[28-30] and as downconverters[31], in solar concentrators[32], in visible light communication[33], electroluminescent diodes,[34-35] photodetectors[36] and in photocatalysis.[37] Among the different members of the vast perovskite family, lead-based halide perovskite (LHP) NCs have been found to be valid alternatives to well-established II–VI, III-V and IV-VI semiconductor NCs (or quantum dots, QDs). We direct the interested readers to recent reviews for a comprehensive comparison of LHP NCs with more traditional QDs.[38-40]

Despite such a rapid advancement in materials synthesis and device applications, studies on the optical properties of perovskite NCs have significantly lagged behind. Many questions pertaining to halide perovskites in general remain unanswered,[41-43] and this is also the case of perovskite NCs. For example, it is now well established that the PLQY of LHP NCs varies depending on the synthetic approach and that even small modifications of a given synthetic protocol can strongly influence their optical properties. The reasons underpinning such a wide variability in their optical properties is that these are strongly related to the stoichiometric ratios of the ions in the perovskite structure,[44] as well as to the types of coating ligands and, more in general, to how the surface is terminated.[45-46] However, the investigation on these aspects is still in its infancy.

Several review articles on the chemistry and optical properties of MHP NCs have been published in the last couple of years.[47-55] This review article seeks to summarize recent developments in colloidal synthetic routes to MHP NCs and their post-synthesis treatments, in order to analyze the strengths and weaknesses of each method. It then gives a snapshot of the current knowledge on their optical properties. Our review is structured into four main sections: i) a brief introduction to MHP NCs is given to aid in understanding their nature, crystal structures and optical properties. ii) We then address in detail the most recent trends in colloidal synthesis of MHP NCs. Our intention is to provide a comprehensive overview of the different synthetic routes for MHP NCs highlighting how the size and shape control can be achieved. iii) We present a broad summary of post-synthesis modifications of MHP NCs, which are applied to either manipulate their optical properties or to increase their stability. iv) The last section deals with recent progress in understanding the optical properties of MHP NCs, such as linear and nonlinear optical features, quantum confinement effects and their exciton binding energies. We end the review with our vision for the future of this field.

## A Brief Introduction to MHP NCs

### History of colloidal synthesis of MHP NCs

The interest in semiconductor NCs was triggered by the discovery of quantum-size effects in the optical spectra of nanometer sized semiconductors in the early 1980s.[56-57] Notably, the quantum confinement effect in MHPs (*i.e.* $CsPbX_3$) was reported long before widespread attention was paid to bulk crystals and thin films of these materials.[58-60] We direct the readers to the recent reviews for a more detailed historical background of MHPs NCs.[38-39] Parallel to the work on classical colloidal semiconductor NCs, a seminal paper was published in 2011 on methyl ammonium lead iodide ($MAPbI_3$) in the form of nanometer-sized crystals as a promising PV material.[61] That work captured the attention of the colloidal chemistry community, which started researching this interesting class of medium-bandgap semiconductors. The first solution-based colloidal approach of MHPs NCs was published only some years after, when Schmidt *et al.* at the end of 2013 were able to prepare $MAPbBr_3$ NCs (as detailed later, in the reverse microemulsion section) having a PLQY of 20%[16]. Soon after (in 2015), Protesescu *et al.* reported a colloidal synthesis of monodisperse $CsPbX_3$ NCs, by adapting the standard hot-injection method, typically employed for classical colloidal QDs (such as CdSe and PbSe),[27] to fit the requirements of the new chemicals and solvents needed to prepare perovskite NCs. The work of Protesescu *et al.* stands out as it evidenced three remarkable properties of LHP NCs: i) the high PLQY (up to 90%) without any specific post-synthesis treatment; ii) a PL with a narrow full width at half maximum (FWHM), that is, below 100 meV; iii) a PL tunability over the entire visible spectral range, which was achieved by varying the nature and ratio of halide ions in the structure.[27] The ease by which these NCs could be synthesized, their interesting properties, as well as the many questions left open by these initial reports have been the main attractors for researchers from various fields.

### Crystals Structure, ionic nature and defect tolerance

The "3D" perovskite structure can be described as formed by three primary ions, with a stoichiometry of $ABX_3$ (Figure 1a, b). For the all-inorganic halide counterparts, the monovalent A-cation can be cesium or rubidium, the divalent B-cation can be lead, tin or germanium, and X are halide ions, that is chlorine, bromine, iodine, or a combination of them.[62] The organic-inorganic perovskite compounds have either methylammonium (MA) or formamidinium (FA) as the A cation.[63] The A and B cations coordinate with 12 and 6 X anions, forming cuboctahedral and octahedral structures, respectively. Notably, the Goldschmidt tolerance factor, *t*, has been used extensively to predict the stability of perovskite structures based only on the chemical formula $ABX_3$ and the ionic radii, $r_i$, of the ions (A, B, X): ($t = r_A + r_X/\sqrt{2(r_B + r_X)}$).[64] In general, stable 3D perovskite structures are formed when the tolerance factor is within the range 0.76 and 1.13, while, out of this range other perovskite-related structures are stablized.[65] For this reason, in the case of LHP perovskites only a limited number of A-cations, that are Cs, MA, and FA, can give rise to stable structures. Other possible candidates are either too small (Na, K, Rb) or too large (imidazolium, ethylamine, and guanidinium).[8] Perovskites at the edge of the tolerance factor requirement, such as $FAPbI_3$ ($t$ ~ 1) and $CsPbI_3$ ($t$ ~ 0.8), easily undergo phase transition at room temperature (RT) to the more stable hexagonal and orthorhombic phases (Figure 1 c, d), respectively (also called "yellow phases").[8] Perovskite structures are further constrained by the octahedral factor $\mu$, defined as $\mu = r_B/r_X$, which describes the stability of the $BX_6^{2-}$ octahedra, and depends on the B and X ions radii. The stability range for $\mu$ is between 0.442 and 0.895.[66] The tolerance and octahedral factors are currently used to predict the stability of novel possible perovskite combinations (see Figure 1e).[67-68]

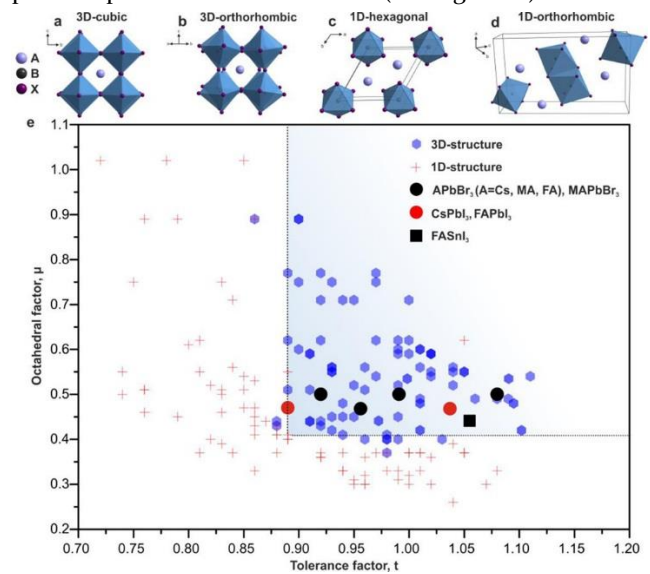

**Figure 1.** Schematic representations of (a) an ideal 3D cubic structure, as observed in α-FAPbI$_3$; (b) an orthorhombically distorted 3D structure, typically reported for CsPbBr$_3$; (c) a 1D hexagonal lattice, found in the yellow phase of FAPbI$_3$; and (d) a 1D orthorhombic structure, found in the yellow phase of CsPbI$_3$. (e) Reported 3D and 1D structures of different all-inorganic and hybrid organic-inorganic ABX$_3$ MHP compounds. The light blue squared area represents the region in which stable compounds are located. The tolerance and octahedral factors were mainly taken from the report of Travis et al.[66] Reproduced from ref.[67] Copyright 2017, American Chemical Society.

Perovskite compounds are mostly held together by ionic bonding, and this is one of the reasons that rationalize the ease by which highly crystalline NCs can be fabricated even at low temperatures.[69] One of the key features of MHPs that has been ascribed to their success as high-performance semiconductors is their defect tolerance, which is an ability to retain the electronic structure of the pristine material even in the presence of a large concentration of defects. The reported defect-tolerance of MHP, , is corroborated by first-principles density-functional theory (DFT) calculations of the formation energy of point defects and their effect on the electronic structure.[14] For perovskites, the defect chemistry and physics is still not well understood, but vacancy-related defects are considered to have energies close to, or inside, the energy bands (see the section "optical properties" for further details).[39, 69]

### Different phases (3D, 2D, 0D and double perovskites)

Most research in lead halide perovskite NCs has been focused, so far, on NCs with the 3D APbX$_3$ crystal structure and composition. On the other hand, the general reactivity of this class of halide perovskites and their intrinsic toxicity has stimulated research in various directions. First, the high ionicity and structural instability of LHP NCs, which limits their range of applications, come nevertheless with a bonus: the facility with which the APbX$_3$ lattice can be reorganized into other phases. This has stimulated an extensive investigation on NCs with other structures and compositions, also defined as perovskite-related structures, such as Cs$_4$PbX$_6$ (so-called 0-dimensional structures) and CsPb$_2$X$_5$ (2D). While the 3D APbX$_3$ structure is characterized by corner sharing [PbX$_6$]$^{4-}$ octahedra with the A$^+$ cations filling the voids created by four neighboring PbX$_6^{4-}$ octahedra (resulting in a cubic or pseudo-cubic), the PbX$_6^{4-}$ octahedra in A$_4$PbX$_6$ structures (0D) are completely decoupled in all dimensions and the halide ions are no longer shared between them (Figure 2a-c).[70-71] Recently, layered perovskites have come into intense scrutiny. CsPb$_2$X$_5$ emerged as the 2D version of lead halide perovskite materials with tetragonal phase which consists of alternating Cs$^+$ and [Pb$_2$X$_5$]$^-$ polyhedron layers, similar to that of layered double hydroxides (Figure 2d).[72] Another type of 2D perovskites is the A$_2$PbX$_4$ phase, which is made of layers of corner-sharing [PbX$_6$]$^{4-}$octahedra alternating with layers of bulky cations (Figure 2e).[73]

The toxicity of lead (and its bioaccumulation in the ecosystem) is actually the major limitation of APbX$_3$ NC systems, and has urged research into materials with comparable optoelectronic properties, such as Cs$_2$SnI$_6$ NCs,[74-78] however with very limited success to date. Cs$_2$SnI$_6$ crystallizes in the face-centered cubic (*fcc*) structure and the unit cell is composed of four [SnI$_6$]$^{2-}$ octahedra at the corners and the face centers and eight Cs$^+$ cations at the tetragonal interstitials (Figure 2f). The Cs$_2$SnI$_6$ structure is a perovskite derivative and is obtained by removing half of the Sn atoms at each center of the [SnI$_6$] octahedron at regular intervals.[74] For this reason, the structure is also referred to as "vacancy ordered double perovskite". In the search for lead-free metal halide compounds, two main strategies are pursued: either the "simple" substitution of Pb$^{2+}$ cations with other less toxic divalent metal ions of the same group IV, such as Sn or Ge,[79] or the replacement of every two divalent Pb$^{2+}$ ions with one monovalent M$^+$ and one trivalent M$^{3+}$ cations, i.e. 2Pb$^{2+}$ → B$^+$ + B$^{3+}$, generating quaternary A$_2$B$^+$B$^{3+}$X$_6$ systems (also named "double perovskites", Figure 2g).[80] The diversity of halide materials related to LHPs are explored by introducing other transition or post-transition metals, such as Fe$^{3+}$ and Bi$^{3+}$.[81-85] Cs$_3$M$_2$X$_9$ (M= Fe$^{3+}$, Bi$^{3+}$) crystallizes in the hexagonal space group P6$_3$/mmc and it consists of face-sharing M$_2$Br$_9^{3-}$ octahedral dimers with Cs serving as bridging atoms between the dimers. Another type of such dimer-like structure is observed in antimony-based halide compounds which consists of bioctahedral (Sb$_2$Br$_9$)$^{3-}$ clusters surrounded by cesium cations(Figure 2h and i).[77]

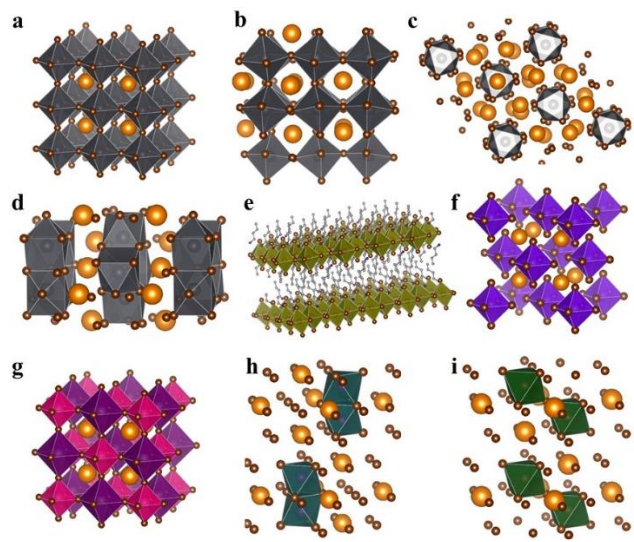

**Figure 2.** Schematic representation of different metal halide structures: (a) cubic-phase ABX$_3$ (3D); (b) pseudo-cubic ABX$_3$ (3D); (c) A$_4$BX$_6$ (0D); (d) AB$_2$X$_5$ (2D); (e) A$_2$BX$_4$ (2D); (f) A$_2$BX$_6$ (0D); (g)A$_2$B$^+$B$^{3+}$X$_6$ (3D); (h) and (i) A$_3$B$_2$X$_9$ (0D) .

### Colloidal Synthesis

To prepare high quality MHP NCs in terms of control over size, shape and quality of their optical properties, many efforts have been devoted to develop reliable and straightforward synthetic strategies, which can be classified either as top-down or bottom-up approaches. The top-down cases comprise fragmentation and structuring of macroscopic solids, either mechanically (e.g., ball-milling in the presence of surfactants[86]) or chemically (e.g. chemical exfoliation,[23] etc.), whereas the bottom-up route starts with molecules and ions and proceeds via gas- or liquid-phase chemical reactions. Among all bottom-up approaches, the liquid-phase approach has been proven as an efficient route to fabricate well-defined colloidal MHP NCs (Scheme 1).[49, 87] In this review, we focus on two main liquid-phase methods for the synthesis of colloidal MHP NCs: the hot injection (HI) and the ligand assisted re-precipitation (LARP) methods, which are by far the most developed ones.[48, 88] Briefly, the HI route requires high temperatures and an inert atmosphere, which inevitably increase the cost and could limit the output in mass-production.[89] To overcome these two potential limitations, the LARP method can be employed as a more cost-effective alternative as it delivers high quality perovskite NCs in ambient atmosphere and at RT.[90]

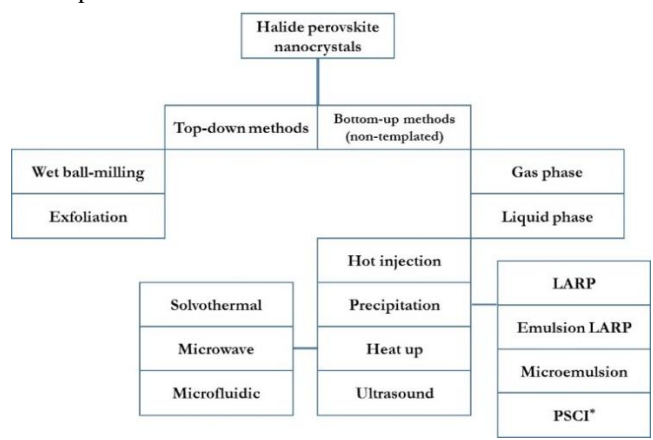

**Scheme 1.** An outline of the various methods employed in the synthesis of MHP nanocrystals. *: polar solvent controlled ionization

### The Hot Injection Strategy

The first HI method was developed two and half decades ago for the synthesis of cadmium chalcogenide NCs.[91] This approach is based on the rapid injection of a precursor into a hot solution of the remaining precursors, ligands and a high boiling solvent."[92-94] The HI method enables, in general, the synthesis of small NCs with a narrow size distribution by attaining a separation between the nucleation and growth stages.[95] More in detail, immediately after the injection, a rapid nucleation burst takes place with the simultaneous formation of small nuclei. A rapid depletion of monomers terminates the nucleation stage, after which the nuclei continue growing (with ideally no new nuclei forming), leading, over time, to the evolution of a NCs population characterized by a narrow size distribution. This is true if the reaction is stopped when it is still in the size focusing regime, *i.e.* when there are still plenty of monomers in the growth environment.[96] Key parameters, which allow to control the size, size-distribution and shape of colloidal NCs synthesized by the HI method are: (i) ratios of the surfactants to the precursors; (ii) the injection temperature of the cations or anions precursor; (iii) the reaction time; and (iv) the concentration of precursors. In 2015, Protesescu *et al.* extended the HI approach to the colloidal synthesis of cesium LHP NCs ($CsPbX_3$, X=Cl, Br and I)[27]. $CsPbX_3$ NCs were obtained by injecting Cs-oleate into a hot solution (140-200°C) of $PbX_2$ (X= Cl, Br, I) salts, which served both as the $Pb^{2+}$ and $X^-$ source, dissolved in octadecene (ODE), carboxylic acids and primary amines (See Figure 3).

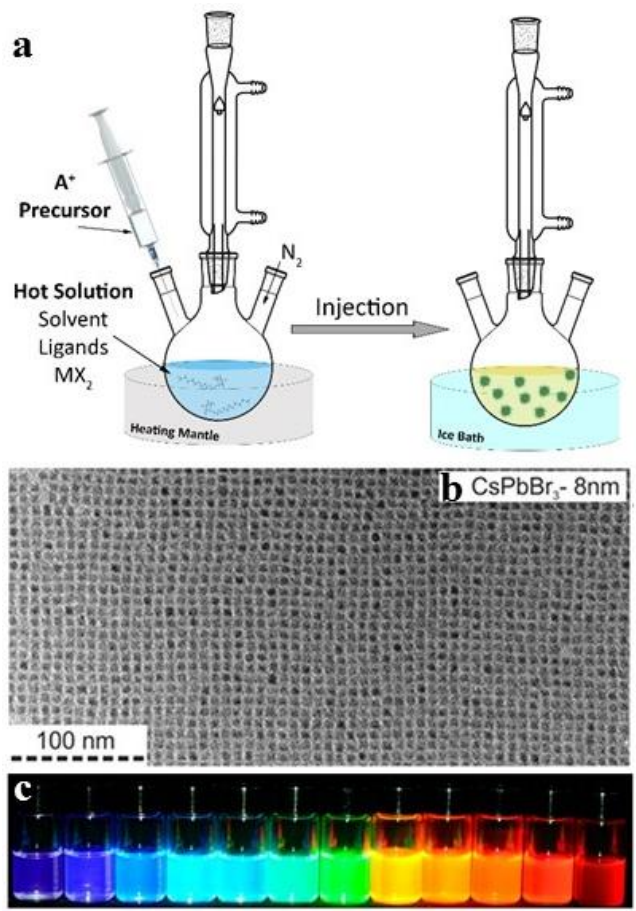

**Figure 3.** (a) Schematic illustration of the HI method used for the synthesis of colloidal MHP NCs. (b) A typical TEM image of $CsPbBr_3$ NCs obtained with the hot injection (HI) strategy and (c) Colloidal perovskite $CsPbX_3$ (X = Cl, Br, I) NC dispersions (in each vial the NCs have a different halide composition) in toluene under a UV lamp (λ = 365 nm), Adapted from ref. [27] Copyright 2015, American Chemical Society.

Equal ratios of amines and acids were observed to result in the formation of monodisperse NCs whose size could be adjusted by varying the reaction temperature. Mixed-halide perovskite NCs could also be conveniently synthesized by simply adjusting the ratios of lead halide salts ($PbCl_2/PbBr_2$ or $PbBr_2/PbI_2$). The PL emission of the resulting NCs could be finely modulated across the entire visible spectrum (410-700nm) by varying the halide composition or by tuning the size of NCs. The HI method was further extended to $MAPbX_3$ (X=Br, I) NC systems by replacing Cs-oleate with a methylamine solution.[97] $MAPbBr_3$ and $MAPbI_3$ NCs were successfully obtained by varying the relative amount of oleylamine (OLA) and oleic acid (OA) capping ligands.

The HI protocol was soon extended to synthesize perovskite-related lead halide based materials. For instance,

$Cs_4PbX_6$ (X = Cl, Br, I) NCs were successfully prepared by working under $Cs^+$ and OLA rich conditions compared to those used in the conventional $CsPbBr_3$ NCs synthesis.[98] The resulting NCs were nearly monodisperse and their size could be further tuned from 9 to 37 nm.[99] Various other groups reported the synthesis of $CsPb_2Br_5$ NCs using octylamine and OA as surfactant in excess of $PbBr_2$ precursor.[100-101]

In order to get insights into the growth kinetics of $CsPbX_3$ NCs produced via the HI approach, Lignos et al. employed a droplet-based microfluidic platform.[102-103] Following the in-situ absorption and photoluminescence of $CsPbX_3$ NCs, they revealed that the entire nucleation and growth took place in the first 1-5 seconds of reaction, highlighting the extremely fast reaction kinetics. Slightly different results were reported by Koolyk et al. who investigated the growth kinetics of $CsPbBr_3$ and $CsPbI_3$ NCs by taking aliquots at different stages of the reaction and analyzing them by transmission electron microscopy (TEM).[104] What they observed is that in $CsPbI_3$ NCs the size focusing regime lasted for the first 20 s, followed by a subsequent size de-focusing regime. On the contrary, $CsPbBr_3$ NCs did not exhibit any size focusing regime, being characterized, from the very beginning, by broadening of the size distribution that persisted during the whole reaction time of 40s.[104] Udayabhaskararao et al., later on, monitored the growth of $CsPbX_3$ NCs at different stages by electron microscopy, eventually proposing a two-step growth mechanism:[105] during the first step, $Pb°$ NCs form, subsequently acting as seeds onto which $CsPbX_3$ NCs nucleate. In the second stage, the growth of NCs occurs through self-assembly and oriented attachment. However, the authors did not provide any evidence of the initial formation of $Pb°$ NCs. Also, it is now well established that the observed $Pb°$ clusters on the surface of NCs, when they are observed under an electron microscope, are formed by the electron beam irradiation.[106-107] Therefore, with so much evidence that such $Pb°$ clusters are formed only during observation under an electron microscope, this mechanism can be dismissed.

Overall, a large variability in the quality of the manufactured MHP NCs reported throughout the literature revealed that, in contrast to classical colloidal heat-up or hot-injection techniques, used to produce II-VI (e.g., CdTe, CdSe, CdS and HgS), IV-VI (e.g., PbSe), III-V (e.g., InP and InAs) or ternary I-III-VI (e.g. $CuInS_2$) QDs, the research community still does not fully understand the MHP NCs nucleation and growth processes. The fact that the nucleation and growth of MHPs NCs are fast and not separable in time is likely due to the extreme ease by which these ionic crystals are formed in solution. The challenge in controlling the nucleation and the growth with a high ensemble uniformity has been one of the major obstacles impeding the exploration and the utilization of the properties of $CsPbX_3$ NCs.[108] However, a recent study has demonstrated that this challenge can be addressed by working under conditions of thermodynamic equilibrium instead of kinetic control.[109]

An important shortcoming of colloidal MHP NCs (and of halide perovskites in general) is their high solubility in polar solvents which translates into a poor stability under ambient atmospheric conditions (with variable humidity, heat, and/or light), with the consequent drop of their PLQY.[90] An example in this direction is represented by the work of Yuan et al.:[110] they have shown that perovskite NCs are particularly sensitive to anti-solvents used in their cleaning. Starting from $CsPbBr_{1-x}I_x$ NCs, they found that different anti-solvents, like isopropanol, n-butanol or acetone, can change the NCs' halide composition, and, thus their overall optical properties.[110] This aspect of post-synthesis treatment will receive considerable attention in the "surface, shape and phase post-modifications" section of this review. For now, we will discuss those works that have taken the instability issue into account already at the synthesis stage and have attempted to adopt countermeasures. In order to counterbalance the stripping of halide ions which accompanies the use of polar antisolvents, Woo et al. found that the stability of $CsPbX_3$ NCs can be considerably enhanced, without drastic drop of PLQY, by working under halide rich conditions.[111] In that work, $ZnBr_2$ was used as an extra source of Br ions in addition to the $PbBr_2$ precursor. The resulting NCs had halide rich compositions (Cs:Pb:Br = 1.0:1.2:3.4) compared to the one synthesized in absence of metal bromide (Cs:Pb:Br = 1.0:1.0:2.8). This approach was defined by the authors as the first successful attempt to achieve in situ stabilization of $CsPbX_3$ NCs (where X is Br or I) by inorganic passivation.[111] This definition of inorganic passivation is however not entirely acceptable, as the NCs at the end were still passivated with organic molecules.

One important drawback of the HI methods described so far is that they rely on the use of metal halide salts as both cation and anion precursors, limiting, thus, the possibility of working with the desired ions stoichiometry.[27, 112] In order to overcome such restrictions Liu et al. developed the so called "three-precursors" HI approach for the synthesis of $CsPbX_3$ (X = Cl, Br, or I) NCs.[113] Their novelty consisted of using $NH_4X$ (X = Cl, Br, or I) and PbO as sources of halide and lead ions separately, instead of conventional $PbX_2$ (X = Cl, Br, or I) salts.[113] Also in that case, as already observed by Woo et al.,[111] $CsPbBr_3$ NCs synthesized under Br-rich conditions (that is, by employing an excess of $NH_4Br$) had superior optical properties and remarkable stability, as they endured the purification step.[113]

Yassitepe et al. further extended the three precursor HI approach to synthesize OA-capped $CsPbX_3$ NCs by eliminating alkylamines from the synthesis.[114] In their approach, Cs-acetate and Pb-acetate were reacted with quaternary alkylammonium halides, such as tetraoctylammonium halides (TOA-X), which cannot form, even in the presence of protons, protonated ammonium species. It was observed that the absence of oleylamine considerably speeded up the growth kinetics, enabling the synthesis of $CsPbX_3$ NCs at lower temperature (i.e. 75°C). $CsPbBr_3$ NCs obtained by this approach exhibited PLQY up to 70%, and enhanced colloidal stability. The method however failed to produce

CsPbI$_3$ NCs of similar quality and stability.[114] The three-precursors HI approach was later adopted and modified by Protesescu et al. for the colloidal synthesis of FAPbX$_3$ NCs.[67, 115-116] Briefly, FAPbBr$_3$ NCs were prepared by reacting FA and Pb acetates with oleic acid in ODE and, subsequently, injecting oleylammonium bromide. It was further observed that the final product contained a 5-10% of NH$_4$Pb$_2$Br$_5$ byproduct that might have formed upon the thermal decomposition of FA$^+$ to NH$^{4+}$ during the synthesis.[115] Phase pure FAPbI$_3$ NCs with better optical quality were successfully synthesized in 2017 by the two precursor HI method by reacting FA-oleate with PbI$_2$ complex in presence of oleic acid and oleylamine in ODE with excess FA (FA:Pb = 2.7) at 80 °C[67].

Although the three-precursors HI approach allows to work with the desired stoichiometry of ions (as the halide and the metal cations sources are not linked anymore) its potential versatility is limited by a series of disadvantages. First, the fact that the synthesis of CsPbI$_3$, CsPbCl$_3$, and MAPbX$_3$ (X=Br, I) NCs has not been reported using this strategy suggests a poor reactivity of the halide precursors in the reaction conditions of that approach.[67, 115, 117] Also, a non-negligible amount of undesired secondary phases, ascribed to the decomposition of the alkylammonium halide precursor, were observed in the synthesis of FAPbX$_3$ NCs.[115] In order to compensate the limitations of the three precursors HI approach, new methods were recently reported by our group[44] and by Creutz et al.[118] Both strategies are based on the use of highly reactive halide precursors which can be conveniently injected into a solution of metal carboxylates, used as cation precursors, dissolved in ODE and ligands (OLA and OA). The injection immediately leads to the nucleation and the growth of the NCs. Imran et al. demonstrated how benzoyl halides, used as halide sources, can be used to prepare the entire family of all-inorganic and hybrid lead halide perovskite NCs (CsPbX$_3$, MAPbX$_3$, FAPbX$_3$ X = Cl$^-$, Br$^-$, I$^-$)[44] with good control over the size distribution and phase purity. Along a similar line, Creutz et al. used silyl halides to synthesize Cs$_2$AgBiX$_6$ NCs (these materials will be covered in the "Pb-free metal halide perovskite NCs" section).[118] Of particular relevance is that these strategies enable one to work with desired cations/anions ratios, and, more specifically, in a halide-rich environment. In some systems, this was observed to strongly increase the PLQY of the resulting NCs: for instance, CsPbCl$_3$ NCs synthesized with either the LARP or the two-precursors HI approach were characterized by significant nonradiative carriers recombination,[119-120] whereas the use of an excess of benzoyl chloride (halide rich conditions) boosted their PLQY up to a record value of 65%.[121]

### Size and shape control of lead halide perovskite NCs by the HI approach

Size and shape control was demonstrated in the HI strategy mainly by varying the ligands combination, their ratios and the reaction temperature. As a general trend, it was observed that working with OLA and OA at low reaction temperatures, in the 90-130 °C range, tended to strongly favor anisotropic growth of NCs, producing quasi 2D geometries, usually referred to as "nanoplatelets" (NPLs).[122] On the other hand, high reaction temperatures (i.e. 170-200°C) and long reaction times lead to nanowires (NWs).[123] In order to investigate the role of ligands on the morphology of CsPbBr$_3$ NCs, Pan et al. carried out a comprehensive study by systematically varying the chain length of alkyl amines and carboxylic acids used in the reaction.[124] In one series of experiments, while keeping fixed the amount of OLA, they employed different carboxylic acids (at 170°C). An increase in the average edge length of CsPbBr$_3$ nanocubes from 9.5 to 13 nm was observed by decreasing the chain length of the carboxylic acids. Also, working with OA and lowering the reaction temperature to 140°C, NPLs with a thickness of 2.5nm and 20 nm wide were formed (see Scheme 2). In a second series of experiments, while fixing the amount of oleic acid, different alkylamines were tested at 170°C: in all the experiments, the authors observed the formation of NPLs except when using OLA, which could still lead to the formation of NPLs, although only at lower reaction temperatures (140°C).

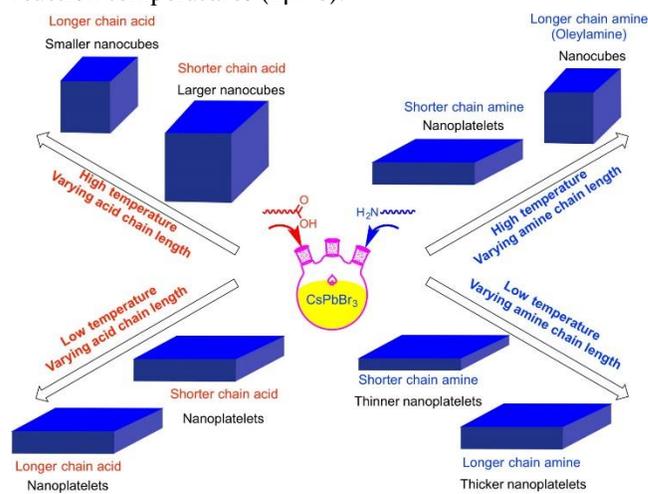

**Scheme 2.** Shape and size control of CsPbBr$_3$ NCs in the HI approach. Reproduced from ref. [124] Copyright 2016, American Chemical Society.

While many works aimed to control the lateral size of lead halide based nanocubes, efforts were also made to synthesize anisotropic nanostructures, such as NPLs and NWs with control over their dimensions. Song et al. reported the HI synthesis of atomically thin CsPbBr$_3$ nanosheets (NSs) with thickness of 3.3 nm and edge length of about 1 μm using dodecylamine and oleic acid and prolonging the reaction time up to 3 hours (see Figure 4a).[125] The same year (2016), Shamsi et al. reported the synthesis of CsPbBr$_3$ NSs with tunable lateral dimensions from 200 nm up to few micrometers while keeping their thickness to few unit cells (see Figure 4b).[126] This was achieved by employing short chain ligands, namely octylamine and octanoic acid, in addition to the conventional OA and OLA. In parallel, in another work Imran et al. demonstrated the synthesis of CsPbBr$_3$ NWs with the width being tunable down to few unit cells (see Figure 4c) via the HI approach.[127] Green emitting CsPbBr$_3$ NWs with 10–20 nm width (hence non-confined) were prepared by employing octylamine and OLA only (no carboxylic acid was used).[128] The diameter of the NWs could be decreased from 10 to 3.4 nm by introducing a short chain carboxylic acid (octanoic acid, or hexanoic acid).[127] Zhang et al. later reported a synthesis strategy to prepare thin[129] and ultrathin[130] (width 2.2 ± 0.2 nm and length up to several microns) CsPbBr$_3$ nanowires (Figure 4d). The NWs were prepared by using OLA, OA and dodecyl amine as a ligands via HI approach and stepwise purification was carried out to enhance the yield.[20]

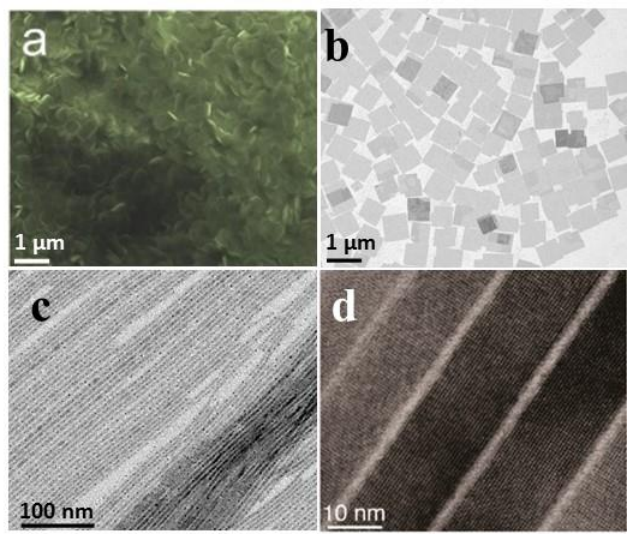

**Figure 4.** (a) SEM[125] and (b) low resolution TEM pictures of CsPbBr$_3$ NSs.[131] (c) Low[127] and (d) high resolution TEM micrographs of CsPbBr$_3$ NWs.[129] Panel (a) is reproduced from ref. [125] Copyright 2016 Wiley-VCH. Panels (b), (c) and (d) are reproduced from refs. [129,131, 127,] Copyright 2016 American Chemical Society.

To better elucidate the way alkylamines and carboxylic acids interact with each other before and during the HI synthesis, Almeida et al. performed an in depth investigation on the interplay between oleylamine and oleic acid and how their relative concentration was affecting the size, size distribution and shape of CsPbBr$_3$ NCs.[108] It was found that, by increasing the concentration of ligands, the precipitation temperature of PbBr$_2$ could be significantly enhanced from 195 to 290°C, allowing to conduct syntheses of CsPbBr$_3$ NCs at higher temperatures. It was revealed, by NMR analysis, that the concentration of oleylammonium species could be increased not only by raising the concentration of oleic acid, but also by lowering the reaction temperature. The concentration of oleylammonium species was found to determine the shape of the final NCs: a high concentration of oleylammonium species (achievable in a high acidic environment) favors anisotropic growth of the NCs, whereas a low concentration leads to the formation of nanocubes (Figure 5). This evidences that oleylammonium species, and, more in general, primary alkyl ammonium species, when able to compete with Cs$^+$ ions for the surface passivation of the NCs, generate platelet shaped particles or even layered structures. In the same study, it was observed that Ostwald ripening could be suppressed by reducing the concentration of ligands (to the minimum amount needed to solubilize PbBr$_2$ at a given temperature), and, as a result, CsPbBr$_3$ nanocubes from 4.0 nm to 16.4 nm with narrow size distribution (8 to 15%) could be prepared (Figure 5). Apart from varying the length of ligands and the amine/acid ratio, the control over the size of CsPbBr$_3$ NCs was also demonstrated by employing extra halide sources such as alkylammonium bromide or ZnBr$_2$ salts.[132-133] In the first case, the size of CsPbBr$_3$ nanocubes could be finely tuned from 17.5nm to 3.8nm by varying the amount of OLA-HBr, while keeping the reaction temperature and the ligand concentration fixed.[132] In the second case, Dong et al. demonstrated an excellent control over the size and size distribution of CsPbX$_3$ NCs by adjusting the reaction temperature and the ZnX$_2$/PbX$_2$ ratio in the reaction mixture.[133] Of particular interest here is that an excess of ZnX$_2$ was found to strongly influence the surface passivation of the resulting NCs, eventually leading to a high PLQY.

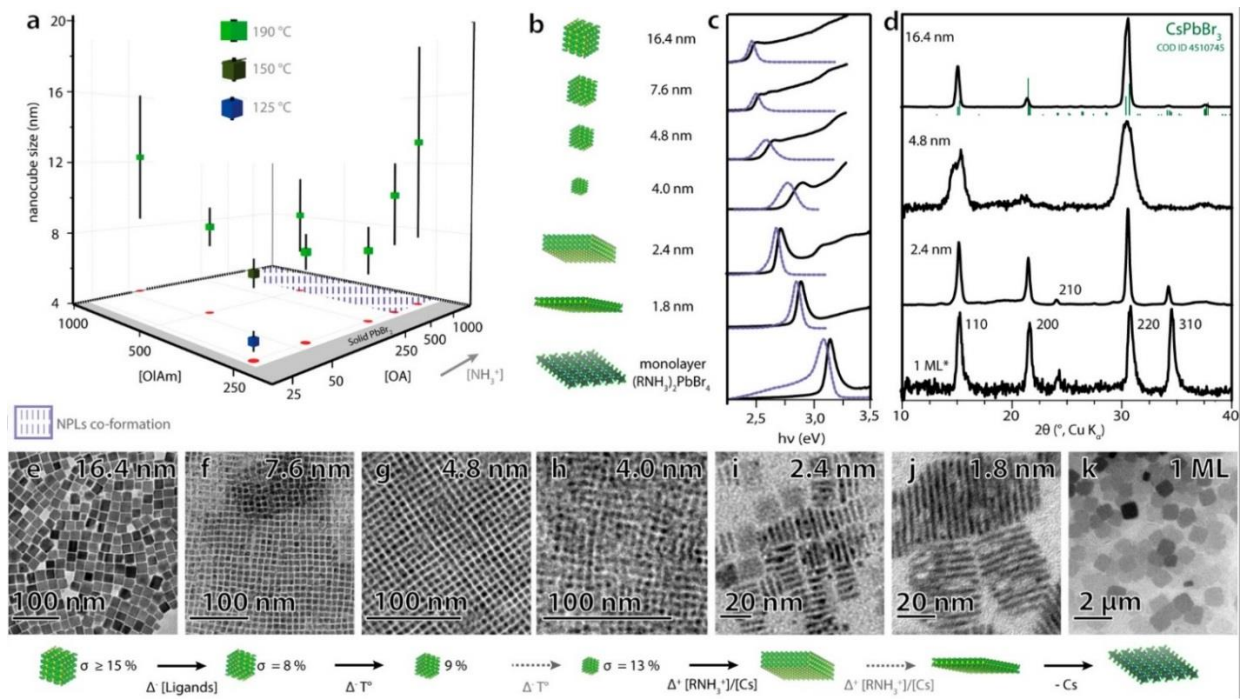

**Figure 5.** (a) Variation of the size of CsPbBr$_3$ nanocubes depending on the concentration of oleylamine (OlAm), OA and the reaction temperature (vertical bars represent size distributions). (b) Illustration of the different CsPbBr$_3$ nanostructures obtained using Olam and OA as ligands together with their corresponding: (c) absorbance (black lines) and photoluminescence (blue dashed line) spectra; (d) XRD patterns; (e-k) TEM images. Reproduced from ref. [108] Copyright 2016, American Chemical society.

A disadvantage of the HI method is that it is, in fact, a two-step approach as it requires, in the first step, the preparation of a complex (e.g. the Cs-oleate) and, in the second step, the actual reaction. Another disadvantage is that the Cs/FA-oleate precursors are solid at RT, and often require a pre-heating step (up to 100 °C) prior to their injection. It is also known that this strategy is hardly up-scalable (the injection of a large amount of precursor at high temperature results in a remarkable drop in the temperature as well as an inhomogeneous nucleation) preventing its use for a large scale production.[27] To overcome these limitations, different groups have developed alternative routes which rely on the same precursors, ligands and solvents used for the HI route, with the difference that all the chemicals are mixed together in one-pot and reacted using a heating mantle (heat-up or solvothermal approaches), or by ultrasonication or microwave irradiation.

Chen *et al.* reported the solvothermal synthesis of both CsPbX$_3$ nanocubes and NWs.[134] Briefly, CsPbX$_3$ nanocubes were synthesized by directly mixing precursors (such as cesium carbonate and lead halides salts) together with ligands, and the resulting mixture was heated up in an autoclave at the desired temperature for a certain amount of time. Ultrathin CsPbBr$_3$ NWs were instead obtained when pre-dissolved precursors (such as Cs-oleate and lead halide dissolved in ODE by using OA and OLA) were used. In 2016, Tong *et al.* first reported a single step ultra-sonication assisted synthesis to produce CsPbX$_3$ NCs with tunable halide composition, thickness and morphology (Figure 6).[120] The same authors, later on, further extended such procedure, prolonging the reaction time to produce CsPbX$_3$ NWs.[135] Similarly, CsPb(Br/I)$_3$ nanorods (NRs) were also prepared by adjusting the ratios of ligands (OLA/OA) and reaction temperature.[136]

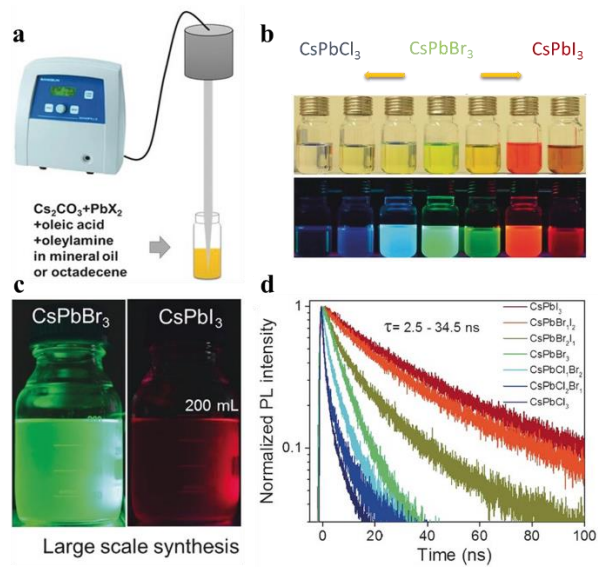

**Figure 6.** Single step ultra-sonication method: (a) schematic illustration of CsPbX$_3$ NCs synthesis; (b) colloidal dispersions of CsPbX$_3$ NCs with different halide compositions in hexane under room light (top) and UV light (bottom, $\lambda_{ex}$=367 nm); (c) photograph (under UV light) of CsPbBr$_3$ and CsPbI$_3$ NCs solutions obtained by scaling up the reaction, (d) PL decay curves of the samples shown in panel (b). Reproduced from ref. [120] Copyright 2016, Wiley-VCH.

In 2017 Pan *et al.* introduced the microwave irradiation strategy for the synthesis of CsPbX$_3$ NCs with tunable morphologies.[137] Shamsi *et al.*, that same year, employed a variant to this approach to produce quantum-confined blue-emitting CsPbBr$_3$ NPLs with a unimodal thickness distribution.[138] In this method, a certain amount of isopropanol can trigger the nucleation of unreactive precursors at RT. The NPLs were then grown by simply heating the solution in a microwave oven.[138] Recently, Liu *et al.* further optimized the microwave based strategy by introducing diethylene glycol butyl ether along with ODE, in order to enhance the absorption capacity of the microwave irradiation. It was observed that the dimensionality of CsPbBr$_3$ NCs could be tuned from cubes, NRs and NSs by adjusting the relative amounts of ligands, solvents and PbBr$_2$ salt.[139] Liu *et al.* and Long *et al.* later adopted the same strategy to prepare CsPbX$_3$ NCs.[140-141] Ye *et al.* proposed alternative one-step approaches for the synthesis of inorganic CsPbX$_3$ NCs:[120, 142] briefly, PbX$_2$ and CsCO$_3$ precursors were added to the mixture containing ODE, OA and OLA and the reaction mixture was heated up to 100°C for 15-30 minutes. In the same year, Yang *et al.* reported a heat-up approach for the large scale synthesis of ultrathin CsPbBr$_3$ NPLs with tunable dimensions by varying the reaction kinetics.[143] The thicknesses of the CsPbBr$_3$ NPLs was increased from 1.3 to 3.1 nm by increasing the temperature from 100 to 180 °C. In 2017, Zhai *et al.* adopted the solvothermal approach[134] to synthesize CsPbBr$_3$ NPLs and to transform them into Cs$_4$PbBr$_6$ NCs.[144]

All approaches discussed so far mainly rely on a binary ligands system composed of carboxylic acids (mainly OA) and alkyl amines. It is now a well-established fact that the surface of LHPs NCs is dynamically stabilized with either oleylammonium halide or oleylammonium oleate.[103, 289] However the final ligands composition strongly depends on the processing conditions, for instance upon the use of polar solvents during washing cycles the ammonium ligands are more prone than carboxylate groups to be detached from the surface, and this eventually modifies the final PLQY.[124] To address this issue Krieg *et al.* proposed a new capping strategy based on long chain zwitterionic molecules (i.e. 3-(N,N-dimethyloctadecylammonio)propanesulfonate). These molecule bind strongly to the NCs surface resulting in improved chemical durability of the material.[145] In particular, this class of ligands allows for the isolation of clean NCs with high PLQYs of above 90% after four rounds of precipitation/redispersion along with much higher overall reaction yields of uniform and colloidal dispersible NCs.[145]

### Mixed A/B cations engineering of ABX$_3$ NCs by HI

Inspired by the opportunity to produce LHP NCs with mixed halide composition, which allows for the tunability of their resulting band gap, various groups started to explore the possibility of preparing ABX$_3$ NCs using mixed A and B cations. For instance, in 2017 Amgar *et al.* and Wu *et al.* reported the synthesis of Cs$_x$Rb$_{1-x}$BX$_3$ by adjusting Cs$^+$ and Rb$^+$ precursor's ratios in the HI synthesis.[146-147] Interestingly, NC samples with higher fraction of the small Rb$^+$ ions had a higher band gap. In the same year, Wang *et al.* reported that the partial substitution of Pb$^{2+}$ with Sn$^{4+}$ ions not only enhanced the stability of the NCs, but also their optical properties.[148] Protesescu *et al.* and Wang *et al.*, almost in parallel, reported a HI based synthesis of mixed organic inorganic FA$_x$Cs$_{1-x}$PbBr$_{3-x}$I$_x$ and FA$_x$Cs$_{x-1}$PbI$_3$ (0 ≤ x ≤ 3) perovskite NCs, NSs and NWs.[67, 149] Both reports concluded that the introduction of the FA cation along with Cs in the A site considerably enhances the stability of these compounds. Vashishtha *et al.* demonstrated that conventional monovalent A cations, such as Cs, Rb, MA and FA, could be replaced by Tl$^{3+}$ ions. Tl$_3$PbX$_5$ NCs (X= Cl, Br, I) and TlPbI$_3$ NCs were prepared using the standard HI approach by replacing Cs-oleate with Tl-oleate.[150] The HI of Tl-oleate into the PbX$_2$ solution (130-175°C) resulted in faceted spheroidal Tl$_3$PbX$_5$ (X= Br, I) NCs with orthorhombic crystal structure whereas Tl$_3$PbCl$_5$ crystallize in the tetragonal phase.[150]

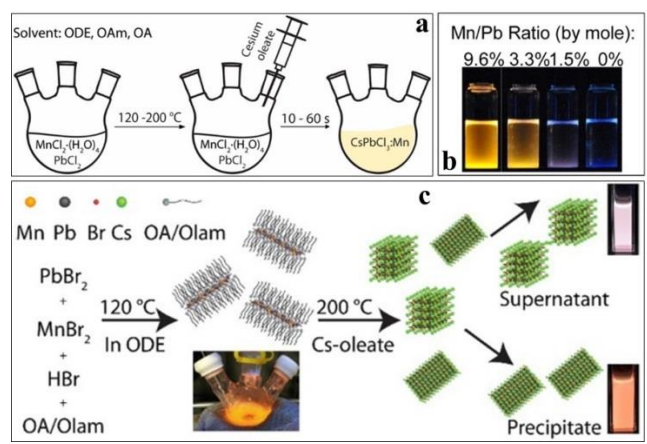

**Figure 7.** Illustrations of the HI approaches used for the preparation of (a) Mn-doped CsPbCl$_3$ (Reproduced from ref. 158 Copyright 2016 American Chemical Society) and (c) Mn-doped CsPbBr$_3$ NCs (Reproduced from ref. 160 Copyright 2018 American Chemical Society)). (b) Photograph of Mn doped CsPbCl$_3$ NCs with different Mn-content illuminated by a UV lamp (365 nm). Reproduced from ref. 157 Copyright 2016 American Chemical Society.

By simple modifications of the standard HI approach, lead halide based NC systems were successfully doped with either Mn$^{2+}$, Bi$^{3+}$ or rare earth (RE) ions in order to modify their optical properties.[151-156] Liu *et al.* and Parobek *et al.* almost simultaneously reported the HI synthesis of Mn$^{2+}$ doped CsPbCl$_3$ NCs with fine control over the doping content. In both cases, the incorporation of Mn$^{2+}$ ions was achieved by simply employing MnCl$_2$, in addition to PbCl$_2$, in the conventional HI method reported previously for the synthesis of CsPbX$_3$ NCs (Figure 7a, b).[157-158] Later on, Adhikari *et al.* further optimized the inclusion of Mn$^{2+}$ ions in CsPbCl$_3$ NCs by using RNH$_3$Cl in addition to MnCl$_2$ salt.[156, 159] The authors claimed that the alkyl ammonium chloride precursor could enable a precise control over the morphology as well as the Mn incorporation in CsPbCl$_3$ NCs. The subsequent year, a similar strategy was adopted and slightly modified by Parobek *et al.* to synthesized Mn doped CsPbBr$_3$ NCs.[160] Instead of using an alkyl ammonium halide precursor, the authors employed HBr that was initially mixed with $PbBr_2$, $MnBr_2$, OA and OLA to form $L_2[Pb_{1-x}Mn_x]Br_4$ (L = ligand) organometallic complexes. Such compounds, exhibiting a strong Mn fluorescence, were subsequently transformed into Mn doped $CsPbBr_3$ NCs by injection of the Cs precursor at high temperature (Figure 7c)[160]. Of particular interest is that doping of $CsPbX_3$ NCs with $Mn^{2+}$ ions was observed to increase their stability under ambient conditions and against thermal annealing.[161-162] Similarly, heterovalent dopants, such as $Ce^{3+}$ and $Bi^{3+}$ ions, were introduced into $CsPbBr_3$ NCs.[163-164] Milstein et al. recently reported Ytterbium-doped $CsPbCl_3$ NCs by the HI synthesis following their previously three-precursors approach for double perovskites.[165]

### Pb-free metal halide perovskite NCs

The first colloidal syntheses of lead free perovskite NCs *via* the HI approach were reported in 2016 by Jellicoe et al.[79] and Wang et al.[74] The former group was able to prepare $CsSnX_3$ NCs by reacting $SnX_2$ salts, dissolved in tri-n-octylphosphine (TOP) and subsequently injecting the resulting solution into a mixture of $CsCO_3$, OA and OLA at 170°C.[79] Wang at al., instead, fabricated $Sn^{4+}$-based perovskite NCs, with $Cs_2SnI_6$ composition, which exhibited a PL peak around 620 nm (2.0 eV) with full width at half-maximum of 49 nm (0.16 eV).[74] The variation of the reaction time enabled tuning the size and shape of $Cs_2SnI_6$ NCs, so that spherical quantum dots, NRs, NWs, nanobelts and NPLs could be selectively prepared (Figure 8).[74] The HI synthesis of Sn based compounds was, later on, slightly modified to access different shapes.[18, 76, 166] For example, Wong et al. demonstrated the synthesis of two-dimensional $CsSnI_3$ NPLs, with a thickness of less than 4nm,[18] by using a combination of long and short chain amines (OLA and octylamine) and a short chain carboxylic acid (octanoic acid). In 2018, Wu et al. reported ternary $CsGeI_3$ NCs by means of HI approaches.[167] NCs were synthesized by simply injecting Cs-oleate into a solution of $GeI_2$ dissolved in ODE, OA and OLA.[167]

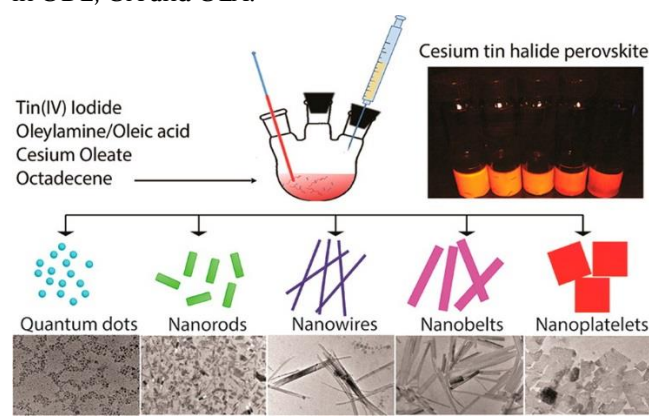

**Figure 8:** Schematic of the synthesis of perovskite $Cs_2SnI_6$ NCs, with the corresponding photographs of the as-prepared $Cs_2SnI_6$ samples under UV light and TEM images of $Cs_2SnI_6$ NCs with different shapes. Reproduced from ref. [74] Copyright 2016 American Chemical Society.

Many attempts have been made, indeed, to prepare Pb-free perovskite NCs. One could summarize them as variations of the initial synthesis of $CsPbX_3$ NCs, by replacing the $Pb^{2+}$ ions with other ions. Depending on the B-cation chosen as a replacement, the perovskite structure can either be retained, or a new structure is obtained. For example, the substitution of two $Pb^{2+}$ ions with one tetravalent ion, such as $Sn^{4+}$, disfavors the formation of the 3D structure and promotes the formation of a oD structure.[168] Similarly, the substitution of three $Pb^{2+}$ ions with two trivalent ions ($Sb^{3+}$) breaks apart the 3D network of metal halide $[MX_6]^{n-}$ octahedra and leads to the formation of a oD structure in which the constituent units are $[Sb_2X_9]^{3-}$ clusters, each made of two corner-sharing octahedra. As the distance between the clusters is larger along one crystallographic axis than along the other two, this structure can also be seen as a layered structure, made of layers of $[Sb_2X_9]^{3-}$ clusters.[168] Along this line, $Cs_3Sb_2I_9$ and $Rb_3Sb_2I_9$ NCs with different morphologies have been prepared using the HI synthesis route.[168-170] First-principles calculations have been extensively used to design stable Pb-free halide perovskites, based on the so-called "cation transmutation" idea, namely, the replacement of two $Pb^{2+}$ ions with one monovalent $M^+$ and one trivalent $M^{3+}$ ions, forming a $A_2M^+M^{3+}X_6$ double-perovskite structure. This strategy has uncovered various stable structures of quaternary halides with various choices for the monovalent and trivalent cations [80]

In 2018, "double-perovskite" $Cs_2AgBiX_6$ (X = Cl or Br) NCs were eventually synthesized by means of HI approaches.[167 118 171-172] Double perovskite in the form of NCs were reported, almost in parallel, by two groups using two different HI synthetic routes: Creutz et al.[118] employed, for the first time, trimethylsilyl halides which were injected at 140°C into a solution of metal acetates precursors (i.e. silver acetate, cesium acetate and bismuth acetate) dissolved in ODE, OA and OLA, triggering, immediately, the nucleation and growth of the NCs (Figure 9); on the other hand Zhou et al. first prepared a solution of $AgNO_3$, $BiBr_3$, ODE, OA, OLA and HBr followed by the injection of Cs-oleate at 200°C.[171]

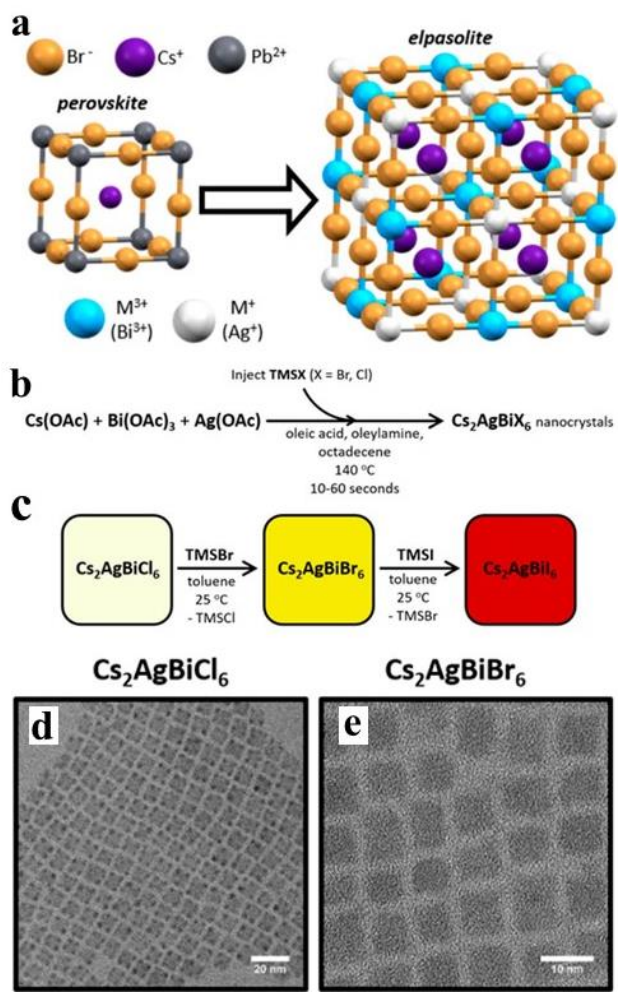

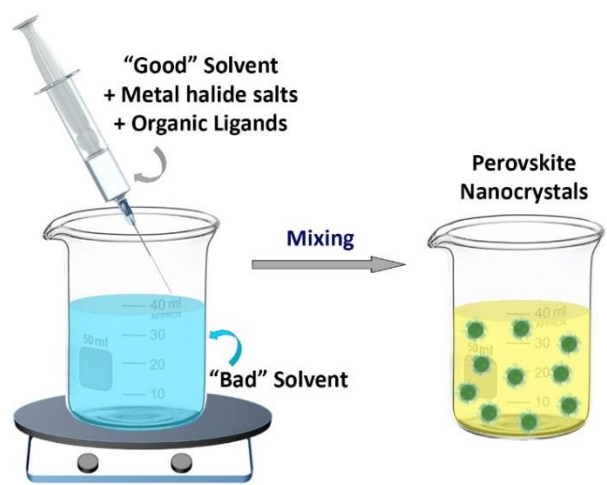

**Scheme 3.** LARP synthesis approach.

**Figure 9.** (a) Structure of double perovskite crystal structure (e.g. $Cs_2AgBiBr_6$). (b) HI reaction scheme used for the synthesis of $Cs_2AgBiX_6$ NCs and (c) their post-synthesis anion exchange reactions using trimethylsilyl halide reagents. TEM images of (d) $Cs_2AgBiCl_6$ and (e) $Cs_2AgBiBr_6$ NCs. Reproduced from ref. [118] Copyright 2018 American Chemical Society.

## LARP approach

The supersaturated recrystallization process dates back to more than 5000 years, when salt was recrystallized in clay pots in South Poland in 3500 BC.[173] This simple process consists in dissolving the desired ions in a solvent, reaching an equilibrium concentration, and, subsequently, in moving the solution into a non-equilibrium state of supersaturation. The supersaturated state can be reached, for example, by varying the temperature (cooling down the solution), by evaporation of the solvent or by addition of a miscible co-solvent in which the solubility of the ions is low. Under such conditions, spontaneous precipitation and crystallization reactions occur until the system reaches an equilibrium state again. This technique was also successfully extended, in the nineties, to prepare organic nanoparticles and polymer dots.[174-175] If this process is carried out in the presence of ligands, thus named ligand assisted reprecipitation (LARP), the formation and growth of crystals can be controlled down to the nanoscale, allowing for the fabrication of colloidal NCs (Scheme 3).

The LARP technique, applied to perovskite systems, simply consists of dropping the desired precursors salts ($MX_2$ (M=Pb, Sn, etc.), CsX, MAX and FAX, where X=Cl, Br, I), dissolved in a good polar solvent (such as DMF, DMSO etc.) into a very poor solvent (such as toluene, hexane) in the presence of ligands (typically alkylamines, carboxylic acid, alkyl thiols). The mixture of the two solvents induces an instantaneous supersaturation, which triggers the nucleation and the growth of perovskite NCs. As in the hot-injection or heat-up methods, described in the previous sections, also in the LARP approach nucleation and growth stages cannot be separated in time.[102] The first reports on the LARP synthesis of hybrid organic-inorganic lead halide based perovskite NCs date back to 2012 when Papavassiliou et al. dropped solutions of $MAPbX_3$, $(MA)(CH_3C_6H_4CH_2NH_3)_2Pb_2X_7$ or $(MA)(C_4H_9NH_3)_2Pb_2X_7$ (X=Br, Cl or I) dissolved in DMF (or acetonitrile) into toluene (or a mixture of toluene and PMMA).[176] What they observed was the formation of luminescent NCs, with sizes of the order of 30-160 nm, and PL intensities much higher than those expected for the bulk $MAPbX_3$ counterparts. It took few more years before the LARP procedure was finally developed to synthesize organic-inorganic $MAPbX_3$ NCs and then extended to $ABX_3$ NC systems (A= MA, FA or Cs; B= Pb, Sn, Bi or Sb; X=Cl, Br, I). We will briefly describe here, for each system, the LARP approaches proposed so far and their current understanding.

### $MAPbX_3$ NCs

The first LARP approach to organic-inorganic $MAPbBr_3$ NCs was demonstrated and systematically studied in 2015 by Zhang et al.[177] In that work, the authors prepared a clear precursor solution by dissolving $PbBr_2$ and MABr salts in alkyl amines, carboxylic acids and DMF. A fixed amount of this solution was, then, dropped into toluene at RT under vigorous stirring to form colloidal NCs. In order to understand the specific role of alkyl amines and carboxylic acids, different amines and acids were tested: n-octylamine, dodecylamine, hexadecylamine, or hexylamine and OA, octanoic acid, or butyric acid. It was observed that the formation of $MAPbBr_3$ NCs could be achieved even without the use of amines, but with no control over the size of the crystals. On the other hand, the exclusion of carboxylic acids from the synthesis resulted in aggregated NCs. Based on these control experiments, the authors concluded that the role of amines was to regulate the kinetics of the crystallization, and thus the NCs' size, while the organic acids were thought to suppress the NCs aggregation. The same approach, in the years that followed, was exploited and rapidly further optimized by different groups.[178-180] Huang et al., for example, demonstrated a better size and size distribution control of the MAPbBr$_3$ NCs by introducing the use of OLA together with OA and varying the reaction temperature (by heating toluene up to 60°C).[178] Arunkumar et al. demonstrated the possibility of doping MAPbX$_3$ NCs with Mn$^{2+}$ simply by adding also MnCl$_2$ in the precursors solution.[179]

Other ligands were subsequently proved to work efficiently in the LARP synthesis of MAPbBr$_3$ NCs. Gonzalez-Carrero et al. showed that OLA and OA could be replaced by 2-adamantylammonium bromide as the only capping ligand to improve the optical properties of the final NCs. Luo et al. demonstrated size tunability of the MAPbX$_3$ (X=Cl, Br or I) NCs by employing two branched capping ligands, (3-aminopropyl) triethoxysilane (APTES) and polyhedral oligomeric silsesquioxane (POSS) PSS-[3-(2-aminoethyl)amino]propylheptaisobutyl-substituted (NH$_2$-POSS). [181] The authors proposed that, compared to straight-chain ligands, APTES and NH$_2$-POSS offer greater control over NCs size, as they are able to protect the formed NCs from dissolution by DMF.[25] Veldhuis et al. introduced the use of benzoyl alcohol as an auxiliary ligand (together with octylamine and OA), which was observed to accelerate the reaction kinetics and to improve the optical properties of the resulting NCs.[182] Luo et al. employed peptides, namely 12-aminododecanoic acid, as the only ligand in the synthesis of MAPbBr$_3$ NCs. Peptides, having both -NH$_2$ and -COOH groups, were found to lead to a good control over the size of the resulting NCs.[183] Minor modifications of the LARP process were also proposed in order to further optimize the process. Shamsi et al., for example, proposed an alternative LARP approach in which PbX$_2$ salts are dissolved in N-methylformamide (NMF), rather than the typical DMF, together with OLA and OA, heated up to 100°C for 10min, and eventually added dropwise at RT to a poor solvent (such as dichlorobenzene or chloroform).[184] The advantage of this approach is that MA$^+$ ions are formed in situ, during the heating step (by a transamidation reaction) with no need for previously synthesizing MAX salts.[184] It is worth mentioning that, the same approach could also yield bulk crystals at RT with no need for an antisolvent (Scheme 4).[184]

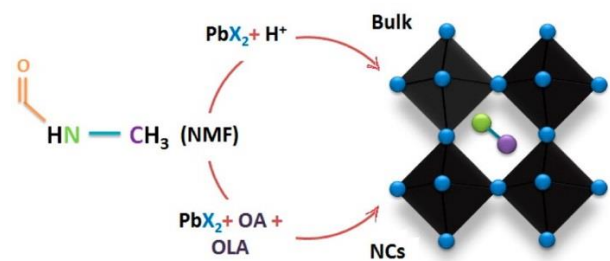

**Scheme 4.** Two different synthetic routes, both employing NMF as the source of MA$^+$ ions, to produce either perovskite NCs or bulk crystals. Reproduced from ref. [184] Copyright 2016, American Chemical Society.

Dai et al. performed the mixing of good and non-good solvents by means of a spray, producing MAPbBr$_3$ NCs with a good size distribution.[185] In this peculiar approach, the precursor's solutions (MABr, PbBr$_2$, OA, Octylamine in DMF) was sprayed into a poor solvent (toluene). The micrometer-size droplets of the sprayed solution is believed to provide a large contact surface area between the two solutions, allowing for a more homogenous mixing.

### Size and shape control over MAPbX$_3$ NCs

Soon after the development of the LARP synthesis of MAPbX$_3$ NCs, efforts were made to optimize the PL of such materials, and, additionally, to control their size and shape. Sichert et al. in 2015 were the first who reported the LARP synthesis of MAPbBr$_3$ NPLs in which no carboxylic acid was employed, but only octylammonium and MA$^+$ ions were used as ligands.[17] The systematic increase of the octylammonium/MA$^+$ ratio was observed to reduce the thickness of the resulting NPLs, with the limiting case being the use of octylammonium only (octylammonium/MA$^+$ →∞) where "single layer" NPLs were observed, similar to what observed in the studies on layered perovskite macrocrystals.[186] Similar results were achieved the year after by different groups. Kumar et al., for example, demonstrated control over the MAPbBr$_3$ NPLs thickness down to one monolayer, working with OA and octylamine.[34] Cho et al., on the other hand, reported a systematic study on the LARP synthesis of MAPbBr$_3$ NPLs in which they used OA and tested many alkylamines having different chain lengths (such as butylamine, hexylamine, octylamine, dodecylamine, and OLA).[187] They revealed that both the chain length and the concentration of alkylamines played a fundamental role in determining the thickness of MAPbBr$_3$ NPLs. In details, a high concentration of alkylammonium cations can efficiently passivate the surface of the MAPbBr$_3$ NCs, precluding their growth along the vertical direction and yielding NPLs of tunable thickness. In this context, longer chain amines have better electrostatic interaction with the perovskite NCs, providing a stronger passivation. In 2017, Levchuk et al., working with OLA and OA, but using chloroform as the bad solvent, extended the thickness-tunability control over MAPbI$_3$ NPLs.[188] Eventually, Ahmed et al. proposed the use pyridine, together with OLA and OA, as an effective co-ligand for finely tuning the thickness of MAPbBr$_3$ NPLs (Figure 10).[189] As emerged from their DFT calculations, pyridine molecules are able to bind to Pb$^{2+}$ ions present on the surface of the growing nanostructures, forming dative N → Pb bonds, slowing down the vertical growth rate and, thus, leading to the formation of 2D nanostructures.[189]

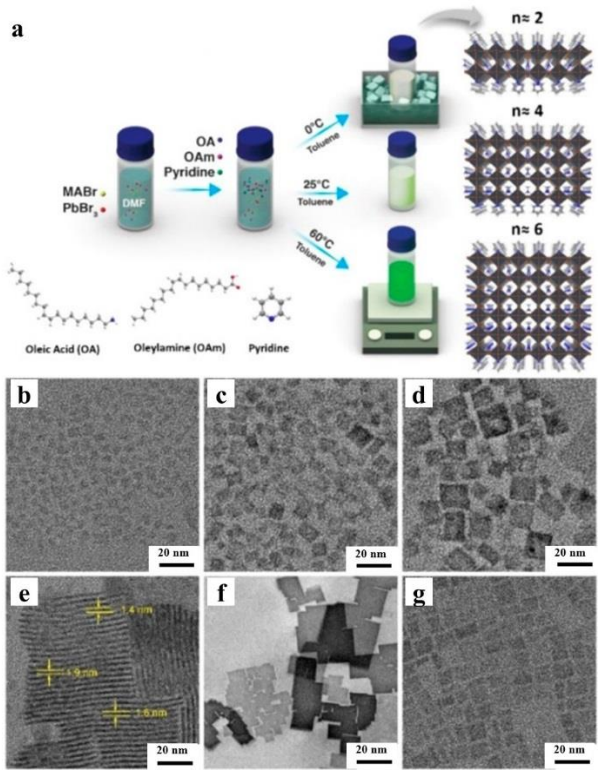

**Figure 10.** (a) Schematic illustration of the LARP synthesis of MAPbBr$_3$ nanostructures using oleic acid (OA), oleylamine (OLA), and pyridine as ligands. Representative TEM images of MAPbBr$_3$ NCs synthesized without (b-d) and with (e-g) pyridine at different precipitation temperatures: 0 °C (b,e), 25 °C (c,f), and 60 °C (d,g), respectively. Adjusted with permission from ref. [189] Copyright 2017, American Chemical Society.

It is important to underline that in all these works no control over the lateral dimensions of the NCs was demonstrated. With the aim of addressing this problem, Kirakosyan *et al.* revealed that the way the precursors solution (ionic salts dissolved in DMF) is added to the bad solvent influences the size and shape of the final NCs.[190] By varying the amount of added precursor's solution (from 1 to 8 drops at a constant rate of ~45 drops/min), they could tune the lateral size of MAPbBr$_3$ NPLs from ~3 to 8 nm, while keeping the thickness unchanged (2.5–3 nm). In the same year, Huang *et al.* performed a systematic study on how the amount of ligands (OLA and OA), the precursors/ligands ratio and the reaction temperature affect MAPbBr$_3$ NCs NCs.[191] Similarly to what reported by Cho *et al.*,[187] a fine tuning of these parameters resulted in a control over the size, and, thus, of the quantum confinement of NCs. More in detail, working at a high ligands/precursor ratio, they observed the formation of small NCs, while by employing a high precursors/ligands ratio they obtained big and polydisperse micro- and nano-crystals. Eventually, at intermediate ligands/precursors ratios, the nucleation and growth of NCs could be better controlled: the synthesis in such conditions produced small NCs that grew over time.

The LARP procedure, initially developed for MAPbX$_3$ NCs, was soon extended to all-inorganic CsBX$_3$ (B=Pb$^{2+}$, Bi$^{3+}$ and Sb$^{3+}$), Cs$_4$PbX$_6$, CsPb$_2$Br$_5$ and organic-inorganic FAPbX$_3$ NC systems. We discuss the achievements reported for the synthesis of each system in separate sections.

### FAPbX$_3$ NCs

In 2016 Weidman *et al.* reported for the first time the LARP synthesis of both all-inorganic and organic-inorganic ABX$_3$ (A= Cs, MA or FA; B= Pb, Sn) NPLs having 1 or 2 monolayers thickness.[192] In their work, the precursors (AX and BX$_2$ salts) were dissolved in DMF together with octylammonium and butylammonium halides (used in a 1:1 ratio). The precursor solution was added dropwise to toluene under vigorous stirring at RT to trigger the immediate formation of the NCs. To achieve colloidal stability and thickness homogeneity an excess of ligands was used (in a 10:2:1 ligands/BX$_2$/AX).[192] Few months later, Perumal *et al.* reported the synthesis of FAPbBr$_3$ NCs. The focus in their work was mainly to achieve bright PL emission rather than a tight control over the particles' shape.[193] In their approach, FABr and PbBr$_2$ are dissolved in DMF, and added dropwise into a solution consisting of toluene, butanol, octylamine and OA under constant stirring.

The first LARP report on all the FAPbX$_3$ (X = Cl, Br, I) NC systems was reported by Levchuk *et al.* who used a modified version of the synthesis of MAPbX$_3$ NCs that they published some months before,[188] with minor modifications.[194] The synthesis relies on the rapid injection of a precursors solution, prepared by dissolving PbX$_2$ and FAX (X = Cl, Br, I) salts in DMF, OA and OLA, into chloroform at RT (Figure 11 a, b). Depending on the OLA/OA ratio, the authors could produce either nanocubes or NPLs with control over their thickness (ranging from 2 to 4 monolayers, Figure 11 c,d). Interestingly, the use of toluene was observed to prevent the formation of FAPbI$_3$ NCs, and to lead to the immediate agglomeration of FAPbBr$_3$ or FAPbCl$_3$ NCs.

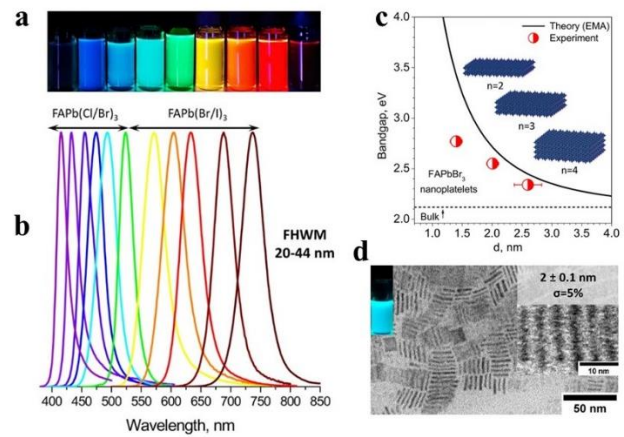

**Figure 21.** (a) Photograph of FAPbX$_3$ NC dispersions under UV-light and (b) the corresponding PL emission curves. (c) Theoretical effective mass approximation and experimental band gaps of FAPbBr$_3$ NPLs as a function of their thickness. (d) TEM characterization of vertically stacked FAPbBr$_3$ NPLs.

Almost in parallel, Minh *et al.* reported the synthesis of FAPbX$_3$ nanocubes introducing PbX$_2$–DMSO complexes as novel precursors.[195] In their approach all the precursors (the PbX$_2$–DMSO complex and FAX) were dissolved in DMF together with OLA, and the corresponding solution was added into a mixture of toluene and OA. In this case, the formation of FAPbX$_3$ NCs is believed to occur via an intramolecular exchange reaction in which alkylammonium halides can replace DMSO molecules in the starting PbX$_2$–DMSO complex. Good control over the size distribution of the NCs was demonstrated by varying the relative amount of OLA.[195] Eventually, Kumar *et al.* reported a slightly modified LARP procedure to get FAPbBr$_3$ NPLs with very high PL QY to be used in LED devices.[196] The difference here is that FABr and PbBr$_2$ salts are dissolved in ethanol and DMF, respectively to form two different polar solutions, which were added simultaneously to a mixture of toluene, OA and octylamine.[196]

### All-inorganic CsPbX$_3$ NCs

Until 2015 the synthesis of all-inorganic CsPbX$_3$ NCs could be achieved only by the hot-injection approach.[27] In 2016 Li *et al.* showed for the first time that all-inorganic perovskite systems could be produced in the form of colloidal NCs also by the LARP approach at room temperature (RT).[197] The synthesis resembles the one used for the organic-inorganic MAPbX$_3$ NC systems, with the only difference being the use of CsX rather than MAX in the precursors solution.[197] Also in this case the inorganic salts, CsX and PbX$_2$, are dissolved in DMF together with OLA and OA and added to toluene to achieve the instantaneous formation of NCs. Few months later, Seth *et al.* further engineered this approach to achieve control over their morphology.[198] What they found is that the shape of the NCs could be controlled by varying the "non-good" solvent from toluene to ethyl acetate, the relative amount of ligands and the reaction time: the use of ethyl acetate promotes the formation of quasi-cubic QDs, NPLs or nanobars, while toluene can be used for the preparation of nanocubes, NRs or NWs.[198] The authors tentatively attributed these different morphologies to the different ligands/non-polar solvent interaction. Being ethyl acetate more polar than toluene, it can act both as a solvent and a nucleophile, provoking the detachment of a fraction of OLA molecules from the surface of the growing nuclei, and, thus, leading to the oriented attachment of NCs (Figure 12 a). On the other hand, in toluene, if a proper amount of OLA is used, the NCs surfaces become more protected in all directions, which consequently prevents both their attachment/merging and growth. Conversely, in the presence of a small amount of OLA in toluene, the NCs can start growing anisotropically at longer reaction times, in the form of NRs and NWs, most likely because of incomplete passivation of some facets (Figure 12 b).

In 2017, Kostopoulou *et al.* reported the LARP synthesis of CsPbBr$_3$ NWs with micron-size lengths.[199] The key feature of their approach consists in the use of anhydrous solvents and a low temperature: the precursors solution (CsBr, PbBr$_2$, DMF, OLA and OA) is dropped into anhydrous toluene kept in an iced water bath. Immediately after the injection, the resulting product consists in small-length, bullet-like NRs, which evolve after 24h at RT into NWs of 2.6 nm width, and finally, after one week, to thicker NWs having a diameter of 6.1 nm (Figure 12 c-e). Eventually, in 2018, Zhang *et al.* voluntarily introduced water in the LARP synthesis of CsPbBr$_3$ NCs demonstrating its influence on the growth rate and the shape of the resulting crystals.[200] The authors suggested that both H$_3$O$^+$ and OH$^-$ may act as surface ligands with higher activity as compared to oleylammonium and oleate species, leading, consequently, to different growth directions of the perovskite NCs.[200]

### Mixed A-cations APbX$_3$ NCs

The interest on tuning the spectral response (*i.e.* the bandgap) of APbX$_3$ NCs motivated various groups not only to vary the halide composition of those materials, but also the A cations composition.[201-202] Several LARP strategies have been reported so far for the synthesis of LHP NCs with control over the composition of the A site. Mittal *et al.*, for example, successfully tuned the bandgap of APbBr$_3$ NCs from 2.38 to 2.94 eV by varying the composition of the A site from pure MA to pure ethyl ammonium (EA), that is, synthesizing (EA)$_x$(MA)$_{1-x}$PbBr$_3$ NCs.[203] In 2017 two different groups reported the synthesis of Cs$_{1-x}$FA$_x$PbX$_3$ NCs:[204-205] Xu *et al.* mixed Cs and MA ions to form MA$_{1-x}$Cs$_x$PbBr$_3$ NCs,[206] while Zhang *et al.* reported the LARP synthesis of mix-organic-cation FA$_x$MA$_{1-x}$PbX$_3$ NCs achieving a continuously tunable PL emission from 460 to 565 nm.[207]

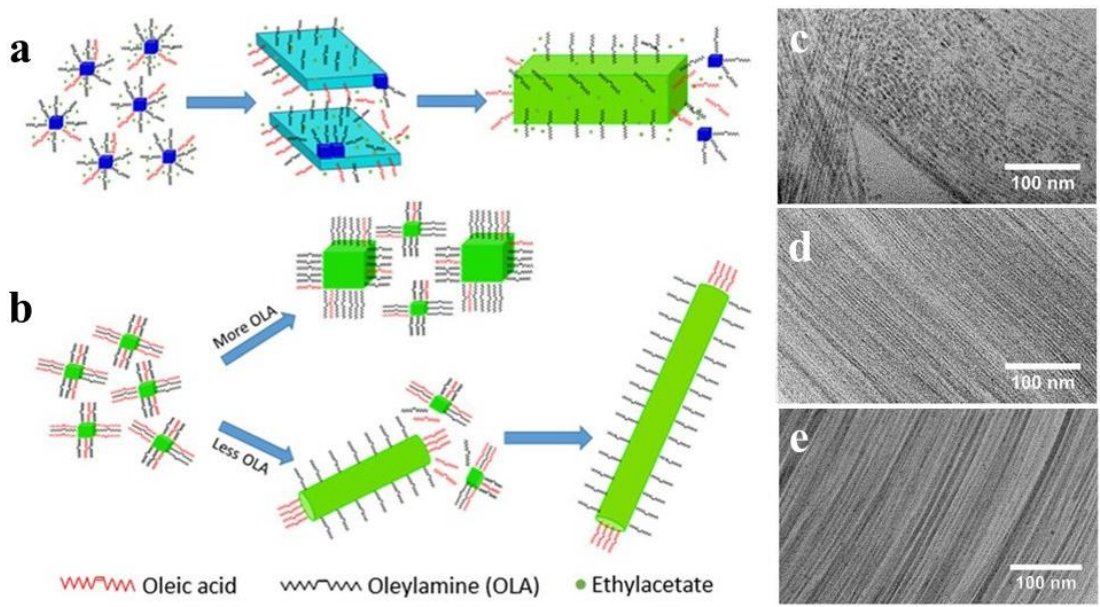

**Figure 32.** Sketch of the mechanism proposed for the formation of (a) NPLs (cyan) and nanobars (green) in ethyl acetate and (b) larger nanocubes, nanorods and NWs from smaller nanocubes in toluene. Reproduced from ref. [198] Copyright 2016 Macmillan Publishers Limited. (c-e) TEM images of $CsPbBr_3$ NWs at (c) 0 day, (b)1 day and (c) 7 days after leaving the colloidal toluene-based solution at RT without stirring. Reproduced from ref. [199] Copyright 2017, Royal Society of Chemistry.

### $CsPb_2Br_5$ NCs

Wang *et al.* in 2016 reported the first colloidal synthesis of perovskite-related $CsPb_2Br_5$ NPLs[208] adopting the LARP approach proposed by Sichert *et al.*.[17] In their work, they prepared two different precursors solutions in DMF, one containing $PbBr_2$ and hexylammonium bromide and the second one CsBr, that were added to toluene to trigger the formation of the final NCs.[208] The authors also reported that such synthetic procedure could not be used for the preparation of pure phase $CsPb_2Cl_5$ and $CsPb_2I_5$ NCs by substitution of bromide salts ($PbBr_2$, CsBr and HABr) with the corresponding chloride or iodide ones.[208]

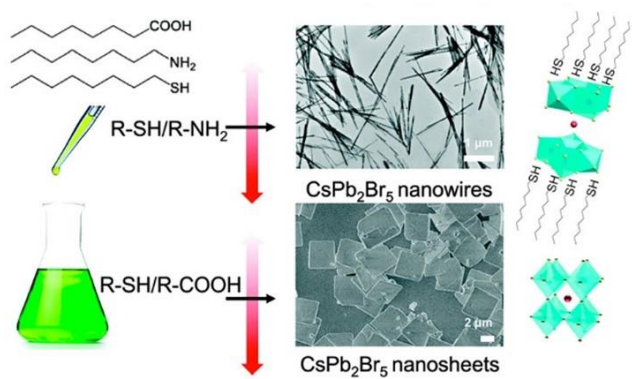

**Figure 13.** LARP synthesis of $CsPb_2Br_5$ NWs and NSs. Reproduced from ref. [209] Copyright 2017 Royal Society of Chemistry.

Ruan *et al.* subsequently optimized the LARP strategy in order to control the morphology of $CsPb_2Br_5$ NCs.[209] This was achieved by introducing, for the first time, alkyl thiols (octanethiol) as ligands together with either alkylamines (OLA) or carboxylic-acids (Figure 13).[209] More precisely, under their experimental conditions $CsPb_2Br_5$ NWs were obtained by employing octanethiol and OLA, while $CsPb_2Br_5$ NSs required the use of alkyl-thiols and carboxylic-acids.[209] The year after, the same authors demonstrated that their approach could be generalized for the direct synthesis of tetragonal $CsPb_2X_5$ (X = Cl, Br, I) NWs, even in mixed halide compositions.[210]

### Pb-free perovskite-related NCs

Compared to the high number of works reporting colloidal synthesis of lead free perovskite NCs by the HI approach, these compounds have been produced by the LARP strategy only in few cases. Leng *et al.* proposed, in 2016, the LARP synthesis of $MA_3Bi_2X_9$ (X=Cl, Br, I) perovskite-related NCs,[211] and, recently, that of $Cs_3Bi_2X_9$ (X=Cl, Br, I) NCs.[81, 84] In their first work, octane was employed as the bad solvent, while DMF and ethyl acetate were used as good solvents for MAX and $BiX_3$ salts, respectively.[211] In the second work, the authors used DMSO to solubilize both CsX and $BiX_3$ salts and ethanol to trigger the recrystallization (Figure 14a, b).[81] As in the case of Pb-based perovskite NCs, the amount of surfactants (octylamine and OA) was observed to play a major role in controlling the LARP synthesis of $Cs_3Bi_2X_9$ NCs, in which, interestingly, the use of toluene as bad solvent, instead of ethanol, did produce unstable NCs solutions.

Zhang *et al.* in 2017 adopted the LARP method to prepare blue emitting $Cs_3Sb_2Br_9$ NCs. In that case $SbBr_3$, CsBr salts were dissolved in DMF or DMSO in the presence of OLA to form a clear precursors solution, which was dropped into a mixture of octane and OA (Figure 14c,d).[77]

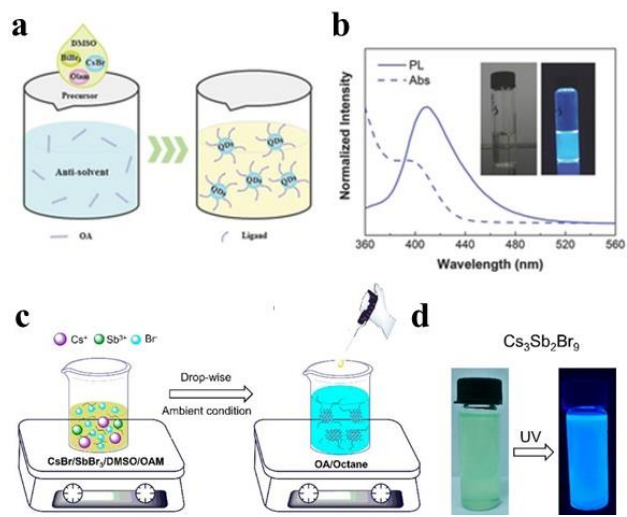
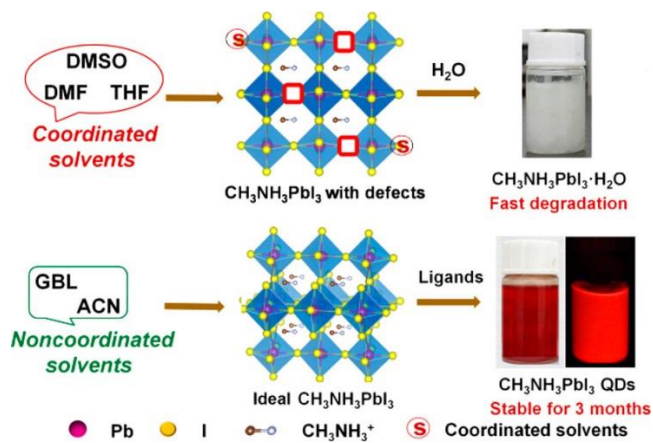

**Figure 14.** Illustration of the LARP approaches used for the synthesis of (a) $Cs_3Bi_2Br_9$ and (c) $Cs_3Sb_2Br_9$ NCs. Photographs of vials containing colloidal (b) $Cs_3Bi_2Br_9$ and (d) $Cs_3Sb_2Br_9$ NCs dispersions with and without UV light excitation. (b) Absorption and PL spectra of $Cs_3Bi_2Br_9$ NCs solution. Panels (a,b) are reproduced with permission from ref.[81] Copyright 2017 WILEY-VCH. Panels (c,d) are reproduced with permission from ref.[77] Copyright 2017 American Chemical Society.

### Disadvantages (or weak points) of LARP

While enabling the direct synthesis of many different perovskite systems at RT under air, the LARP method was also observed to have some weak points. Perovskite NCs are very sensitive to polar solvents, hence polar solvents that are normally used in the LARP, for example DMF, could easily degrade and even dissolve the $CsPbX_3$ NCs, especially the $CsPbI_3$ ones.[177, 212-213] Precursors-polar solvent interactions, indeed, were observed to play an important role in the formation of defective perovskite NCs. To elucidate this point, also called the "solvent effect", Zhang *et al.* investigated the effects of the use of different polar solvents in the crystallization of $MAPbI_3$ NCs.[214] What they proved is that $PbI_2$ generates stable intermediates in coordination solvents like DMSO, DMF, and THF, while it does not form complexes when dissolved in non-coordinating solvents, like γ-butyrolactone and acetonitrile.[214]

**Figure 15.** Effects of the solvents on the crystal structure of resulting $CH_3NH_3PbI_3$ NCs: the use of coordinated solvents (top) leads to the formation NCs crystal defects, which are prone to degradation under humidity; noncoordinated solvents (bottom) allow for the formation of "defect-free" and stable NCs. Reproduced from ref.[214] Copyright 2017 American Chemical Society.

These different binding motifs have, in turn, different impacts on the crystal structure of the synthesized NCs: the strong bonding between $PbI_2$ and coordinating solvents leads to the formation of defective $MAPbI_3$ NCs containing residual solvent molecules on the surface and iodine vacancies in the bulk; when non-coordinating solvents are employed, $PbI_2$ units are able to crystallize into defect free $MAPbI_3$ NCs (Figure 15). Furthermore, DMF and DMSO, typically polar solvents used in this approach, have a higher boiling point and are toxic, which are features that are very disadvantageous to large-scale production.

### Alternative approaches: Emulsion LARP, Microemulsion and Polar Solvent Controlled Ionization

While the "typical" LARP synthesis relies on the use of miscible polar and non-polar solvents, the emulsion techniques are based on the use of non-miscible liquids. The precursor's salts are dissolved in two different immiscible solvents with ligands and mixed together to form an emulsion. The reaction does not occur until, in a second step, a third solvent, also called demulsifier (which is miscible with both the two previous solvents), is added, immediately driving the system to a state of supersaturation, with the consequent precipitation and nucleation of the $APbX_3$ NCs.[215, 216] We will refer to this process as emulsion LARP. On the other hand, it is also known that ligands in the polar-non polar systems can self-assemble into ordered structures called micelles.[215-216] In the microemulsion technique, such micelles are used as nanoreactors in which the precursors can diffuse in and locally react to form the NCs. The process is termed "microemulsion approach" if the micelles form within a bulk polar phase where the hydrophobic carbon chains are turned inward to help stabilize the oil phase. Opposite conditions characterize the "reverse microemulsion" approach, in which the main solvent is non-polar and the micelles form with the hydrophilic head groups turned inward to help stabilize the aqueous phase (reverse micelles).[217] In such syntheses, NCs nucleation and growth is controlled by the diffusion of the precursors from the main phase to the interior of the micelles.[217] Once the NCs are formed inside the micelles, the addition of a demulsifier enables the precipitation and the cleaning of the product. It is worth pointing out that, in some cases, the difference between the emulsion LARP approaches and the microemulsion (or inverted microemulsion) one is very subtle. Indeed, if the emulsion is not properly characterized before the addition of the demulsifier, it is not always clear if the NCs nucleate inside the micelles before being precipitated of if their formation takes place in the very same moment the demulsifier is added to the solution.

### Emulsion LARP

In 2015, Huang *et al.* demonstrated for the first time that the emulsion LARP strategy could be used for the synthesis of MAPbBr$_3$ NCs.[218] More in details, they first prepared a solution of MABr and PbBr$_2$ in DMF that was mixed with hexane, OA and octylamine to form an emulsion. The subsequent addition of tert-butanol or acetone, acting as demulsifiers, was used to initiate the recrystallization process with the formation of the NCs (Scheme 5).

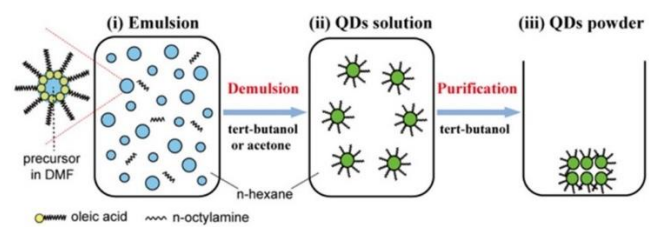

**Scheme 5.** Steps characterizing the emulsion LARP synthesis: (i) formation of the emulsion, (ii) demulsion by adding a demulsifier with the concomitant formation of perovskite NCs, (iii) purification of the NCs. Reprinted with permission from ref. [218] Copyright 2015 American Chemical Society.

The year after, Sun *et al.*, in parallel with Li *et al.*,[197] reported the emulsion LARP synthesis of CsPbX$_3$ NCs.[219] They employed a solution of Cs-oleate dissolved in ODE, which was added to a solution of PbX$_2$ in DMF together with an alkylamine and a carboxylic acid. Such emulsion was, then, added dropwise to toluene to start the nucleation process. In the same study, the authors tested different combinations of alkylamine and carboxylic acids, and found that the systematic variation of the length of the alkyl chains of both the amine and the carboxylic acid resulted in the variation of the NCs' shape/size, similar to what observed for organic-inorganic systems.[187] The use of hexanoic acid and octylamine was observed to produce spherical QDs; oleic acid and dodecylamine shaped the NCs into nanocubes; acetate acid and dodecylamine promoted the formation of NRs, while OA and octylamine produced NPLs (Figure 16). Akkerman *et al.* synthesized CsPbBr$_3$ NPLs using a similar approach with some modifications: they solubilized Cs-oleate in ODE together with OA, OLA and HBr, while PbBr$_2$ was dissolved in DMF. The two solutions were mixed together forming an emulsion which was inert until acetone (which is miscible with both the two previous solvents) instantaneously induced the formation of perovskite NPLs.[19] Interestingly, the thickness of the NPLs could be finely tuned, down to only 3-5 monolayers, by varying the amount of HBr. A similar approach was then used to produce CsPbBr$_3$ NC inks for high-voltage solar cells.[29]

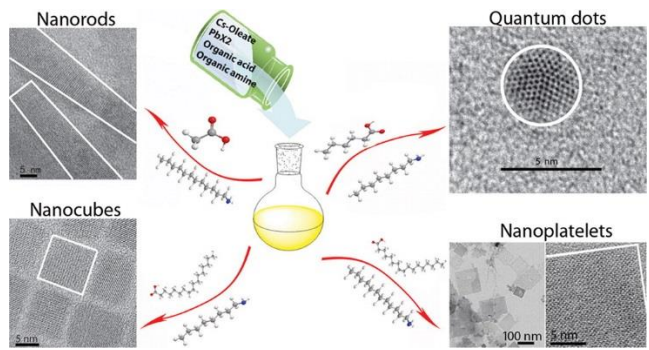

**Figure 16.** Different shapes of CsPbB$_3$ NCs that can be achieved via the emulsion LARP approach at RT by varying the organic acid and amine ligands. Reprinted with permission from [219] Copyright 2016 American Chemical Society.

The same strategy was then used to synthesize ultrathin CsPbBr$_3$ nanowires and MAPbX$_3$ nanostructures.[220] [221] The latter were synthesized by Liu *et al.* who prepared an oil phase solution by mixing OA, ODE and hexadecylamonium halide, and an aqueous phase by dissolving MAX and PbX$_2$ in different volumes of DMF. After mixing the two phases at RT, chloroform or acetone were used as demulsifiers, to quench the reaction and to collect the resulting nanostructures. Interestingly, the amount of DMF used in the aqueous phase was observed to control the shape of the final nanostructures (Figure 17).[221]

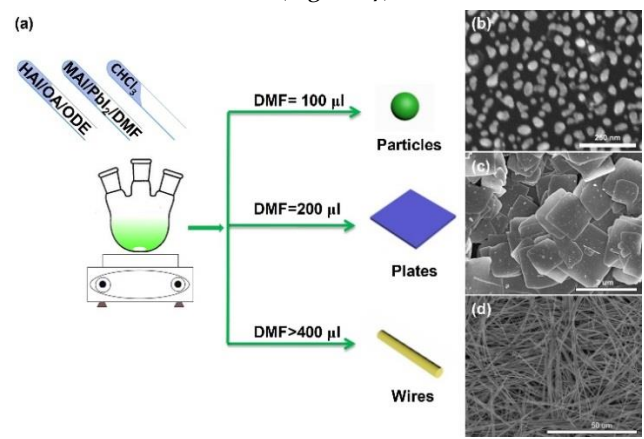

**Figure 17.** Schematic representation of the emulsion LARP synthesis of shape-controlled MAPbI$_3$ perovskites by employing hexadecylamonium (HA) and varying the DMF volume, together with (b-d) the corresponding SEM pictures. Reprinted with permission from ref. [221] Copyright 2017 Elsevier.

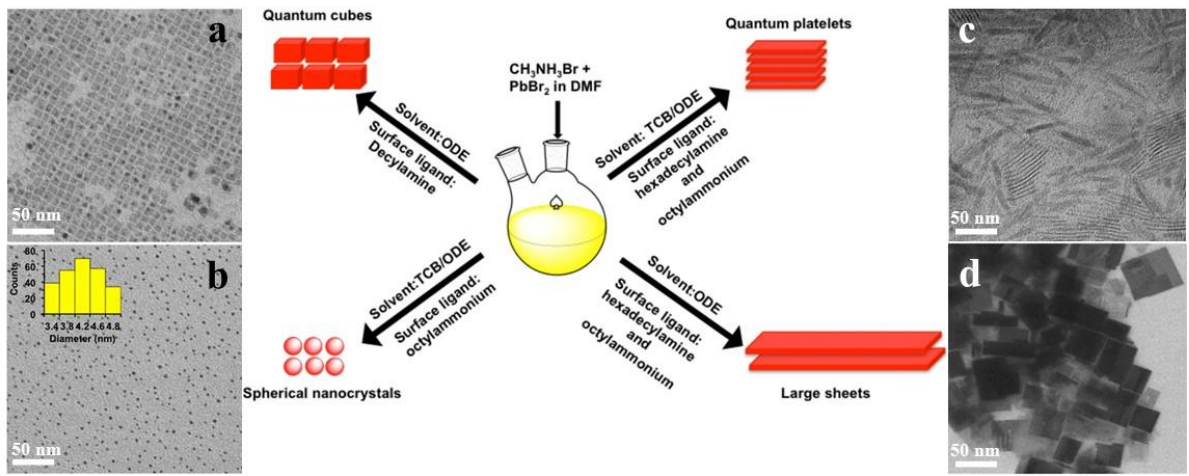

**Figure 18.** Shape control in the reverse microemulsion synthesis of colloidal MAPbBr$_3$ NCs. By controlling nucleation and growth parameters (solvents, surface ligands, and temperature), the shape of the NCs can be systematically changed. Oleic acid is present in each synthesis. (a-d) The TEM pictures of the corresponding nanostructures are also reported. The scale bars are 50nm. Adapted with permission from ref. [222] Copyright 2017 American Chemical Society.

In 2017 Kim Y-H et al. extended the emulsion LARP approach to FAPbBr$_3$ NCs. This was achieved by "simply" dissolving FABr and PbBr$_2$ salts in DMF to form a precursor solution which was added to a mixture of a desired alkylamine, OA and hexane, yielding a milky emulsion.[223] The subsequent addition of tert-butanol into the emulsion solution caused the formation of perovskite NCs. The OA was observed to improve the stability of the FAPbBr$_3$ NCs, suppressing their re-aggregation, while the amine ligands could prevent the formation of large crystals (> micrometer) by controlling the kinetics of crystallization. More specifically, FAPbBr$_3$ NPs synthesized using n-butylamine had a diameter of ~10 nm, but those synthesized using n-hexylamine and n-octylamine had smaller diameter (d < 10 nm).

### Reverse microemulsion

The first reported approach for the synthesis of perovskite NCs in the form of MAPbBr$_3$ QDs was a reverse microemulsion technique developed at the end of 2013 by Schmidt et al.[16] In their synthesis, octylammonium bromide was dissolved in a warm mixture (80°C) of OA and ODE, followed by the addition of MABr and PbBr$_2$, previously dissolved in a small amount of DMF. The mixing resulted in the formation of a reverse microemulsion in which NCs nucleated and grew. Eventually, the precipitation and the cleaning of the product NCs was achieved by the addition of acetone.[16] The same procedure was observed to work for the synthesis of MAPbI$_3$ NCs, even though with less control over the quality of the final NCs (in terms of size distribution and optical properties). In a subsequent work, the same group further improved the optical properties of these QDs by avoiding the use of OA and by finely adjusting the molar ratio between the bromide salts and the amount of non-polar solvent (ODE).[224] Tyagi et al. adopted the same method to produce a mixture of NCs and NPLs, from which they could selectively precipitate and analyze MAPbBr$_3$ NPLs.[225] The subsequent year MAPbBr$_3$ were produced in different morphologies by optimization of such approach. Yuan et al., for example, demonstrated that MAPbBr$_3$ NCs can be synthesized with control over the thickness down to one monolayer:[226] they first prepared a transparent precursor solution by dissolving PbBr$_2$ and the desired amount of ammonium salts (methyl ammonium and alkylammonium bromides) in DMF. The injection of such solution into vigorously stirred hexane at RT led to the formation of the NCs, which were eventually precipitated and washed by the addition of acetone.[226] In this case, control over the size of the NCs was achieved by using different ligands: benzylamonium ions were employed to make confined NPLs (1 or 2 monolayers), while thicker nanostructures were obtained using octadecylammonium bromide. On the other hand, Teunis et al. reported the synthesis of MAPbBr$_3$ nanoclusters[227] and NWs.[228] This was done by adding a solution of MABr in DMF and, in a second step, a solution of PbBr$_2$ in DMF into a hot (90°C) mixture of ODE and alkylammonium bromide salts. The reaction was then allowed to run for a specific amount of time (3 or 4 min) to produce the NCs, which were precipitated and collected by the addition of acetone. In order to get nanoclusters, the alkylammoinium bromide ligands used were hexylamine and diaminododecane,[227] while NWs were produced using octylammonium bromide only.[228] The year after, the same group demonstrated good shape control over MAPbBr$_3$ NCs via the reverse microemulsion approach.[222] In that work, a systematic variation of the solvent of the oil-phase, the temperature and the ligand environment resulted in the formation of NPLs, NSs, nanocubes or spherical nanoparticles (Figure 18).[222]

In 2016 Chen et al. reported the reverse microemulsion synthesis of oD Cs$_4$PbBr$_6$ microcrystals employing DMF and n-hexane as the "aqueous phase" and the "oil phase", respectively:[229] PbBr$_2$ and CsBr were dissolved in DMF together with OLA and OA. This solution was then introduced into hexane under stirring, leading over time to the formation of the Cs$_4$PbBr$_6$ microcrystals (Figure 19).[229]

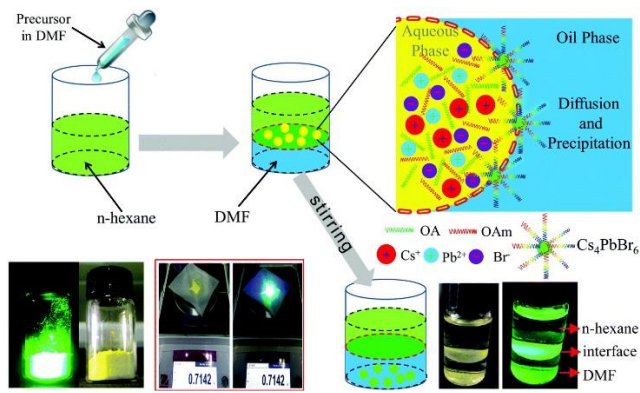

**Figure 19.** Reverse microemulsion synthesis of $Cs_4PbBr_6$ microcrystals. The inset shows the products prepared on a large scale without and with UV (365 nm) irradiation. Adapted from ref. [229] Copyright 2016 Royal Society of Chemistry.

Zhang *et al.* adopted the same strategy and revised it in order to provide better control over the size distribution and, more importantly, over the optical properties of oD $Cs_4PbBr_6$ NCs.[230] In their new approach, a solution of $PbBr_2$, hydrogen bromide (HBr) in DMF, OA and OLA was injected into a solution of cesium oleate in hexane and OA under stirring. The introduction of HBr as an extra source of $Br^-$ ions helped to better control the anion/cations ratio in the synthesis.[230] Moreover, the separation of Pb and Cs precursors in DMF and hexane, respectively, allowed for the control over the nucleation rate of the NCs. Indeed, in this way the nucleation rate is limited by the slow release of $Cs^+$ ions from the oil to the aqueous phase.[230] Importantly, they also demonstrated control over the size of the NCs by tuning of the micelle size, which, in turn, depends on the amount of OA used to stabilize the microemulsion.[230] Thanks to the versatility of this approach, the same group, soon later, further extended it to the synthesis of 3D ($CsPbBr_3$) and 2D ($CsPb_2Br_5$) NC systems (Figure 20).[231] They key parameter for the fine tuning of the final stoichiometry, and thus, of the crystal structure of the final NCs, was the Cs:Pb:Br feed ratio.

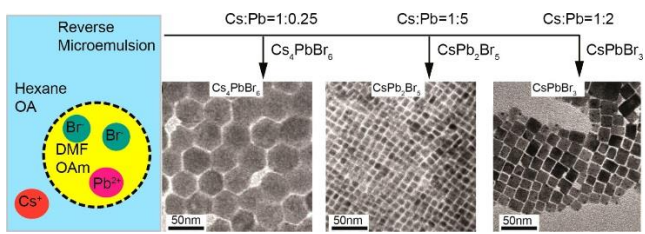

**Figure 20.** Illustration of the micelle structure formed in the reverse microemulsion process, which comprises an "oil" phase with n-hexane, and an "aqueous" phase with DMF. TEM images of $Cs_4PbBr_6$, $CsPb_2Br_5$ and $CsPbBr_3$ perovskite NCs formed with the different Cs:Pb:Br feed ratio. Reproduced from ref. [231] Copyright 2017 American Chemical Society.

### Polar solvent controlled ionization

The polar solvent controlled ionization (PCI) could be seen as the "opposite" of the LARP technique.[232] In the LARP strategy the starting precursors are dissolved, and thus ionized, in polar solvents from the beginning of the synthesis (state C in Scheme 6). The polar solvent is then transferred into a large amount of nonpolar solvent (state B), inhibiting the solubility of the ions and, thus, inducing the rapid crystallization due to supersaturation (state D).[232] On the other hand, in the PCI method all the precursors are initially not ionized as they are in the form of metal oleates or alkylammonium halides dispersed in a non-polar solvent (like hexane) (state A).

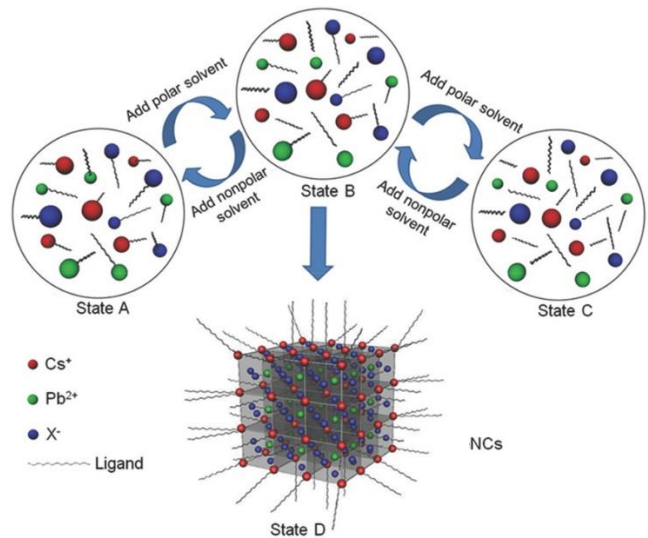

**Scheme 6.** The states which characterize the polar solvent controlled ionization and the LARP approaches. Reproduced from ref. [232] Copyright 2018, WILEY-VCH.

The subsequent addition of a polar solvent can be used to increase the degree of ionization of the precursors (state B) until reaching a critical concentration of ionized ions that results in the precipitation of the final NCs (state D). The clear advantages of this method is the use of a minimal amount of polar solvents. In 2018 Fang *et al.* demonstrated the synthesis of all-inorganic $CsPbX_3$ NCs via the PCI strategy.[232]

### Alternative (Indirect) Synthesis Approaches

Beside the synthetic approaches discussed so far, there have been few two-step syntheses of LHP NCs in which the nucleation of LHPs take places within starting colloidal seeds.[98, 233-234] For example, pre-synthesized $PbI_2$ NCs were used as a colloidal template to form 2D and 3D LHP NCs when reacted with alkyl ammonium iodide or MAI.[233] Also, CsX NCs, which are easy to synthesize and for which size tunability is straightforward, were employed as monodisperse colloidal precursors to make LHP NCs (Figure 21).[234] Interestingly, it has been demonstrated that when CsBr NCs react with Pb-oleate, the CsBr → $CsPbBr_3$ transformation goes through $CsBr/CsPbBr_3$ core/shell NCs as intermediate structures.[234] The ternary oD compound (*e.g.* $Cs_4PbBr_6$ NCs) is another colloidal candidate that transforms into the LHP NCs (*e.g.* $CsPbBr_3$ NCs) by further reaction with $PbBr_2$ (dissolved in toluene/OLA/OA).[98]

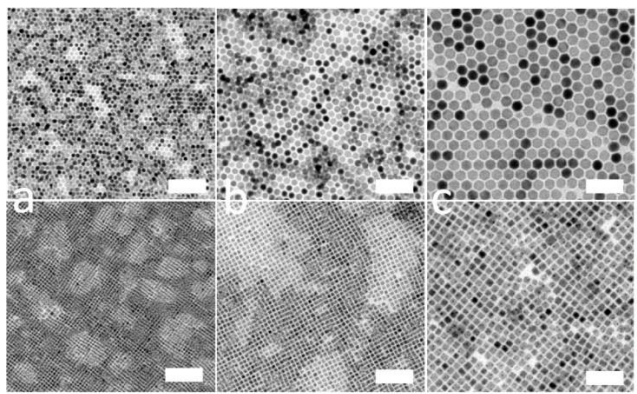

**Figure 21.** (a–c) TEM images of CsBr NCs of three representative sizes (top panels) and the resulting CsPbBr$_3$ NCs (bottom panels) obtained after the addition of Pb-oleate to the former NCs. Scale bars correspond to 50 nm in all panels. Reproduced from ref. [234] Copyright 2017, American Chemical Society

A potential advantage of such indirect syntheses is that a control over the size of LHP NCs is possible by tuning the size of the parent NCs. Also, monodisperse colloidal NCs in principle can serve as precursors towards more complex morphologies of LHP NCs.

## Complex morphologies

A possible approach to improve the stability of MHP NCs is to grow a protective shell in order to isolate them from the environment. Unfortunately, the labile nature of LHP NCs makes them unsuitable to be used as starting points to fabricate more complex morphologies, such as core-shell NCs and other NC heterostructures. However, due to their aforementioned defect-tolerance feature, such complex morphologies have been less in demand than from more traditional NCs, where for example core-shell geometries are needed to achieve high PLQYs.[235-236]

The simplest approach to accommodate MHP NCs (i.e. LHP NCs) into a heterostructure is using halide perovskites as inorganic capping ligands for PbS NCs.[237-240] This strategy allows to modify the surface of PbS NCs and grow LHPs as shell via a similar approach as the LARP one,[241-242] reducing the rate of Auger recombination in such infrared-emitting NCs.[243] Recently, a few strategies have been proposed to grow metal or semiconductor domains onto LHP NCs under mild conditions. In a very simple approach, the surface capping ligand (i.e. OLA) acts as a reducing agent for Au(III) cations (wich is previously introduced to the LHP NCs solution by adding the AuBr$_3$ salt) and, consequently, forming Au⁰ onto the CsPbBr$_3$ NCs.[244] In such hybrid structures, the absorption spectrum shows little dampening of the excitonic band as well as an absorption tail extending to longer wavelengths (similar to the modification of CdSe NCs with Au nanoparticles[245]). In addition, it has been observed that the PLQY decreases (e.g. from 80% to 60%) after the formation of Au domains, possibly due to the excited-state interaction of LHP and Au NCs (Au domains serve as charge recombination sites).[244] As the corner facets are expected to be less passivated than lateral ones, Au domains grow mainly at the corners of LHP NCs (Figure 22a). Interestingly, the addition of a proper amount of 1-dodecanethiol can remove Au segments and consequently restore the PL of LHP NCs.[246] As shown in Figure 22b, the consecutive modulation of CsPbBr$_3$/Au heterostructures by the addition of AuBr$_3$ and thiol ligands is reflected in the alternation of weak and strong PL emission.[246]

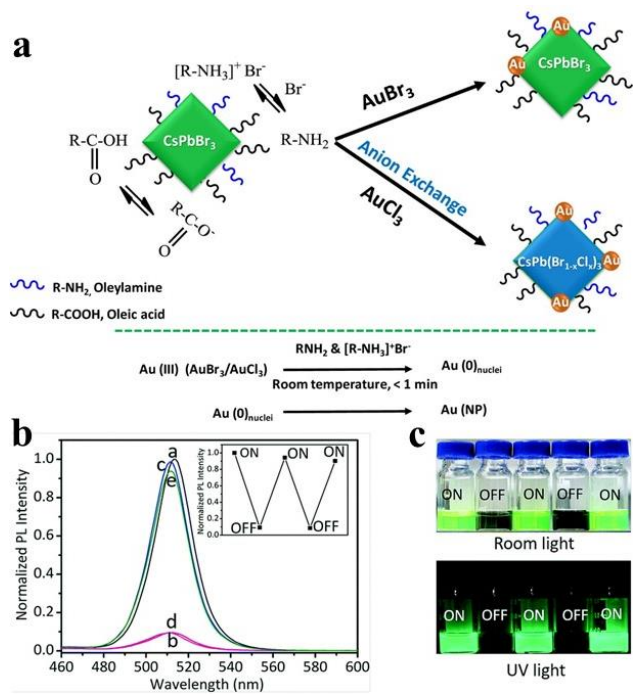

**Figure 22.** (a) Influence of ligands on the reduction of Au(III) at the surface CsPbBr$_3$ NCs to form Au–CsPbBr$_3$ hybrid structures. Reproduced from ref. [244] Copyright 2016 American Chemical Society; (b) PL spectra of CsPbBr$_3$ NCs before(a), and after the sequential addition of AuBr$_3$ (b and d) and 1-dodecanethiol (c and e). (c) Photographs depicting the modulation cycles of the system under room light and UV light, respectively. Reproduced from ref. [246] Copyright 2018 The Royal Society of Chemistry.

It is worth noticing that the addition of Au(III) to a LHP NCs dispersion not always results in the formation of Au-LHP heterostructures.[244, 247] In fact, the addition of AuBr$_3$ solubilized in EtOH/toluene to LHP NCs results in a Pb$^{2+}$ → Au$^{3+}$ cation exchange reaction producing Cs$_2$Au$^I$Au$^{III}$Br$_6$ NCs with a tetragonal crystal structure and a band gap of about 1.6 eV.[247]

Differently from the previous cases, significant PL enhancement has been reported when Ag domains were grown onto CsPbBr$_3$ NCs.[248] Ag nanoparticles display strong local surface plasmonic resonance absorption at ~ 410 nm (higher than the absorption onset of CsPbBr$_3$) and therefore can enhance the UV light absorption, and further enhance the PL of the perovskite NCs.[248-249]

In order to improve the stability of LHP NCs, different attempts to cover them with inorganic shell materials were also reported.[250] By adding proper amounts of zinc stearate and 1-dodecanethiol at high temperature (e.g. 120°C) to pre-synthesized CsPbX$_3$ NCs, ZnS domains could be grown on their surface, boosting their photostability.[250]

While 3D LHP NCs (e.g. $CsPbBr_3$ NCs) are extremely sensitive to water, treating $Cs_4PbBr_6$ NCs dispersed in a nonpolar solvent with water can trigger the 0D → 3D transformation.[251] This method was then exploited in a new sol-gel approach for the surface modification of $CsPbBr_3$ NCs at a single particle level.[252] In this method, the starting point is a $Cs_4PbBr_6$ NCs/hexane solution, to which tetramethoxysilane is added as a silica precursor. The addition of a small amount of water triggers two processes: i) stripping CsBr from the NCs, so that the 0D NCs are transformed into 3D $CsPbBr_3$ NCs; ii) removing OLA/OA from the surface of LHP NCs in contact with the water/hexane interface (Figure 23). After the ligands removal, the silica can deposit onto LHP NCs via a sol-gel method, resulting in highly stable $CsPbX_3$/oxide Janus-like NCs.[252] The same strategy has been used for the synthesis of $CsPbBr_3/Ta_2O_5$ NCs.[252]

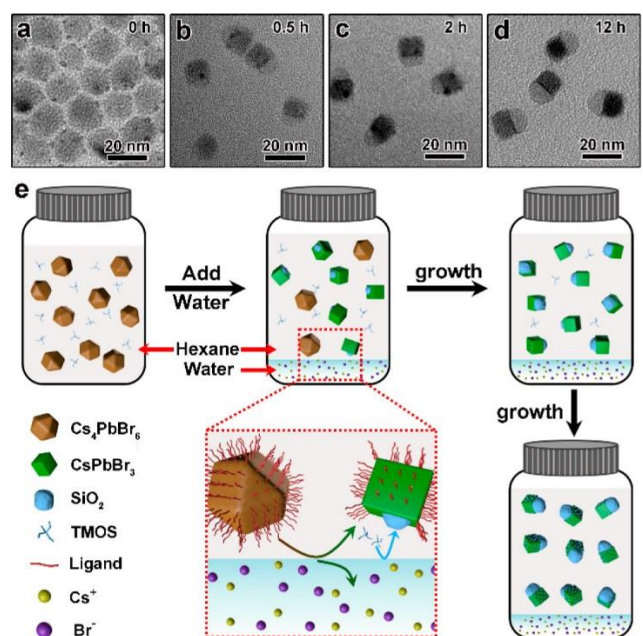

**Figure 23.** TEM images of the $CsPbBr_3/SiO_2$ Janus NCs obtained at different reaction times: (a) 0 h, (b) 0.5 h, (c) 2 h, and (d) 12 h. (e) Schematic illustration of the formation process Reproduced from ref. [251] Copyright 2017, American Chemical Society.

Although such hybrid structures were characterized by an improved stability against degradation by air, water, and light irradiation, for a complete protection of LHP NCs from the environment (moisture, air, etc.) a core-shell architecture would be more favorable. Octylammonium lead bromide,[253] amorphous $CsPbBr_x$,[254] $TiO_2$[255] and recently $SiO_2$[256] have been reported as shelling materials for LHP NCs. The $TiO_2$ shell not only efficiently protects the NCs from degradation, but also facilitates charge carrier transfer.[255] As a result, the $TiO_2$ layer coated $CsPbBr_3$ NCs exhibit excellent water stability for at least three months with the size, structure, morphology, and optical properties remaining unchanged.[255] $CsPbBr_3/TiO_2$ core/shell NCs (Figure 24) can be simply synthesized by dissolving titanium butoxide in $CsPbBr_3$ NC toluene solution, followed by a calcination process at 300°C.[255]

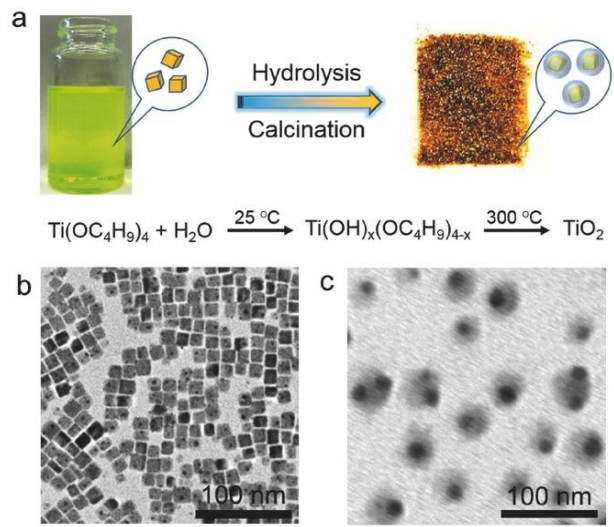

**Figure 24.** (a) Sketch of the fabrication process employed in the production of $CsPbBr_3/TiO_2$ core/shell NCs. TEM images of (b) $CsPbBr_3$ NCs and (c) $CsPbBr_3/TiO_2$ core/shell NCs after calcination at 300 °C for 5 h. Reproduced from ref. [255] Copyright 2018, WILEY-VCH.

### Anion exchange

The major contribution to the valence band maximum (VBM) of LHP materials comes from the halide orbitals.[40] Thus, systematic variation of the halide composition of such compounds allows for a fine adjustment of the VBM (Cl→ Br→ I or I→ Br→ Cl). One of the most effective ways to change the halide composition of preformed LHP NCs is the post-synthetic anion exchange (AE) reaction (Figure 25a, b), which has been shown to be a versatile tool to tune the optical properties of LHPs in both bulk and nanoscale.[257-259] Differently from what observed in cation exchange reactions in II-VI colloidal semiconductor NCs, the AE process in LHP NCs is completely reversible, suggesting that the solvation energy of the exchanging ions is not driving the process.[21-22, 260-261] Also, the fast AE rate observed in such systems is indicative of the high mobility of anions in MHPs, which, indeed, are known to have high ionic conductivities.[262] In general, the presence of two types of mobile ions, that is, the A-cations and halides anions, in both organic and inorganic lattices, gives rise to a broad spectrum of lattice dynamics.[263-264] A key factor affecting the mobility of halide ions (as the major ionic carriers) is the high intrinsic concentration of halide vacancies in such materials,[265] and the associated high mobility of halide ions, which is ultimately responsible for the hysteresis in the current-voltage curves from films of these materials.[7] The activation energies ascribed to ion/defect migration have been experimentally determined, and they range from ~0.1 eV to ~0.6 eV.[266-267] In addition, it is well documented that, among the various ions and defects,[268] halides and halide-vacancies are the fastest migrating species.[269] These two intrinsic factors (low defect formation energy and fast motion of ion/defect) facilitate the halide exchange reaction which, in turn, can be easily exploited to tune the LHP NCs emission spectra (in the range of 1.8 to 3.0 eV, Figure 25c) by simply exposing the NCs to specific amounts of the desired halide precursor.[40, 62, 270-271] Interestingly, AE enables the PL tuning from blue (410 nm) directly to red (690 nm) or red to blue (Cl→ I or I→ Cl) without going through intermediate colors. This can be explained considering the large difference between the ionic radii of Cl⁻ and I⁻ anions, which does not promote the stability of CsPb(Cl/I)$_3$ solid solutions.[22]

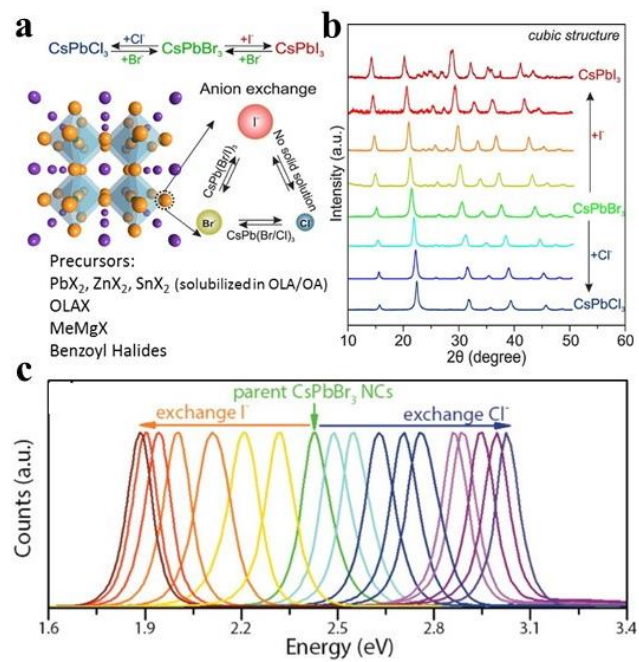

**Figure 25.** (a) Schematic of the AE reaction involving CsPbX$_3$ NCs. (b) XRD patterns and (c) PL spectra of CsPbX$_3$ (X =Br, Cl, I) NCs prepared by AE from CsPbBr$_3$ NCs. Reproduced from ref. [272-273] Copyright 2015 American Chemical Society.

In practice, one of the simplest approaches consists in mixing two different pre-formed CsPbX$_3$ NCs solutions having different halide compositions: the two populations of NCs tend to blend forming a homogeneous (Br/I) or (Br/Cl) composition, which is in between the two parent NCs.[21-22] The most accurate way to control the extent of the AE in LHP NCs is to start from preformed NCs (CsPbBr$_3$ NCs) and to expose them to different types of halide sources, such as OLA-X, PbX$_2$ or SnX$_2$ (dissolved in OLA/OA/solvent), benzoyl halides, *etc*.[21-22, 44, 274-276] Overall, the size and the shape of the parent LHP NCs are preserved, while their compositions is successfully tailored in a desired range.

The *in situ* synthesis of halide anions *via* reductive dissociation of the solvent molecules (*e.g.* dihalomethane) is another less common approach to perform AE reactions.[277] In this technique, the reaction begins with the photo-induced electron transfer from CsPbX$_3$ (X = Cl, Br) NCs to dihalomethane solvent molecules, producing halide ions *via* a reductive dissociation (Scheme 7).

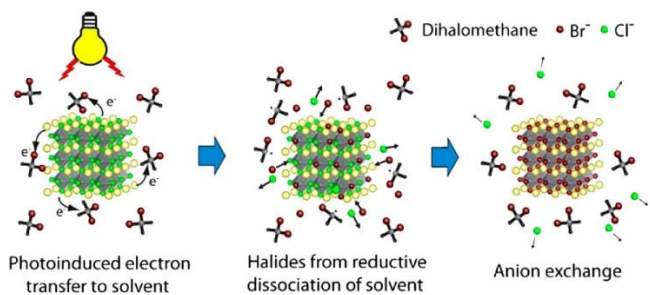

**Scheme 7**. Proposed mechanism for the photoinduced AE process of perovskite NCs in the presence of dihalomethane as the solvent. reproduced from ref. [277] Copyright 2017 American Chemical Society.

The halide ions generated *in situ* near the surface of the NCs drive the AE reaction efficiently, as long as the CsPbX$_3$ NCs are photoexcited above the bandgap. The extent of the anion-exchange reaction can be precisely controlled by adjusting either the photon dose or the wavelength of the excitation light that self-limits the reaction when the light becomes off-resonant with the absorption of the NCs. Salts added to the NC solutions can also act as halide sources. Some salts, for example PbX$_2$[21] and LiX$_2$,[278] lead either to a slow and an incomplete exchange (with the resulting NCs having broadened PL spectra), while other salts, such as ZnX$_2$, trigger a fast (reversible) exchange at RT.[279-280]

It was initially hypothesized that the NCs produced by AE possessed a homogeneous composition, at least in the case of the CsPbBr$_3$ → CsPbI$_3$ transformation, with the intermediate steps of the transformation being CsPb(Br/I)$_3$ homogeneous alloys[21-22] (Scheme 8, path A).[281] However, further optical[275] and quantitative X-ray photoelectron[282] spectroscopy analyses indicated that the intermediate exchanged structures have a radial gradient compositions, with the concentration of the iodide ions being higher at the surface (Scheme 8, path B).

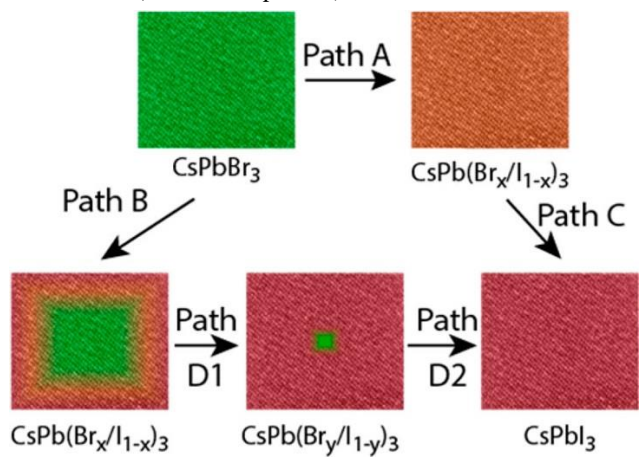

**Scheme 8.** Different possible AE pathways involving a CsPbBr$_3$ NC and iodide anions. Paths A+C describes the CsPbBr$_3$–CsPbI$_3$ conversion having as intermediate CsPb(Br+I)$_3$ alloyed structures. In the transformation depicted in paths B+D1+D2 core@graded-shell heterostructures form as intermediate steps, with CsPbI$_3$ being the final product. reproduced from ref. [282] Copyright 2017 American Chemical Society.

AE reactions can be also performed in the solid state by immobilization of LHP NCs on fine powders of KX (X = Cl, Br, or I).[283] This was achieved by Guhrenz et al., who first added pure KX salts to LHP NC solutions in hexane, and subsequently removed the solvent under vacuum. Upon evaporation of the solvent the resulting AE was observed to take place entirely in the solid state.[283] Even sintered NC thin films can undergo halide exchange when submerged in a heated lead halide solution,[284] a process that can greatly facilitate the post-synthetic tuning of the optical properties of the LHP NCs toward applications. AE can also take place at the solid/gas interface by treating NC thin films with volatile halide precursors: organic ammonium halides with low molecular weights are normally used (such as methylammonium halides or butylammonium halides) because they slowly evaporate (decomposing to amine and hydrogen halides) under mild heating (~100 °C).[285]

A striking application of the AE has been demonstrated in the fabrication of multicolor $CsPbX_3$ (X = Cl, Br, I, or alloy of two halides) NW heterojunctions with a pixel size down to 500 nm with the PL tunable over the entire visible spectrum (Figure 26a).[286] In practice, after transferring an individual $CsPbBr_3$ NW to a clean $SiO_2$/Si substrate, a thin layer of poly(methyl methacrylate) (PMMA) was spin-cast. A selected area of the PMMA was removed by e-beam lithography to expose a specific part of the NW. Then the substrate was dipped into an OLA-X (X = Cl, I) solution. During the reaction, the PMMA layer remains intact. After the AE reaction, the PMMA layer was completely removed by washing with chlorobenzene and hexane.[286]

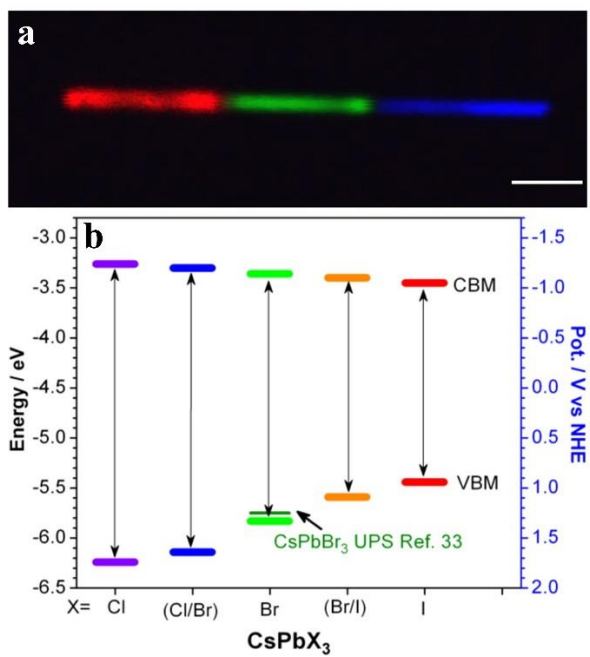

**Figure 26.** (a) Confocal image of a three-color heterojunction NW. Blue, green and red represent the PL emissions at 410-450 nm, 500-550nm and 580-640nm, respectively. Reproduced from ref. [286] Copyright 2017 National Academy of Sciences (b) Band edge energies of $CsPbX_3$ NCs extracted from cyclic voltammetry data. reproduced from ref. [287] Copyright 2016, American Chemical Society.

Very recently, a further step forward in controlling AE reactions was made by Ravi et al. who demonstrated AE can be suppressed on $CsPbX_3$ NCs passivated with $PbSO_4$–oleate.[288] This kind of protection could be utilized to deposit LHP NCs of different compositions as tandem layers to engineer effective harvesting of light.[288] Finally, it is also worth mentioning that supplementary cyclic voltammetry studies showed that the conduction band minimum (CBM) also undergoes slight changes over halide exchange and shifts systematically with halide composition (Figure 26b), though the shift is small (0.19 eV).[287] This indeed highlights the dependency of band edge states on the halide composition of $CsPbX_3$ NCs, shedding light on designing appropriate interfaces in both new heterostructured materials and optoelectronic devices.

### Cation exchange

Cation exchange (CE) reactions, in which the cations of preformed NCs are substituted with new cations in solution, have emerged as particularly powerful tools to achieve precise control over NCs composition and for accessing novel nanostructures.[260] Such transformations allow one to use NCs synthesized by conventional methods as templates for the production of nanoheterostructures or alloyed NCs while preserving the size and shape of the parent crystals.[289] In LHP NCs, in principle, both A and B-cations can be partially or totally replaced via CE reactions.[67, 290-292]

The band gaps of LHPs are weakly dependent on A-cations as the latter do not contribute significantly to the density of states near the Fermi level. However, the nature of the A-cation might play an important role in determining the symmetry of the LHPs compounds. For example, if compared with inorganic cations (e.g. $Cs^+$), organic cations (e.g. $MA^+$ or $FA^+$) are more dynamic and are able to bind PbX octahedra through hydrogen bonds. Therefore, when organic cations replaced inorganic ones, via CE, they can induce symmetry changes in LHP NCs, ultimately changing the vibrational modes.[65] Further, exchanging the A cations in LHP NCs can lead to the variation of: I) the tolerance factor and consequently the stability of NCs; II) the PL peak position (due to the tilting of the B-X-B bond) and life time; III) the dimensionality of LHPs. As an example, in $CsPbI_3$ NCs, $Cs^+$ ions can be partially or fully exchanged with $FA^+$ cations (by using FA-oleate as a precursor) offering long term stability and near-IR emission to the resulting NCs.[67] Similarly, $FAPbX_3$ NCs can be also obtained through a solid–liquid–solid CE reaction by adding a FA-acetate salt to a solution of $MAPbX_3$ NCs.[291] Starting with 2D perovskites, the replacement of octadecylammonium ($OA^+$) with $Cs^+$ or $FA^+$ in $OA_2PbBr_4$ microplatelets leads to the reorganization of the lattice. In this process, $[PbBr_6]^{4-}$ octahedra become corner-sharing and form green emitting sheets having a 3D LHP phase, with high crystallinity and excellent optoelectronic properties.[292-294] The mechanism of such 2D → 3D transformation, in which A-cations replace organic cations in the layered structures, is still not known, however, its driving force is believed to be the formation of an extended 3D octahedral network.

The band gap of MHPs is known to widen with the increase of the electronegativity of both B and X atoms and with the decrease of the unit cell volume.[295] Therefore, post-synthetic B-cation exchange can be used as an alternative route to tune the optical properties of LHP NCs. Considering that $Pb^{2+}$ ions are surrounded by 6 halide anions, forming $[PbX_6]^{4-}$ octahedra, which are enclosed by 8 A-cations, it is easy to understand that CE of the B cations in LHPs is rather difficult. However, partial CE between LHP NCs (*e.g.* $CsPbBr_3$ NCs) and different metal cations has been observed when employing $MBr_2$ solutions (*i.e.* $SnBr_2$, $ZnBr_2$ or $CdBr_2$ dissolved in OLA/toluene).[296] The resulting alloyed $CsPb_{1-x}M_xBr_3$ NCs show a blue-shift of the absorption and emission spectra, while preserving high PLQYs (>50%) and narrow PL line widths (80 meV) of the parent NCs. The observed blue-shift was explained considering a lattice contraction induced by "guest" $MBr_6$ octahedra, which are electronically decoupled from the $PbBr_6$ framework. The contraction of the $PbBr_6$ octahedra leads to shorter Pb–Br bonds and, therefore, to stronger interactions between Pb and Br orbitals. Since the CBM is composed of antibonding combinations of Pb(6p) and Br(4p) orbitals, a stronger interaction between the two atoms results in a shift of the CBM to higher energies, and, thus, in a widening of the band gap.

Similar to the AE, the CE process was tentatively explained as being facilitated by the presence of halide vacancies in the parent $CsPbBr_3$ NCs.[296] Briefly, the exchange between $M^{2+}$ and $Pb^{2+}$ ions takes place through a halide vacancy-assisted migration, i.e., OLA molecules in solution remove $Br^-$ ions from the NCs' surface and $PbBr_2$ units one after another, leaving vacancies behind. Subsequently, entering-$MBr_2$ species occupy the vacant sites, and from there diffuse inside the lattice.[290] A schematic representation of such mechanism is given in Scheme 9.

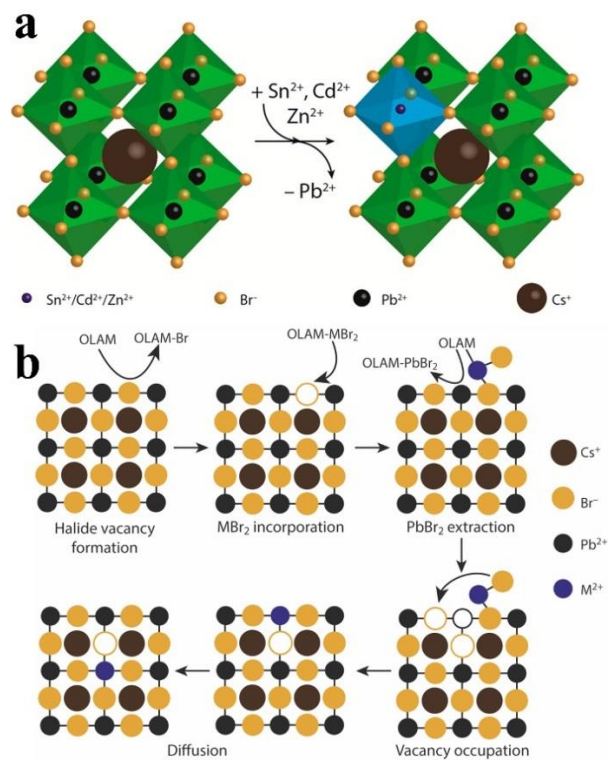

**Scheme 9.** (a) Partial $Pb^{2+} \rightarrow M^{2+}$ (M= Sn, Cd, Zn) CE in $CsPbBr_3$ NCs and (b) the corresponding proposed reaction mechanism. Reproduced from ref. [296] Copyright 2017 American Chemical Society.

Similar to the aforementioned cases, CE can also be employed as an alternative approach to produce alloyed $CsPb_xMn_{1-x}Cl_3$ perovskite NCs with the preservation of the shape and crystal structure of the parent $CsPbCl_3$ NCs.[297-298] Interestingly, the exchange between $Pb^{2+}$ and $Mn^{2+}$ can be almost completely reversible[297] by treating "as prepared" $CsMnCl_3$ NCs in $PbCl_2$ solution dissolved in OLA/OA/ODE, following the reaction:

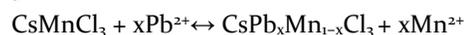
$$CsMnCl_3 + xPb^{2+} \leftrightarrow CsPb_xMn_{1-x}Cl_3 + xMn^{2+}$$

However, it is also not possible to obtain pure $CsPbCl_3$ through CE from $CsMnCl_3$. This is perhaps related to the balance between the inward diffusion of $Pb^{2+}$ and the outward diffusion of $Mn^{2+}$.[297] These alloyed NCs severely suffer from photo oxidation. Interestingly, to protect such systems from photooxidation and thermodegradation, a one-step dual ion exchange reaction was proposed by Xu *et al.*, who prepared $CsPb_xMn_{1-x}Cl_3$ NCs embedded in a KCl matrix: in their approach, preformed $CsPbBr_3$ NCs undergo both cation and anion exchange when exposed to $MnCl_2$ and KCl salts.[298] In such transformations, the partial exchange of Pb by Mn has a large probability to occur as a concomitant result for opening the rigid halide octahedron structure around Pb during halide exchange. This unique exchange feature results in a series of unusual phenomena, including long reaction time, core–shell structured intermediate structures with triple emission bands, and dopant molecules composition-dependent doping process(Figure 27).[299] Since the diffusion of large $MnCl_2$ species from the surface to the core of the NCs is quite difficult, the exchange requires extremely long reaction times (e.g. 40 h).

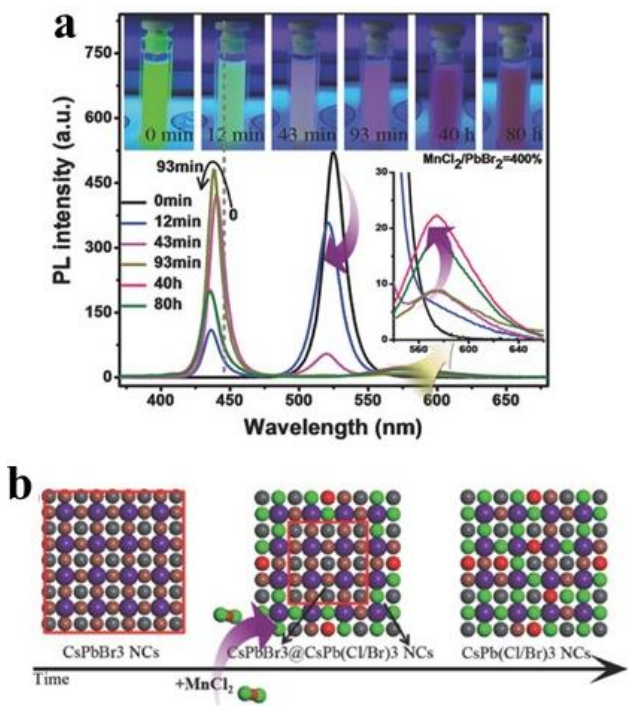

**Figure 27.** (a) Temporal evolution of the PL emission of CsPbBr$_3$ NCs after the addition of MnCl$_2$, together with (inset) the corresponding photographs taken under an UV lamp (365nm). (b) Schematic representation of the ion exchange process from pure CsPbBr$_3$ NCs to XXXX NCs obtained by the addition of MnCl$_2$. Reproduced from ref. [299] Copyright 2017 WILEY-VCH

In a similar fashion, homogeneous Cs(Pb$_x$Mn$_{1-x}$)(Cl$_y$Br$_{1-y}$)$_3$ NCs can be synthesized via a post-synthetic cation-anion co-substitution reaction by mixing pre-prepared colloidal CsPbBr$_3$ and CsPb$_{1-x}$Mn$_x$Cl$_3$ NCs in hexane.[300] The AE rate, which is much faster than that of the CE, can promote the Mn$^{2+}$ substitution in CsPb(Cl$_y$Br$_{1-y}$)$_3$ systems rich in Br.[301] Following a similar rationale, white light emitting Cs(Pb$_{1-x-z}$Zn$_z$)(Cl$_y$Br$_{1-y}$)$_3$:xMn$^{2+}$ NCs were obtained by controlled mixing of as-prepared CsPb$_{1-x}$Mn$_x$Cl$_3$ NCs and ZnBr$_2$ (dissolved in OLA/hexane).[301]

## Surface, shape and phase post-modifications

Apart from AE and CE reactions, several post-synthetic treatments have been investigated in order to modify the optical and the structural properties of LHP NCs, and to increase their performance once implemented in devices. Those modifications, which are "chemically" triggered, can be classified into three categories: surface, shape and phase (and/or mixed) post-modifications.

### Surface chemistry of MHP NCs

Surface post-treatment is a convenient method to enhance the optical properties as well as making LHP NCs more robust in ambient conditions.[302-303] Similar to the internal bonding in LHPs, the bonding of ligands to the surface has a strong ionic character and, in solution, it is highly dynamic, resulting in the facile desorption of surface capping molecules during the isolation and purification steps. In addition, the proton transfer between oleate species in solution and oleylammonium capping ligands can lead to the neutralization of surface-bound oleylammonium species, which will subsequently leave the A-site position from the surface of LHP NCs.[114] To prevent such ligand loss, the surface of LHP NCs can be opportunely modified either in solution[304-306] or via a solid-state process.[303, 307] Before going into the details of these surface treatments, we summarize here the main features of the possible surface terminations that characterize MHP NCs. As shown in Scheme 10a, b, MHP NCs can be terminated by either AX[308-309] or MX$_2$[45, 309] rich surfaces. It is worth mentioning that different electronic and optical properties of MHPs correspond to specific surface-terminations.[309] For instance, a theoretical investigation indicated that the CsI-terminated γ-CsSnI$_3$ material exhibits a higher electron mobility if compared to the SnI$_2$-terminated counterpart, thus highlighting the importance of the surface termination in MHPs.[310] As regarding CsPbBr$_3$ NCs, for example those reported by Protesescu et al.[311], it was initially assumed that the oleylammonium cations bind to surface bromide anions, presumably through a hydrogen bridge or via electrostatic interactions. On the other hand, surface cesium or lead ions can be passivated only by bromide or oleate anions, resulting in a NC(X)$_2$ binding motif.[312] However, De Roo et al. demonstrated via Nuclear Overhauser Effect Spectroscopy (NOESY) NMR analysis that OA does not bind to CsPbBr$_3$ NCs, ruling out the oleylammonium oleate termination and proposing oleylammonium bromide as a capping layer (Scheme 10c).[312] This model is also in accordance with the fact that highly ionic CsPbBr$_3$ NCs prefer ionic ligands over those that bind with a more covalent character, such as lead oleate. However, further experimental works evidenced that the Cs/Pb ratio in CsPbBr$_3$ NCs is typically lower than 1 (especially in confined systems).[19, 313] This finding suggested that the surface of small NCs is terminated by PbBr$_2$ and surface Cs$^+$ cations are replaced by oleylammonium ions (Scheme 10d).[313]

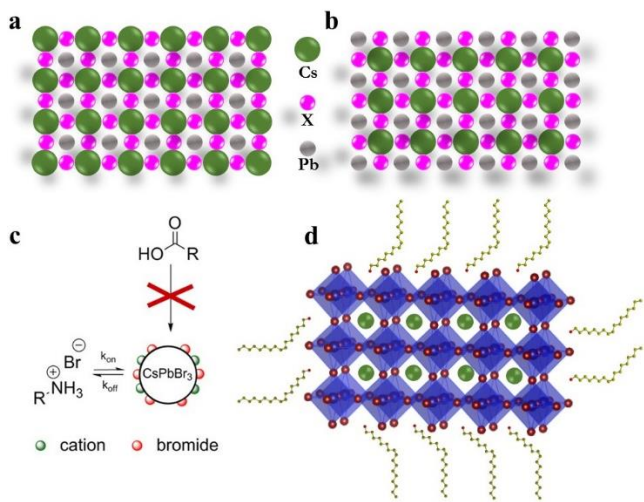

**Scheme 10.** Schematic illustration of: CsPbX$_3$ NCs being (a) CsX or (b) PbX$_2$ surface terminated; (c) the dynamic surface stabilization of a CsPbBr$_3$ NC by oleylammonium bromide (OA is not part of the ligand shell). Reproduced from ref. 312 Copyright 2016, American Chemical Society; (d) a LHP NC passivated by OLA$^+$ ions which replace surface Cs$^+$ ions.

Furthermore, DFT calculations shown that such OLA$^+$→Cs$^+$ substitution does not require much energy, stabilizing the system by the formation of three hydrogen bonds between the –NH$_3^+$ moiety and the surrounding Br$^-$ ions on the surface of the NCs.[314] In such system, the occurrence of Br vacancies may leave naked Pb atoms on the surface of the NCs and act as surface traps that may be responsible for the lower PLQY.[305]

The addition of PbBr$_2$ (dissolved in both an amine and a carboxylic acid) to as-prepared LHP NCs was demonstrated to induce a strong ligand binding at the surface[26, 312] as well as introducing excess amount of Br$^-$.[26] These two effects essentially promote the surface defect passivation and, consequently, result in a substantial enhancement of the PLQY.[26]. Interestingly, the non-radiative pathways of carriers recombination can be also effectively reduced by passivating the under-coordinated Pb ions with small molecules such as thiocyanate, leading to LHP NCs with a near-unity PLQYs.[305] It is worthwhile to mention that in all these post-treatments, the size and the shape of parent NCs are preserved.

### Ligand Exchange

Di-dodecyl dimethyl ammonium bromide (DDAB, which is relatively shorter than OLA) has been reported as a promising X-type ligand[315] to replace OLA on the surface of LHP NCs, leading to highly stable LHP NC films.[304] In order to achieve such exchange, the role of OA is crucial to desorb protonated OLA, otherwise adding DDAB directly to the purified NC solution would cause NC degradation.[304] Potassium-oleate (K-oleate) has been also introduced to protect and passivate the surface of LHP NCs, which led to an enhancement of the optical properties and photo- and thermal stability of LHP NCs.[316]

In all these protocols, the presence of insulating long-chain aliphatic molecules impairs the exploitation of such NCs in optoelectronics applications. To overcome this issue, conventional short-chain solid-state ligand-exchange procedures have been proposed.[317] Unfortunately, the low stability of LHP NCs in such processes (*i.e.* LHP NCs are not stable when exposed to many polar antisolvents), limits their effectiveness. It has been found that *short aromatic* ligands can trigger the precipitation of LHP NCs in a proper solvent, such as benzene.[307] This could be attributed to the partial ligand exchange on the surface that results a low steric repulsive force and therefore facile sedimentation. This observation suggested that short aromatic ligands (*e.g.* benzoic acid or 4-phenylbutylamine) in an appropriate (co)solvent (*e.g.* octane and benzene) can be utilized as a safe medium to remove the long-chain aliphatic molecules present on the surface of the as-synthesized NCs.[307]

### Phase transformations

It must be considered that the introduction of "new" ligands to as-prepared LHP NC solutions not only modifies the surface of such NCs, but also can trigger their phase transformation. Typically, phase transformations in LHP NCs are accompanied by shape evolution. The most common example is represented by the 3D→0D (CsPbBr$_3$ → Cs$_4$PbBr$_6$) NCs transformation.[318-322] This is triggered by the extraction of PbBr$_2$ from the 3D CsPbBr$_3$ perovskite NCs operated by the excess amines.[70, 318] It has been reported that by adding alkyl-thiol ligands, the size uniformity and chemical stability of the as-transformed Cs$_4$PbBr$_6$ NCs can be greatly improved.[321] Intriguingly, the 3D→0D transformation can be reversed by various methods, such as the addition of OA,[319] PbBr$_2$[98] or by chemical reaction with Prussian Blue[323], making these reactions appealing for reversible patterning.[318] Another engrossing example is the 3D→2D (CsPbBr$_3$ → CsPb$_2$Br$_5$) NCs phase transformation.[324-328] Such conversion can occur either in PbBr$_2$-rich conditions (CsPbBr$_3$ + PbBr$_2$ → CsPb$_2$Br$_5$)[322, 324] or by destabilizing/removing Cs$^+$Br$^-$ (2CsPbBr$_3$ → CsPb$_2$Br$_5$ + Cs$^+$Br$^-$) from the starting 3D NCs.[325] In practice, it has been found that the latter goes through a more complicated path when it is driven by ligands.[325] As illustrated in Scheme 11, when employing DDAB, such phase transformation starts with a ligand exchange step. Subsequently, the DDAB induces the initial formation of [PbBr$_3$]$^-$ and [Pb$_2$Br$_5$]$^-$ complexes, as evidenced from the optical analysis of the exchanging solution: absorption bands at 320 and 345 nm appear at the early stages of the process. The interaction between DDAB and CsPbBr$_3$ NCs results in the quick exfoliation of the starting material, in which Cs$^+$ ions are exchanged with alkylammonium ions:[325]

i) nCsPbBr$_3$ + DDA$^+$ → (n-x)CsPbBr$_3$ + xDDA$^+$[PbBr$_3$]$^-$ + xCs$^+$

ii) 2xDDA$^+$[PbBr3]$^-$ + xCs$^+$ → x[CsPb$_2$Br$_5$] + xDDA$^+$Br$^-$ + xDDA$^+$

As the system equilibrates, the lead halide complexes reorganize to form CsPb$_2$Br$_5$ NCs in solution.[325]

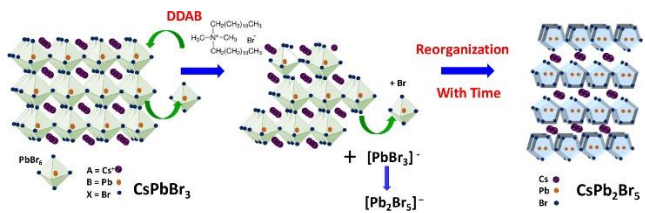

**Scheme 11.** The transformation of cubic CsPbBr$_3$ perovskite NCs into tetragonal CsPb$_2$Br$_5$ NSs which takes place by the progressive exfoliation of the former. Reproduced from ref. 325 Copyright 2018 American Chemical Society.

Interestingly, the shape of the original 3D NCs is completely lost upon the phase transition, which, in fact, leads to the formation of 2D NSs. In addition to ligands, other external agents can induce these shape/phase transformations.[39, 138, 329-331] It has been demonstrated that heat,[323, 326] light,[327] pressure[332] and water[251, 328] can also cause phase-change of perovskite NCs in the solid state. As an example, the thermal annealing facilitates the Cs$_4$PbBr$_6$ → CsPbBr$_3$ transformation, through a physical extraction of CsBr,[323] or the CsPbBr$_3$ → CsPb$_2$Br$_5$ transformation *via* the decomposition of the initial material (triggered by the desorption of weakly bound carboxylates ).[326]

### Self-Assembly

Thanks to post-synthesis techniques it is also possible to induce the self-assembly of as-prepared NCs by which individual components arrange themselves into an ordered structure.[333] Various methods have been demonstrated to lead to 1D or 3D LHP NC superlattices, such as solvent/ligand-assisted,[334-335] templated,[336] and drying-mediated[40] self-assembly. Compared to the PL emission of the starting NCs, the superlattices are generally characterized by a red shifted PL, due to the cooperative emission of LHP NCs. Significant features of the cooperative emission in such superlattices are the modification of the radiative lifetime (e.g. from 148ps to 400ps for 3D superstructures of CsPbBr$_3$ NCs) and their superfluorescence (short, intense bursts of light).[337] It must be noted, however, that ordered assemblies of these materials have yet to achieve the richness of structures that can be created using conventional NCs. For example, multinary superstructures have yet to be demonstrated.

### Composites

Suffering from the poor stability, the perovskite materials are sensitive to moisture, light, and temperature due to their low formation energy.[338-339] This chemical instability is the bottleneck that limits all technological applications where the NC processing into a composite or layer is required and where NCs are exposed to an external source of energy for a long time.[340] Several chemical methodologies have been reported to ameliorate the stability of LHP NCs by combining them with other classes of materials such as different types of polymers,[340-354] silica/alumina,[355-365] Cs$_4$PbBr$_6$,[366-369] graphene/2D materials,[370-373] and others.[374-382] Organic polymers with low oxygen and moisture transmission rates have been widely investigated to serve as effective matrices for embedding LHP NCs.[54, 270] The resulting nanocomposites can be fabricated mainly through two routes: i) the *in-situ* preparation of LHP NCs-polymer composites; ii) the post-encapsulation of LHP NCs into polymers. The *in-situ* fabrication strategy basically consists in the LARP or HI synthesis of LHP NCs in which extra polymerizabile species are added to the reaction mixture. For example, poly(maleic anhydride-alt-1-octadecene)[340] or methacrylic acid (MtA)[354] can be simply integrated into the standard HI approach, resulting in polymer- or monomer-passivated (encapsulated) LHP NCs. Such surface-modified LHP NCs exhibit the same crystal structure, shape and similar emission peak as those prepared in the same conditions by the standard synthesis protocol, with no polymers added.[340, 354] This approach may need further steps to polymerize the capping monomers. For example, the addition of 2,2-azobis(isobutyronitrile) (radical initiator) together with methyl methacrylate and methacrylisobutyl polyhedral oligomeric silsesquioxane leads to the copolymerization of MtA-LHP NCs under UV-light irradiation (Scheme 12).[354]

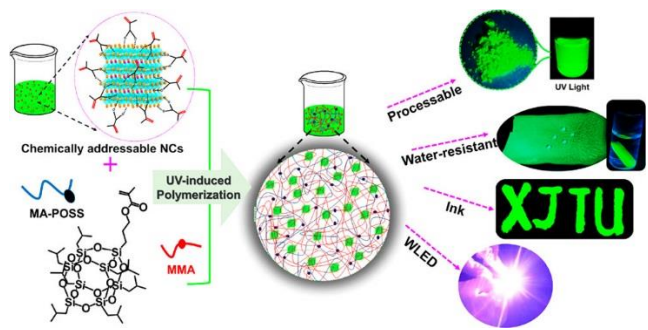

**Scheme 12.** Copolymerization of chemically addressable NCs (methacrylic acid-capped) with POSS-appended methacrylate monomer (MA-POSS) and/or methyl methacrylate (MMA) to produce polymer composites. Reproduced from ref. 354 Copyright 2018 American Chemical Society.

More complex systems (e.g. ternary graphene oxide–polymer–CsPbX$_3$ NC composites) can be achieved via the HI process by functionalizing the constituent monomers and/or adding extra components (e.g. graphene, *etc*), thus delivering robust emissive materials.[383] Other examples are represented by the LARP synthesis of LHP NCs using DMF as the solvent and poly-vinyl pyrrolidone (PVP) as the only capping ligand.[353] PVP can be physically adsorbed on the surface of the product NCs not only protecting them, but also making them compatible with polystyrene to fabricate water-resistant CsPbX$_3$@microhemispheres.[353] Recently, a universal "LARP related" strategy has been proposed by Xin *et al.* to prepare LHP NCs@polymer composites with many different polymers, such as poly(methyl methacrylate), poly(butyl methacrylate), and polystyrene in a one-pot chemical reaction.[349] In this approach, all required precursors (PbX$_2$ and CsX) are dissolved in DMF together with ligands (OLAM and OA), and a mixture of ethylene dimethacrylate, (2,2'-azobis(2-methylpropionitrile) and a chosen monomer was used as the bad solvent (instead of toluene).³⁴⁹ The as-prepared solutions underwent UV or thermal polymerization without any tedious purification and separation of LHP NCs (Scheme 13).³⁴⁹

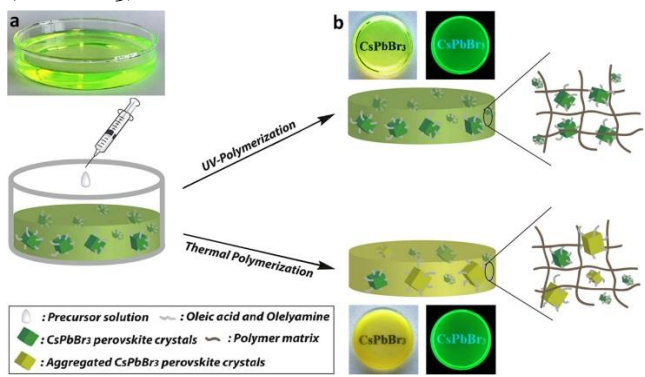

**Scheme 13.** One-pot strategy used to prepare perovskite–polymer composites (CsPbBr₃–polymer or MAPbBr₃–polymer): (a) formation of perovskite crystals in liquid monomers. The digital picture under room light illustrates a dispersion of CsPbBr₃ crystals in styrene; (b) the subsequent UV- or thermal-initiated polymerization leads to the formation of perovskite–polymer composites. Representative disks (under room and UV light) are shown in the photos. Reproduced from ref. ³⁴⁹ Copyright 2018 American Chemical Society.

A very similar procedure was devised for preparing CsPbBr₃ NCs/ethylene vinyl acetate composites in which a EVA/toluene solution was employed as the bad solvent.³⁸⁴

LHP NCs have been also embedded into different polymer matrices using two step approaches in which pre-synthesized samples are exposed to desired polymers: poly(methyl methacrylate)²⁷, ³⁴⁸, ³⁵¹ (insulating polymer), polystyrene³⁴¹, ³⁵² (one of the most widely used plastics), poly(styrene-ethylene-butylene-styrene)³⁴¹ (which is highly ductile), poly(lauryl methacrylate)³⁴¹ (a highly viscous liquid) and poly(acrylic acid)³⁸³ (a weak anionic polyelectrolyte). The so called "swelling–shrinking" technique is one of the simplest strategies to achieve mixing between pre-synthesized colloidal NCs and the desired polymers.³⁵² This approach is based on the fact that the crosslinked polymers swell in a good solvent and shrink in a theta solvent (a solvent in which polymer coils act like ideal chains).³⁸⁵ For example, toluene and hexane are good and theta solvents for polystyrene, respectively. Thus, LHP NCs@PS composites can be easily obtained, without inert gas or baking/heating operations, by simply dispersing LHP NCs and polystyrene beads in toluene and, subsequently, adding hexane in order to achieve shrinkage of the polymer.³⁵² Air stability and water resistance are the most remarkable characters of reported perovskite@polymer composites.

Alternative matrices for stabilizing LHP NCs are oxides materials, such as silicon oxide (silica, $SiO_2$)³⁵⁶⁻³⁵⁸, ³⁶¹, ³⁶³⁻³⁶⁴, ³⁸⁶⁻³⁸⁸, titanium oxide (titania, $TiO_2$)³⁷⁷ and aluminium oxide (alumina, $Al_2O_3$),³⁵⁵, ³⁵⁹, ³⁶² as they present very low diffusion rates of ions or atoms (environmental effects). In a straightforward method, a precursor solution of LHPs (e.g. MAPbI₃ or CsPbBr₃) directly fills the high-porous matrix of silica or alumina, without the use of colloidal stabilization.³⁵⁵, ³⁸⁶ This method normally requires a heating step to evaporate solvents and, consequently, to form the LHP crystals. It is worth noting that the pore size of the matrix determines the size and therefore the optical properties of the LHPs.

The formation of LHP@MO$_x$ heterostructures was also achieved by *in-situ* (or one-step) approaches. Examples are the encapsulation of MAPbBr₃³⁸⁸ or CsPbBr₃@Cs₄PbBr₆³⁶⁴ NCs in a silica matrix by using the standard LARP approach and adding (3-aminopropyl) triethoxysilane (APTES) to the precursors solution (Scheme 14).

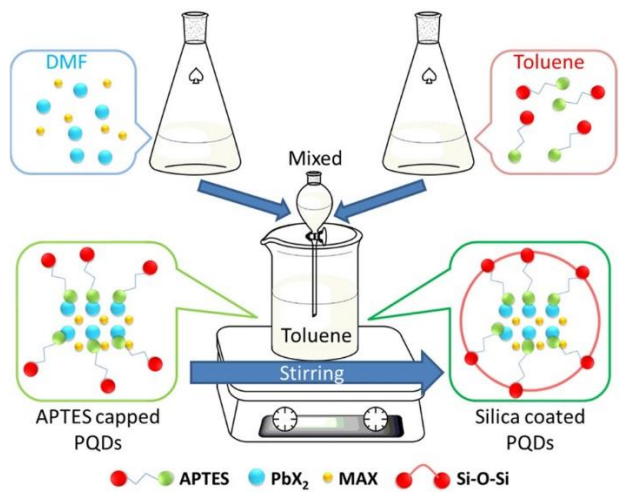

**Scheme 14.** Encapsulation of MAPbX₃ NCs in a silica matrix using a LARP based approach together with triethoxysilane (APTES). Reproduced from ref. ³⁸⁸ Copyright 2018 American Chemical Society.

APTES can also functionalize the surface of silica spheres with amino groups, facilitating the nucleation and growth of LHPs directly on the surfaces of silica particles.³⁸⁹

As regarding post-synthesis approaches, Hu *et al.* observed that adding an appropriate amount of the silica precursor (*i.e.* tetraethoxysilane, TEOS) into a solution of freshly synthesized LHP NCs (prepared by the HI method) led to the formation of micron-sized silica spheres in which LHP NCs are embedded.³⁶¹ However, strong aggregation of NCs in the resulting spheres was observed using this strategy.³⁶¹ Di-sec-butoxyaluminoxytriethoxysilane is an interesting precursor for fabricating silica/alumina monoliths *via* a sol-gel process in toluene.³⁵⁷ This single molecular precursor contains two alkoxide functionalities and a typical Al-O-Si linkage. The compatibility of this molecule with toluene enables the incorporation of pre-synthesized LHP NCs into silica/alumina monoliths, resulting in highly stable and homogeneous systems (no aggregation was observed).³⁵⁷ Polysilazanes (PSZ) are classified as inorganic polymers in which Si and N atoms alternate to form the basic backbone. The nitrogen-based ligand groups of the PSZ matrix (–H₂–SiNH–) can be used as capping agents, and the remaining silazane moiety can be reacted with water to induce a hydrolysis condensation reaction which forms a PSZ inorganic matrix. LHP NCs@PSZ nanocomposites can be synthesized simply by a hydrolysis and condensation reaction of liquid PSZ precursors and a small amount of water, as shown in Scheme 15.[360] Thanks to this approach a $SiN_x$, $SiN_xO_y$, or $SiO_y$ surface protection on LHP NCs can be selectively applied by manipulating the moisture-induced hydrolysis temperature.[360]

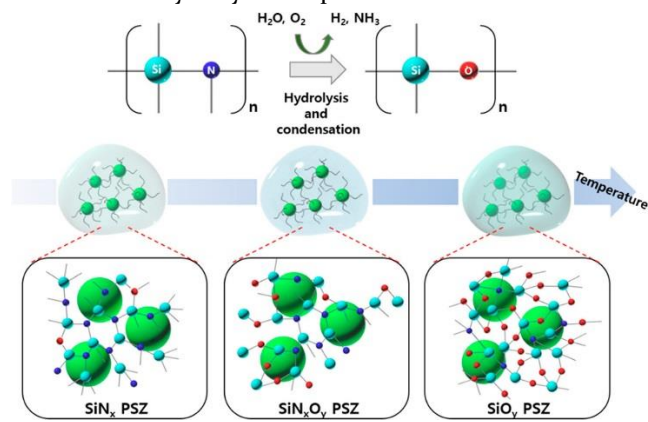

**Scheme 15.** Sol–gel route (moisture-induced hydrolysis), based on the use of liquid PSZ, employed to protect the surface of LHP with either $SiN_x$, $SiN_xO_y$, or $SiO_y$. Reproduced from ref. [360] Copyright 2018 American Chemical Society.

Recently, a few groups have demonstrated other interesting two-step strategies for preparing LHP composites.[375, 390] For example, Zhang *et al.* used a Pb-based metal-organic framework (MOF) as a starting precursor that was reacted with a MABr/n-butanol solution to form NCs@Pb-MOF composites (Figure 28).[390-391] This approach can be exploited as a smart luminescent system towards confidential information encryption and decryption: the Pb-MOF, which is transparent in the visible, can be easily printed and locally converted into luminescent LHP NCs (Figure 28).[390] Almost all LHP@oxide composites show superior thermal, photo, air and humidity stability as well as bright PL.

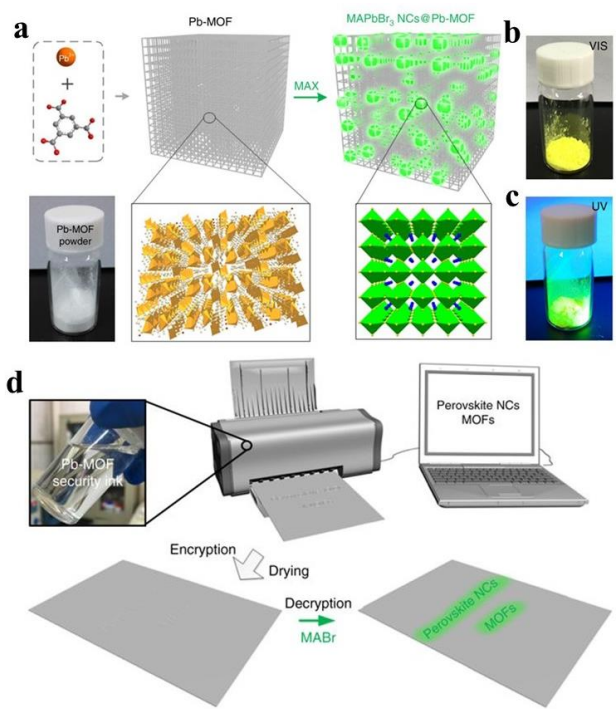

**Figure 28.** (a) Schematic representation of the conversion of a Pb-MOF into a luminescent $MAPbBr_3$/Pb-MOF composite. MAX is a halide salt ($CH_3NH_3X$, X = Cl, Br, or I) and the green spheres in the matrix are $MAPbBr_3$ NCs. The two black boxes show the crystal structures of the Pb-MOF (left) and the $MAPbBr_3$ (right). (b, c) Optical images of $MAPbBr_3$ NCs@Pb-MOF powder under (b) ambient light and (c) 365 nm UV lamp. (d) Illustration of an ink-jet setup based on Pb-MOF that can be information encryption and decryption. Reproduced from ref. [390] Copyright 2017 Macmillan Publishers Limited.

## Optical Properties of Halide Perovskite NCs

### Linear absorption and emission

Since the first demonstration of halide perovskite NCs by Schmidt *et al.* in 2014, showing a remarkable photoluminescence quantum yield (PL QY) of 20%, many groups have focused on improving the optical properties of these NCs for optoelectronic applications in terms of tunability, PLQY, stability and excitonic properties.[16, 392] One of the greatest advantages of MHPs is the ease with which the individual components can be exchanged to tune the bandgap of the resulting material (see previous sections). After the first report on $MAPbBr_3$ NCs, having an absorption onset and a PL emission around 529 nm, it was quickly shown that a modification of the halide composition could be used to shift the optical properties of such NCs throughout the entire visible spectrum, as previously observed for bulk perovskite thin films.[257, 393] Protosescu *et al.* showed that for $CsPbX_3$ (with X= Cl, Br, I) NCs the emission wavelength could be shifted from 410 nm (X = Cl) to 512 nm (X = Br) and then all the way out to 685 nm (X = I), virtually going through all values in between by using mixed halide components (Figure 29).[27]

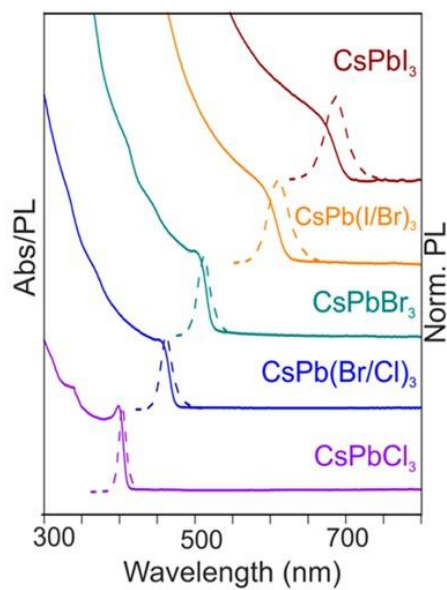

**Figure 29.** Bandgap tuning of CsPbX$_3$ NCs as a function of halide content, as demonstrated by UV-Vis and PL spectra. Adapted with permission from ref [27] Copyright 2015 American Chemical Society.

The reason for this tunability lies in the nature of the electronic structure of halide perovskites, as shown in Scheme 16. The conduction band forms from the antibonding orbitals of the hybridization of Pb 6p orbitals and the outer p orbitals of the halide (5p for I, 4p for Br and 3p for Cl), and it is mostly p-like due to the high density of states (DOS) from the lead contribution. The valence band, on the other hand, forms from the antibonding states of the hybridization of the Pb 6s and the same halide p-orbitals.[394]

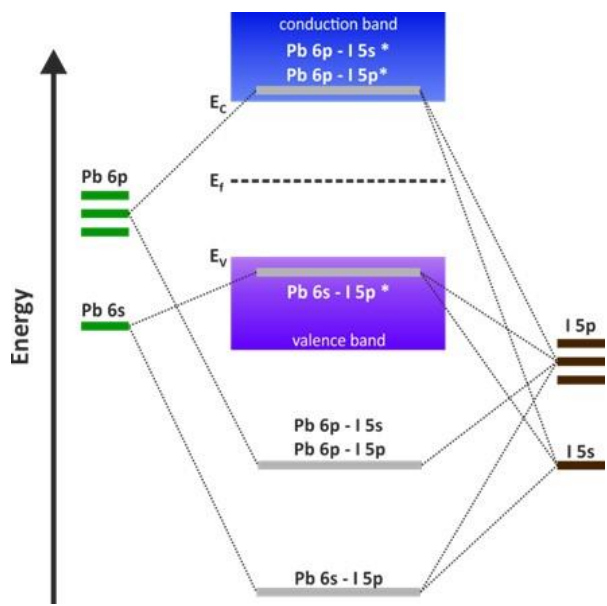

**Scheme 16.** Formation of energetic bands in a lead iodide perovskite material through hybridization of lead and iodide orbitals.

This is in stark contrast to conventional semiconductor materials, such as gallium arsenide, whose bandgaps are formed between bonding and antibonding orbitals. Accordingly, the valence band edge shifts in energy as the halide component is changed, while there are only small changes to the valence band edge.[395] The A-site cation does not contribute significantly to the conduction and valence bands in terms of density of states, yet this component also has a significant effect on the bandgap of the perovskite.[396] Three ions have been shown to satisfy the Goldschmidt tolerance factor when incorporated at the A-site of the perovskite crystal structure, namely the inorganic cesium (Cs$^+$), and the organic MA$^+$ and FA$^+$ cations. As the size of the A-cation decreases (from FA to MA to Cs), the bandgap of the corresponding LHP material blue shifts, due an increased tilting angle of Pb-X-Pb bonds and the concomitant distortion of the cubic crystal structure.[397] Consequently, the emission energies of MA-based perovskite NCs were shown to vary with the halide content from 407 to 734 nm.[177] In FA-based NCs the PL emission shifts further to the red with reported values of 408nm (Cl), 535 nm (Br), and 784 nm (I).[115-116, 195] Furthermore, the B-site cation can also play a role in dictating the final optical properties of MHP NCs. For example when Pb$^{2+}$ is replaced with Sn$^{2+}$ (one of the elements of choice for fabricating potentially environmentally-friendly perovskites[398]) the bandgap and the PL emission of the resulting MHP NCs are strongly red-shifted to 443 nm (Cl) and 953 nm (I), most likely as a consequence of the higher electronegativity of the Sn$^{2+}$ in respect to Pb$^{2+}$.[79] However, the stability of Sn$^{2+}$- and, similarly, Ge$^{2+}$-based perovskite compounds is extremely poor due to a reduced inert electron pair effect which corresponds to a decreased stability of the divalent oxidation state.[399] Consequently, only a single study on Ge-based perovskite NCs has been demonstrated to date.[167] Whether the NCs are synthesized directly, or obtained through a subsequent anion or cation exchange process, their final bandgap or PL emission wavelengths depend only on their final stoichiometry for comparable sizes and shapes.[21-22, 128]

Stokes shifts of LHP NCs are typically small, ranging between 20 and 85 meV.[135, 400-401] Notably, the Stokes shift increases as the size of the NCs is reduced. This was explained by the formation of a confined hole state, which can delocalize across the entire NC, and, thus has size-dependent properties.[309] The PL linewidths of MHPs are another critical parameter, especially for light emitting applications. In most reports, PL spectra are displayed on wavelength scales, with resulting linewidths strongly varying with the halide component (10-12 nm for Cl-perovskites and up to 40 nm for I-perovskites). However, as the wavelength of an energetic transition is related to the energy via: $E = \frac{hc}{\lambda}$, one should use an energy scale for comparing absolute values of linewidths. Indeed, one finds that the linewidths are typically of the order of 70-110 meV and do not vary noticeably with the halide content. These are extremely narrow and comparable with the best CdSe-QD systems currently used in commercially available lighting technologies.[39] In NC ensembles, two effects lead to the observed linewidths: the homogeneous broadening, intrinsic to the perovskite material, and the inhomogeneous broadening which depends on variations among the individual

NCs. Two methods can be used to probe the origin of the broadening of absorption and PL linewidths of NC ensembles, namely either measuring individual NCs, or the so called "four wave mixing" technique. Using these methods, as shown in Figure 30a, it was determined for $CsPbX_3$ NCs that, at RT, inhomogeneous broadening only plays a minor role in shaping the total linewidths. The homogeneous broadening, largely due to the strong electronic-phononic coupling, is the predominant source of linewidth broadening.[402-403] As the temperature is lowered, the number of occupied phonon modes decreases, leading to the concomitant reduction of the homogeneous broadening. At very low temperatures, the linewidths are nearly exclusively inhomogeneously broadened. The effect of linewidth broadening is even enhanced in NCs exhibiting quantum-confinement (Figure 30b).

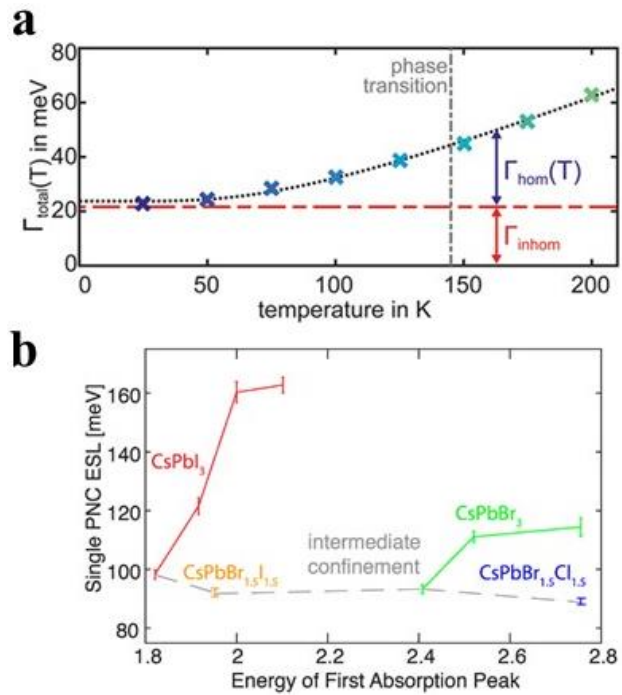

**Figure 30.** (a) Broadening of the 1s exciton level ($\Gamma_{total}$, FWHM of the absorption edge) of bulk-like $MAPbI_3$ NPLs as a function of the temperature, together with the contribution of homogeneous and inhomogeneous broadening. Adapted with permission from ref. [402] Copyright 2017 American Chemical Society. (b) Effective spectral linewidth (ESL) of single nanocubes (PNC) as a function of their respective first absorption peaks. Adapted with permission from ref. [403] Copyright 2017 American Chemical Society.

In terms of PLQYs, LHP NCs typically exhibit extremely high values, even without an electronic surface passivation with wider-gap epitaxial shells, which, conversely, is necessary for chalcogenide-based QDs.[116] Both Cs- and MA-based lead halide perovskites have had reported values of up to 100% PLQY, with typical yields of 80-95 % for bromides and iodides.[25, 46, 120, 305] These high values are a sign of the strong defect tolerance of halide perovskite materials, which results from their electronic structure and the bandgap forming between two antibonding orbitals. Indeed, this characteristic, coupled with the fact that interstitial and antisite defects require very high formation energies, lead to the predominant formation of shallow traps.[404] FA-based NCs are not far behind their Cs- and MA-counterparts, with PLQY values of 70-90%. On the other hand, much lower PLQYs, typically below 20%, were reported for $APbCl_3$ systems.[39, 44, 194] This is likely due to the small size of the chloride anions, and the resulting effects on the crystal structure, or due to the fact that defects in Cl-perovskites are not as shallow as in their Br and I counterparts, and they might act as electronic non-radiative traps.

An additional method to manipulate the optical emission of LHP NCs is represented by the doping with additional ions. This has been investigated in many recent publications focusing on the use of Cs-based Cl/Br perovskites. The introduction of $Mn^{2+}$ ions, which partially replace $Pb^{2+}$ cations in the LHP lattice, leads to a strongly Stokes-shifted emission, with the bandgap given by the perovskite matrix and the emission resulting from atomic states of $Mn^{2+}$ ions, (Figure 31a). Thanks to such features, these systems are particularly appealing for solar concentrators (Figure 31b).[32] A fine-tuning of the $Mn^{2+}$ doping levels can also lead to a controllable dual emission from both the localized $Mn^{2+}$ states and the bandgap recombination.[300, 359] In contrast, the doping of $CsPbBr_3$ NCs with other divalent cations, such as $Sn^{2+}$, $Cd^{2+}$ and $Zn^{2+}$ was shown to lead to a blue shift of the band edge and PL emission.[290] In these cases, however, a significant fraction (0.2 % - 0.7%) of the original Pb ions was replaced by new metal cations, generating, in practice, alloyed NCs. The blue shift in such systems can be rationalized considering that upon alloying, the original perovskite lattice contracts, widening the bandgap.

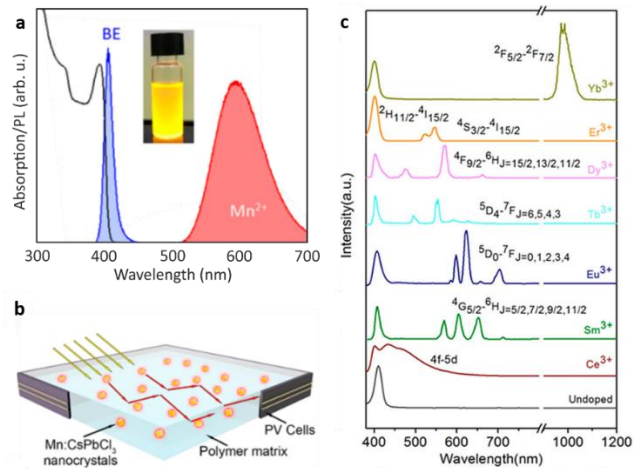

**Figure 31.** (a) Absorption and PL emission of Mn-doped $CsPbCl_3$ NCs: the large Stokes shifts in these systems make them an ideal candidate material for solar concentrators (b). Adapted from ref. [32] Copyright 2017 American Chemical Society. (c) PL curves of $CsPbCl_3$ NCs doped with different lanthanide ions. Adapted with permission from ref. [151] Copyright 2017 American Chemical Society

The alloying $CsPbBr_3$ NCs with $Al^{3+}$ ions (0.2% content) was shown to have similar effects: the resulting NCs had a blue-shifted PL emission centred at 456 nm and a surprisingly high PLQY of 42%.[405] In all of these cases, the perovskite NCs serve as an absorbing host which excites the dopants through energy transfer. Thus, even CsPbCl$_3$ NCs, which have very large absorption cross sections, but typically very low QYs, can be induced to emit strongly, provided the energy transfer efficiency to the dopant ions is high. Furthermore, by specifically selecting the dopant atoms, the emission wavelengths of the resulting NCs can be easily tuned. As demonstrated for the case of doping with Lanthanide ions and shown in Figure 31c, the emission of CsPbCl$_3$ nanocubes could be shifted between 400 and 1000 nm with QYs on the order of 15-35%.[151]

### Quantum confinement

Regardless of the approach, the PLQY of blue-emitting perovskite NCs still lags significantly behind those of red- and green- emitting counterparts. One additional method to obtain blue-emitting LHP NCs, as commonly done for conventional semiconductor QDs, is the exploitation of quantum confinement.[406] In this process one or more dimensions of a NC are shrunk down to or even below the size of the excitonic Bohr radius. In these conditions, the wavefunctions of the charge carriers become confined, and, consequently, the absorption onset blue shifts, as the electron and the hole acquire confinement energies: $E_{onset} = E_G + E_e + E_h$. Upon confinement, the PL of LHP NCs also commonly exhibits a blue shift, although often not as large as the shift of the absorption onset.[17] This is due to the interaction energy of the electron-hole pair, also known as the exciton binding energy, which leads to a modified optical transition energy: $E_{PL} = E_G + E_e + E_h - E_X$. The binding energy in quantum-confined nanostructures can be several times larger than in bulk semiconductor materials.[17] This is due to changes in the dimensionality and the concomitant reduced screening of the Coulomb interaction through the surrounding, mainly weakly polarizable ligands and organic media. In LHP materials the excitonic Bohr radii are extremely small: in CsPbX$_3$ they range from 5 nm (CsPbCl$_3$) to 12 nm (CsPbI$_3$); while in their organic/inorganic hybrid counterparts they are even smaller, in the order of 1.5-3 nm.[27, 393] Consequently, many of the reported LHP NC systems, even those with a high aspect ratio, such as NPLs or NWs/NRs, display only a weak quantum-confinement or no confinement at all.[128, 407-408]

We start discussing, first, LHP NPLs which can be considered effectively colloidal quantum wells. Thanks to the use of long chain surfactants, which are too large to be incorporated into the crystal structure, it is possible to template the growth of LHP NCs forming sheet-like morphologies with a high degree of control over the NPLs thickness.[51, 120, 126] A reduction of the NPLs thickness to below 4 or 5 monolayers leads to a quantum confinement that can be observed in a strong blue shift of the absorption onsets, as well as of the PL emission, by up to 0.8 eV, as shown in Figure 32.[17, 19, 192] In the case of MAPbI$_3$ NPLs, this effect is quite striking: the bulk emission lies in the near-infrared, while the PL of NPLs can be tuned almost throughout the visible range down to the green, in case of single layer thick NPLs.[23] For Cs-based tin halide NPLs it was demonstrated that the PL-emission can be tuned in a wide range in the near-infrared, from 950 nm down to 710 nm.[18]

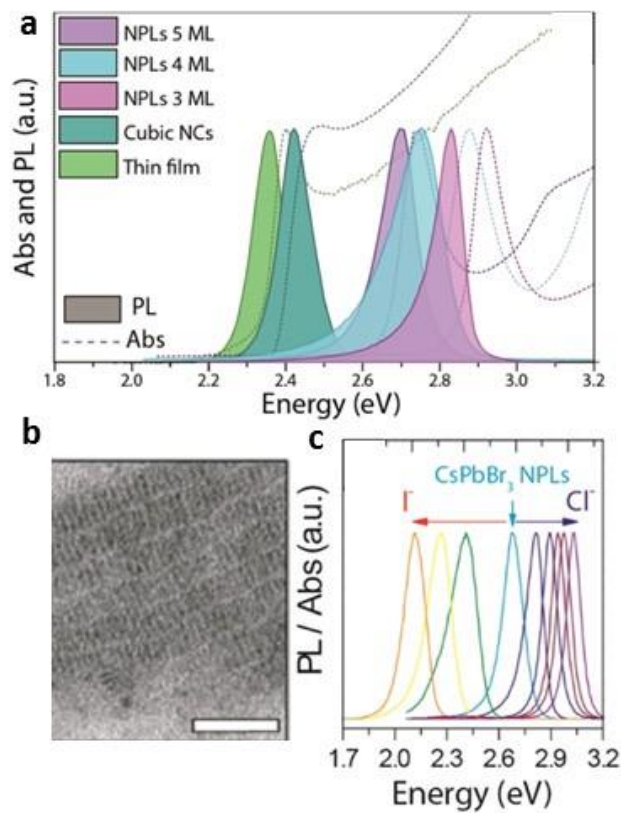

**Figure 32.** (a) PL and absorption curves of CsPbBr$_3$ thin films, cubic NCs, and NPLs with different thicknesses. As the thickness is reduced, pronounced quantum confinement effects can be seen in absorption and photoluminescence spectra. (b) TEM picture of 3 ML thick NPLs. (c) Post-synthesis AE reactions performed on NPLs lead to a wide tuning range of their PL, while retaining their size and shape. Adapted from ref. [19] Copyright 2015 American Chemical Society.

Unfortunately, the large tunability of the PL emission of these strongly confined nanostructures also has a significant drawback in that most NPLs exhibit low QYs, on the order of 5-20% for n ≥ 2 and only 1 % for n=1. This is likely a side-effect of the reduced dielectric screening of the electron-hole pair, which leads to an incomplete polaron formation around the charge carriers, leaving them more susceptible to scattering with defects and polar optical phonons.[51] However, recent publications have shown that this may only be a part of the reason and that it is possible to mitigate some of the drop in the PLQY by limiting the number of surface defects through adequate surface passivation.[409-411] Bohn *et al.*, for example, recently demonstrated that the poor blue emission of CsPbBr3 NPLs could be boosted by a post-synthesis treatment with a PbBr$_2$-ligand solution (with the PLQY increasing from ~7 to ~70%).[412] This work suggests that by carefully limiting the amount of surface defects in halide perovskites NPLs (Pb and Br vacancies in the aforementioned example) it is possible to obtain efficient blue emitting nanostructures which can be integrated into electroluminescent devices.[143, 412-413] Despite this, the QY is not constant for all NPLs, but actually increases with increasing thickness. This is a likely evidence of the dielectric screening effect of the charge carriers.

Importantly, as the thickness of NPLs can only be controlled in an incremental value, the PL and absorption shifts are quantized, according to the number of monolayers (n=1, 2, 3,...) forming the thickness of such nanostructures. Thus, the optical features of a NPL ensemble can be directly ascribed to those of a specific thickness, if the elemental composition is known. When dispersed in organic solvents, the NPLs typically do not interact with each other. However, in the solid state, they tend to stack together, so that the wavefunctions of neighbouring NPLs (with a thickness lower than 4 monolayers) can overlap, leading to electronic coupling and to the formation of minibands, provided that the inorganic spacer layer is thin enough (on the order of 1.5nm).[17, 73, 414]

In contrast to 2D nanostructures, there have been far fewer publications on perovskite quantum wires/rods and quantum dots. This is likely due to the propensity for halide perovskites to form two-dimensional structures, as in the form of the Ruddlesden-Popper perovskites, and a difficulty in obtaining 1D and 0D perovskite NCs with precisely tuned dimensions.[415] One-dimensional NWs of CsPbBr$_3$ were first demonstrated in 2016, with clear quantum confinement effects visible in NWs having a diameter smaller than 10nm (Figure 33).[20, 416] As for the case of 2D NPLs, typical colloidal approaches lead to NWs having a certain thickness distribution, which, in turn, results in multiple absorption and PL peaks.[17, 416-417] It was found that quantum confinement in 1D NWs is stronger, for each given thickness, than for their 2D counterparts. This is to be expected as the NWs are confined in an additional dimension, further blue-shifting PL and absorption onsets. PLQYs were also observed to decrease when reducing the diameter of the NWs, an effect that was attributed to the higher concentration of surface defects in thinner NWs.[20] An alternative explanation for this latter observation was given by an induced strain in NCs, which increases when reducing the dimensionality and leads to the formation of defects inducing a non-radiative recombination.[418] It was also possible to tune the PL emission of these NWs through subsequent halide ion exchange reactions.[20, 134] In colloidal form, NWs are unstable, often merging together, as suggested by a subsequent redshift of their PL emission.[199, 419]

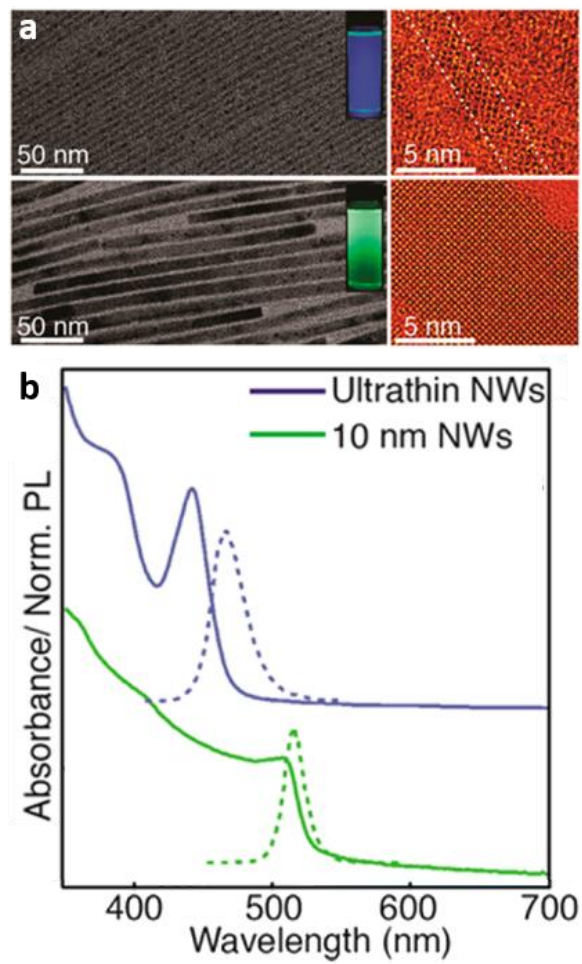

**Figure 33.** (a) TEM and HRTEM images of 10nm and ultrathin (~2nm) CsPbBr$_3$ NWs, and (b) their PL and absorption spectra. Similar to 2D NPLs, pronounced quantum confinement effects can be seen in these spectra. Adapted from ref. [20] Copyright 2016 American Chemical Society.

In 2015, Protesescu et al. demonstrated size-tuning of perovskite nanocubes down to 3.8 nm, which exhibited a noticeable quantum confinement (Figure 34a, b). As for the 1D NWs, though, especially for the smallest nanocubes, the increased width in the PL and absorption spectra indicates that the ensembles are comprised of nanocubes of varying sizes, making a detailed analysis of the quantum confinement effects on 0D NCs difficult.[27] Figure 34 depicts the quantum confinement for Cs-Pb-Br systems of varying dimensionalities. Here, it becomes apparent that for a given size of the confined dimensions there is an increase in the confinement as the dimensionality is reduced from 2D to 1D and, finally, to 0D. To date, only few studies were focused on investigating the quantum confinement in perovskite nanocubes. Unfortunately, while a significant number of publications have been reported on perovskite QDs, there are significant discrepancies between observed PL emission wavelengths and NCs size. The reason behind this was attributed to the fact that the NCs, observed via TEM, were, most likely, the degradation products of larger perovskite NCs formed during the electron beam illumination.[47, 106, 420] Whereas a monolayer constitutes the extreme case of a 2D perovskite system, the smallest possible 0D unit is a single [PbX$_6$]$^{4-}$ octahedron. In solution, these octahedra exhibit molecule-like excitonic absorption and emission spectra, with bandgaps far in the UV between 3.38 eV for PbI$_6$ and 4.37 eV for PbCl$_6$.[421] Balancing the strongly negative charge in the solid state requires a number of cations, for example Cs$^+$ ones, leading in that case to the Cs$_4$PbX$_6$ phase. In this phase, the octahedra no longer share corners, but they are separated by cesium ions, with eight of them surrounding each octahedron. The resulting optical properties do not differ strongly form the solvated octahedra. With charges highly confined in the octahedra, A-cations play nearly no role in determining the spectral position of the optical features, in contrast to the halide ions. Additionally, as to be expected, Cs$_4$PbX$_6$ NCs show no size-dependence in their optical properties.[98] Interestingly, some publications have reported on green-emitting Cs$_4$PbBr$_6$ NCs, despite their strong absorption in the UV.[230] While some works argue that such green PL arises from defect states or from self-trapped excitons, the spectral position of the PL emission, its temperature-dependence and, in some studies, also a strong absorption in the green, all suggest that these Cs$_4$PbBr$_6$ crystals are actually "contaminated" with small CsPbBr$_3$ inclusions, instead.[422] Similar features were also observed in intentionally synthesized Cs$_4$PbBr$_6$/CsPbBr$_3$ composite materials, where the Cs$_4$PbBr$_6$ perovskite acts as the strong absorber and transfers the energy to the emissive CsPbBr$_3$ NCs.[367] On the other hand, this issue is not settled at the moment, as no clear evidence has been found that indeed such CsPbBr$_3$ impurities are always present in samples of green emitting Cs$_4$PbBr$_6$ materials. We would like to point the reader to two review articles from this year focusing exclusively on this subject of 0D perovskites. [cite: 70,71]

One additional feature was observed in corrugated 2D, 1D and 0D microstructure assemblies (Figure 34 d-e). In these microstructures, the individual quantum-confined nanostructures are electronically separated by organic spacer layers, consequently they retain the optical properties of their respective building blocks. An additional featureless and extremely broad PL emission, strongly red-shifted with respect to the expected perovskite emission, was observed and attributed to the formation and subsequent radiative recombination of self-trapped excitons.[423-425] These are produced through the excitation of a perovskite nanostructure, which then undergoes a physical reorientation to minimize the interaction energy with an adjacent nanostructure or within the crystal lattice. This reorientation significantly reduces the energy of the exciton, which is dubbed "self-trapped" as the radiative recombination must occur concomitantly with the physical reorganization of the perovskite structure. This consequently leads to a long-lived emission.[426] Interestingly, some of these structures are extremely efficient for light emission with QYs approaching unity.[427] Yet, more work needs to be focused on understanding the nature of the self-trapping of the exciton and the resulting optical properties of the nanostructured LHP material.

### Exciton binding energies

The exciton binding energy is one of the most important parameters for an optoelectronic material, as it governs whether free electron-hole pairs or predominantly excitons govern the response of such materials when optically (or electrically) excited. Lately, many studies focused on determining the exciton binding energy in bulk LHP and specifically in MAPbI$_3$, as this material exhibits very interesting properties in photovoltaic applications[428]. Generally, it has been agreed that the binding energy of excitons in LHPs is sufficiently small (and the Bohr radius sufficiently large) that they can be considered as Wannier-Mott excitons.

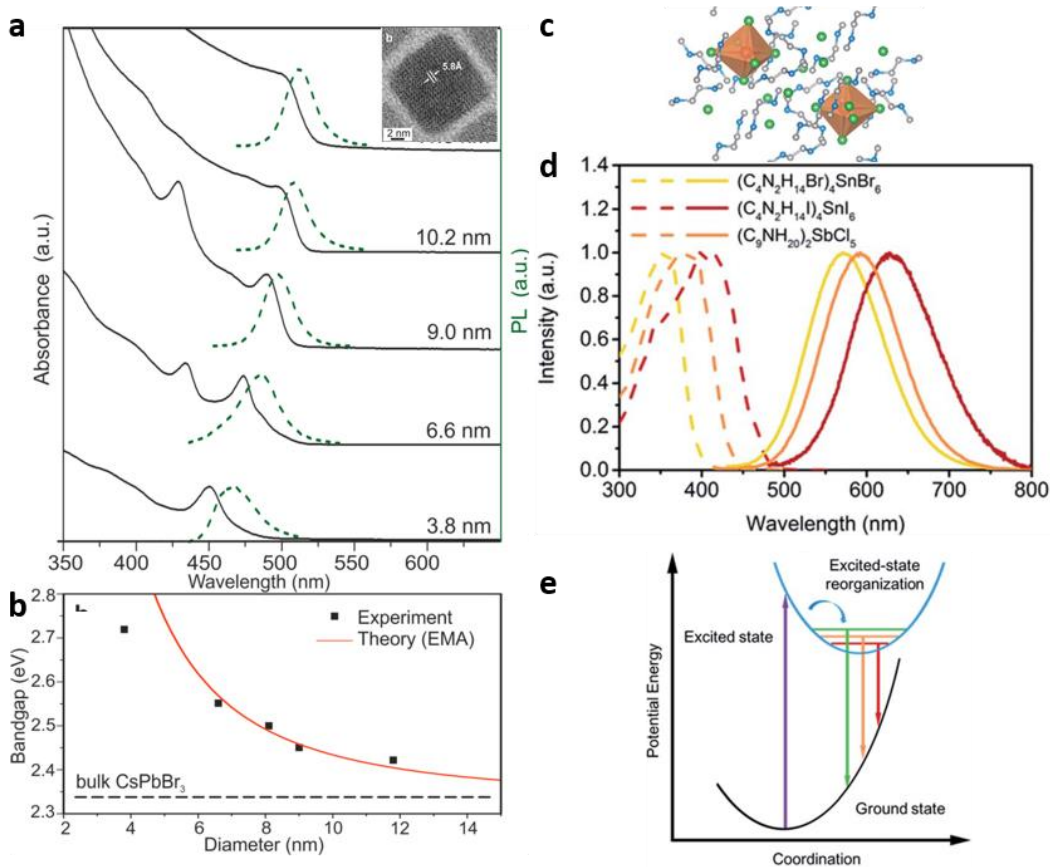

**Figure 34**(a) PL and absorption spectra of CsPbBr$_3$ nanocubes of different sizes together with (b) the corresponding calculated bandgaps. Adapted with permission from ref [27]. Copyright 2015 American Chemical Society. (c) Depiction of corrugated 0D metal halide perovskite structures. (d) Corresponding excitation and PL spectra of these structures exhibiting a strong Stokes shift and a significant broadening of the PL, which are both induced by a self-trapping of excitons. (e) Energy diagram depicting the formation and recombination of self-trapped excitons. Adapted from ref [425] with permission from the Royal Society of Chemistry.

This means that the wave functions of such excitons are hydrogenic and their respective Schrödinger equations can be solved analogously. Unfortunately, despite the plethora of studies on this issue, there are huge discrepancies on the reported values which, in the case of MAPbI$_3$, range from 2 meV to 60 meV at RT.[429] This is not due to the different methods employed in the experiments, as even similar experimental approaches have resulted in a strong discrepancy. Additionally, some methods are at best problematic and more likely inapplicable, as they deliver unreliable or even false values. For example, the dielectric function in LHPs is generally accepted to be a function that is strongly varying with the temperature and frequency. As such, it is also unclear which value of the function needs to be taken into account in order to obtain the binding energy, as this also depends on the relationship between the binding energy and the optical phonon energy of the material.[430] Applying the Elliot model to absorption spectra can also be problematic: due to the broadening of the absorption features and small exciton binding energies, the separation between excitonic and continuum absorption is small and a distinction becomes nearly impossible.[431] Another commonly used method is to calculate the exciton binding energy from the temperature dependence of the PL intensity.[432] This model assumes that bound electron-holes pairs are luminescent, whereas unbound pairs recombine non-radiatively. Consequently, as the temperature is increased, the thermal energy dissociates a larger fraction of excitons, consequentially resulting in a quenching of the PL. However, it is known for halide perovskites (*e.g.* MAPbI$_3$) that there is quite efficient luminescence from free electrons and holes even at RT. Consequentially, this method tends to lead to significantly higher and unreliable values, such as 50 meV for MAPbI$_3$ or 75 meV for MAPbBr$_3$.

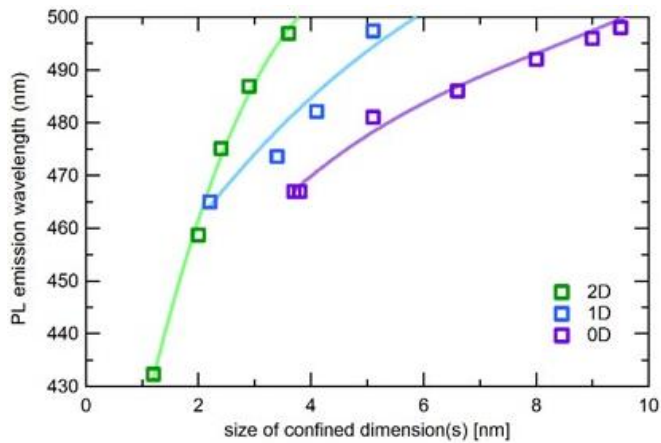

**Figure 35.** Variation of the PL emission as a function of the size of different CsPbBr$_3$ nanostructures: 2D, 1D and 0D. Data extracted from refs. [20, 27, 133, 412, 416]

Unfortunately, there are far fewer studies on the exciton binding energy of LHP NCs than those on their bulk counterparts. For NCs with dimensions significantly larger than the exciton Bohr radius, and thus not exhibiting quantum confinement, the values should be identical to those of their bulk counterpart. Indeed, the study of Protesescu et al. reported calculated values for CsPbX$_3$ NCs of 20 meV, 40 meV, and 75 meV for X= I, Br, and Cl, respectively, which fit into the range of those commonly accepted for the corresponding bulk materials. On the other hand, decreasing the size of LHP NCs should lead to an increase in the exciton binding energy. The most effective systems for determining the effects of quantum confinement on the exciton binding energy turned out to be 2D perovskites, such as Ruddlesden-Popper compounds or NPLs, as for these systems the thickness can be controlled with single monolayer precision.[415] Here, because of the 2D nature of the materials the density of states in the continuum exhibits a step-like appearance and quantum-confinement separates continuum and excitonic states progressively. This facilitates the determination of the exciton binding energy, as for example, the Elliot model becomes easier to apply. Interestingly, for the NPLs and also bulk thin films, values of up to several hundreds to meVs have been determined experimentally by applying the Elliot model to linear absorption spectra (Figure 36 a,b). These values are more than ten times those reported for perovskite bulk materials. [23, 433-434] In the theoretical limit, due to geometric considerations, the binding energy should only be fourfold enhanced: $E_B^{2D} = 4 \cdot E_B^{3D}$. However, this does not consider the dielectric surrounding of the 2D structures. As shown in Figure 36 c, in perovskites the high-dielectric inorganic layers are surrounded by low-dielectric organic ligands. Thus, as the thickness is reduced, the Coulomb interaction between the electron and hole is progressively less screened and, consequently, the binding energy is increased to the observed values.[435] With the excitons playing such a prominent role in LHPs, even at RT, future work needs to focus on understanding the exciton binding energy in LHPs and how it is affected by composition, size and dimensionality.

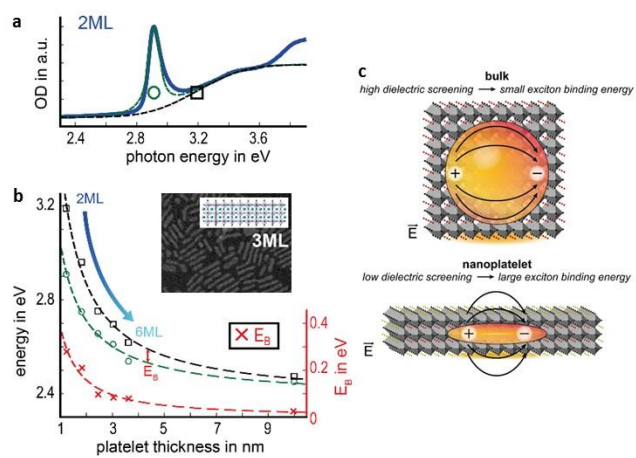

**Figure 36.** Determining the exciton binding energy in 2D LHP NPLs. (a) Optical density of 2ML NPLs dispersed in toluene: the 1s exciton absorption peak (green dot) and continuum onset (black square) are determined by applying Elliot's formula. The exciton binding energy is given by the energy difference of these two levels as a function of NPL thickness. Adapted with permission from ref. [412] Copyright 2018 American Chemical Society. (c) Scheme depicting that a reduced screening of the Coulombic electron-hole interaction in the thin NPLs strongly contributes to a large value in the hundreds of meVs of the exciton binding energy. Adapted with permission from ref. [51] Copyright 2017 American Chemical Society.

### Nonlinear effects

While most of the studies on LHP NCs have focused on their linear properties, several groups have also investigated nonlinear effects in such systems. For high laser power densities, multiple excitations can occur inside a single NC. Within these confined volumes, the excited carriers can interact with each other before decaying. One form of interaction is called the biexciton, in which two excitons bind together to form a coupled state comprising two electrons and two holes. As an exciton is electrically neutral, the binding energy of biexcitons is significantly lower than that of the "simple" exciton, and, consequently, it is often not observable due to its thermal dissociation. However, in bulk and, especially in quantum-confined LHP structures, which typically have large exciton binding energies, biexcitons can be stable at low temperatures, and in some cases even at RT. Makarov et al. first revealed biexcitons in LHP NCs reporting a biexciton energy of approximately 11-12 meV for CsPbI$_3$ and CsPbI$_{1.5}$Br$_{1.5}$ NCs, roughly half of the value of the simple exciton (Figure 37).[436] In more recent studies discrepancies in the biexciton binding energies have arisen, with some groups confirming the earlier values, for pure CsPbBr$_3$ NCs[437] and others reporting significantly enhanced energies.[438-439] Additionally, these were shown to have energies dependent on the excitation power, although the studies failed to provide an explanation for this or the contradictory reports on the binding energies.

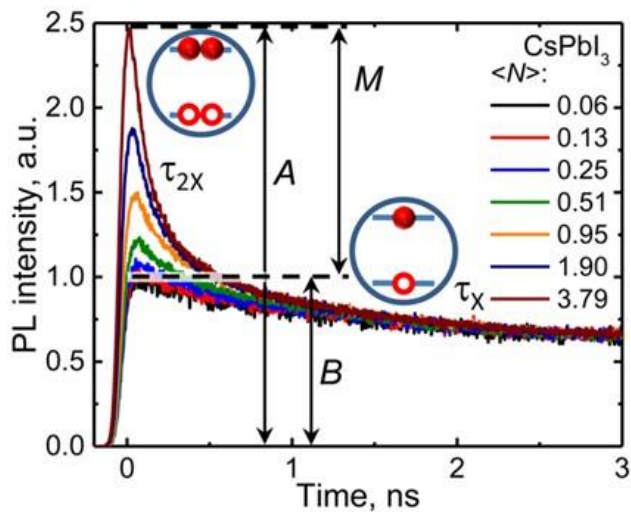

**Figure 37.** Biexctionic decay (lifetime $\tau_{2X}$) in CsPbI$_3$ nanocubes revealed for high pump intensities in time-resolved PL measurements. The late-time tail of the decay curve corresponds to single exciton recombination (lifetime $\tau_X$), while M and B correspond to the intensities of the multiexciton and single exciton signals, respectively. Adapted with permission from ref .[436] Copyright 2016 American Chemical Society.

While there are currently no reports on biexcitons in NPLs or other strongly confined perovskite NCs, there are two publications on bulk perovskite quantum wells.[440-441] In both works, biexciton energies of approximately 40-45 meV were reported for monolayer perovskites, roughly an order of magnitude below those of single excitons. Additionally, Elkins et al. revealed that increasing the thickness of such structures from one to two monolayers, the corresponding biexciton energy was reduced to only 15 meV, becoming insignificant in case of n=3.[440] Consequently, such findings suggest that biexciton energies should be negligible in weakly or non-confined LHP NCs. The ratio of photogenerated biexcitons to excitons seems to depend strongly on the halide composition and the size of NCs, as reported by Utzat et al.[403] The authors revealed that quantum confinement plays a major role here, as it can enhance the electron-hole overlap, increasing, thus, the Auger recombination rate with respect to the radiative decay rate of biexcitons. Indeed, they were able to determine $n_{biexciton}/n_{exciton}$ ratios of 2 % for highly confined CsPbBr$_3$ NCs and 54 % for weakly confined CsPbI$_3$ NCs. Their results illustrate possible applications for the individual types of NCs: I-containing NCs are an excellent choice for high brightness lighting applications, while Cl/Br and strongly confined Br variants are more attuned to single photon applications. Biexcitons in LHPs decay significantly faster than single excitons with time constants being typically below 50-60 ps, which are substantially higher than those of conventional semiconductor QDs.[436, 442] Additionally, as depicted in Figure 38, LHP NCs appear to deviate from the "universal volume scaling" law, which relates the Auger recombination lifetimes to the volume of the NC, irrespective of the material and composition. As the deviation appears for NCs whose diameter is larger than the excitonic Bohr radius, the reason for this discrepancy could be due to these NCs only being in the weak-confined regime (a consequence of the small excitonic Bohr radii in LHPs).[438] Clearly, more work needs to be done to understand the binding energies of biexcitons, their generation and subsequent recombination in LHP NCs.

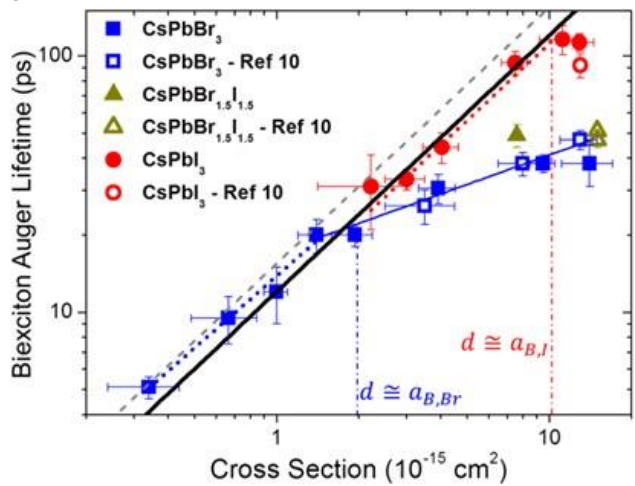

**Figure 38.** Biexciton Auger lifetimes of small LHP NCs in dependence of the NCs size. A deviation from the universal volume scaling law can be observed for NCs when their diameter exceeds the exciton Bohr radius. Adapted with permission from ref. [438] Copyright 2016 American Chemical Society.

In some cases, a photoexcited electron-hole pairs can be photoionized, whereby the electron, or the hole, leaves the NC before a radiative recombination occurs, resulting in a charged NC. A subsequent absorption process can induce the formation of another exciton in the same NC, which can, then, interact with the residual charge with the formation of a "so called" trion.[443-444] This quasiparticle typically recombines non-radiatively through Auger recombination, although radiative decay was also observed in LHP NCs, with emission energies that are slightly redshifted by about 15-20 meV from the excitonic transition energy.[443-444] The lifetimes of trions are typically 4-5 times longer than those of biexcitons, however their occurrence strongly depends on both the surrounding of the NCs and the excitation energy.[436, 438-439] Yin et al. recently reported also higher order quasiparticles, such as charged biexcitons (XX$^-$) – effectively a biexciton with an additional electron – , or doubly charged single excitons (X$^{2-}$) – an exciton with two additional electrons – in CsPbI$_3$ NCs, although only at ultralow temperatures, due to their low binding energies (Figure 39).[445]

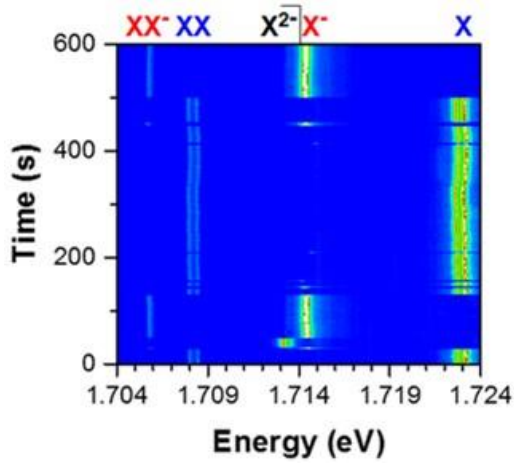

**Figure 39.** Time-dependent PL spectra of a single CsPbI$_3$ NC excited with an average of 1.5 excitons at 4K, revealing not only single exciton recombination (X), but also singly charged (X$^-$) and doubly charged (X$^{2-}$) single excitons, biexcitons (XX) and charged biexcitons (XX$^-$). Adapted from ref. [445] Copyright 2017 American Physical Society.

In that study, the authors observed a doublet emission, split by several hundreds of μeV, originating from the exciton. They further postulated that this large splitting could be caused by lattice anisotropy in the perovskite NCs. In a similarly comprehensive single NC study, Becker *et al.* observed a fine-structure splitting, however in their case into a triplet state (Figure 40).[443]

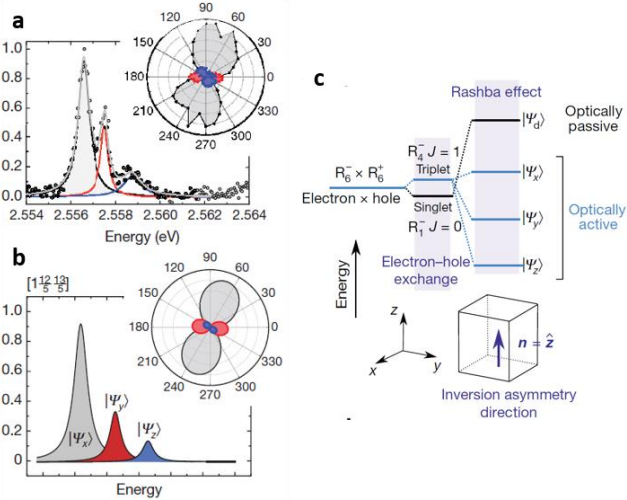

**Figure 40.** (a) Angular-resolved single NC PL measurements reveal a fine structure, which can be reproduced through simulations (b). (c) Energy scheme depicting the splitting of energetic states in cubic NCs. The Rashba effect is introduced to explain the experimental findings. This results in the lowest energetic state being optically active in contrast to common semiconducting NCs. Adapted from ref. [443] Copyright 2018 Macmillan Publishers Limited.

Combining PL measurements with theoretical calculations and employing the effect of Rashba splitting, the authors concluded that in LHP NCs the lowest energetic states are, in fact, emissive triplets. The Rashba effect is a splitting of initially degenerate spin states in k-space in materials exhibiting large spin-orbit coupling and an inversion asymmetry[446]. Such a finding is in stark contrast to conventional semiconductor QDs, whose lowest state is typically dark, rendering them only weakly emissive at low temperatures.[447] Further work needs to determine the exact cause of the Rashba effect and to investigate other materials in which this might play a similarly large role.

Instead of exciting a material with photons having an energy larger than the bandgap ($\hbar\omega \geq E_G$), generating an electron-hole pair, also photons with energy below the bandgap can be used. Provided that these carry at least half of the bandgap energy ($\hbar\omega \geq \frac{1}{2}E_G$), at very high excitation densities a two-photon absorption (TPA) process, via virtual states, can be observed in a given material. This phenomenon is especially interesting for photovoltaic applications, as a larger part of the solar spectrum could be used to generate current, boosting the efficiency of solar cells. TPA was first observed in CsPbBr$_3$ NCs by Wang *et al.* in 2015 with an enormous absorption cross-sections, which is two orders of magnitude larger than that observed in conventional semiconductor QDs.[448] In organic MA-based NCs, the absorption cross-section was determined to be one order of magnitude larger still.[449] In LHP NCs even higher orders of absorption were observed by Chen *et al.* with optical cross sections determined for up to five-photon absorption processes in both MA- and Cs-containing bromide perovskite NCs.[450] Interestingly, the authors found that core-shell structures, comprising a core of MAPbBr$_3$ and a shell of (OA)$_2$PbBr$_4$, to be the most efficient for such higher order absorption processes and suggested possible applications in multiphoton imaging.

Carrier multiplication (CM), also called multiexciton generation (MEG), a process by which one high energy photon is converted into multiple electron-hole pairs, constitutes an additional method to boost the efficiency of photovoltaic devices. This phenomenon has been demonstrated to have a high efficiency in PbSe and CdSe QDs.[451] Unfortunately, in perovskite NCs this process was not observed even at energies as high as $2.65 \cdot E_G$. This could be due to the fact that the effective electron and hole masses are similar. Consequently, due to optical selection rules, the excitation energy is split equally between the hot electron and the hot hole, increasing, in turn, the threshold for carrier multiplication to $\approx 3 \cdot E_G$. For CsPbI$_3$ NCs this corresponds to 5.5 eV or roughly to a 230 nm excitation wavelength.[436] This energy lies far outside the PV-relevant region of the solar spectrum, reducing the interest in investigating the CM process in MHP materials. Interestingly, Manzi *et al.* observed a combined effect of multiphoton absorption leading to multiple excitons generation in thin films of all-inorganic Cs-based LHP NCs (Br and I), as shown in Figure 41.[452]

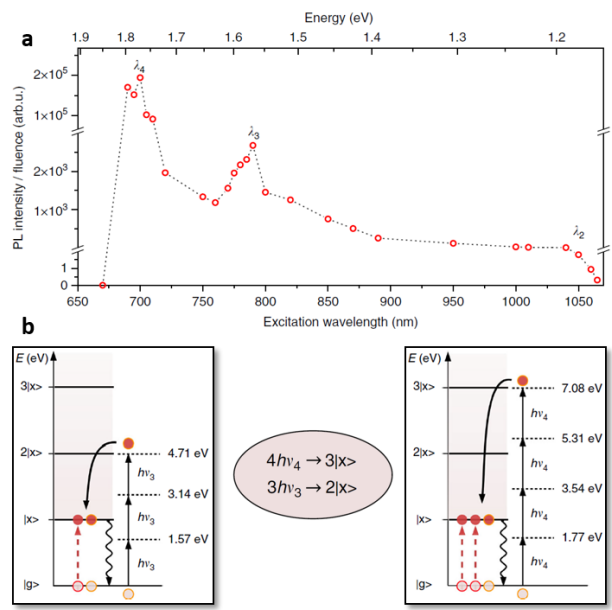

**Figure 41.** (a) Interesting resonances appear in wavelength-dependent below bandgap excitation spectra of LHP NC films. (b) This occur whenever an integer multiple of the exciting photon energy equals an integer multiple of the exciton energy. This is explained with a multiphoton absorption leading to multiple exciton generation. Adapted from ref. [452] Copyright 2018 Macmillan Publishers Limited.

The authors observed a strong nonlinear absorption for a wide range of below bandgap photon energies ($0.5\,E_G < \hbar\omega < 0.8\,E_G$). Importantly they determined the absorption processes at specific energies to be of higher order, so that $p \cdot \hbar\omega = n \cdot E_X$ wherein $E_X$ is the exciton transition energy and $p, n$ are whole integer numbers. This process uses below-band gap photons to create excitons, with the excess energy of the absorption process being turned into multiple electron-hole pairs. As these two processes are the largest source of energy loss in a PV device, the study presents a possibility to enhance the efficiency of a PV-device past the Schokley-Queisser limit. It will be interesting to see in the future whether this phenomenon is constricted to LHP NCs or whether it can be observed in other materials, being consequently a general phenomenon in semiconducting NCs.

## Outlook

This review has covered many aspects concerning the synthesis, post-synthesis treatments and optical properties of metal halide perovskite NCs. While this field is reaching maturity, there are several directions on which we foresee that efforts will be focused in the near future. These will be briefly discussed here.

One exciting direction is the exploitation of the large family of "elpasolite" materials. This search is primarily motivated by the need to move to materials that are environmentally friendlier than Pb-based ones. Even if at least 350 compounds of such family have been synthesized so far, most of them are still unexplored[80]. Indeed, as highlighted by Giustino and Snaith, the possible combinations of A, $B^+$, $B^{3+}$ and $X^-$ ions correspond to a total of more than 9000 hypothetical quaternary halide compounds[453]. Among them, even in the least optimistic case, at least 900 would satisfy the Goldschmidt tolerance factor[453]. We can therefore assume that, to date, more than 600 compounds have not yet been investigated. After a first screening performed via first-principles calculations, Zhao et al. in 2017 already proposed 11 non-toxic candidates as promising absorbers to replace $APbX_3$ in PV solar cells[80]. Among those, $Cs_2InSbCl_6$ and $Cs_2InBiCl_6$, have shown theoretical maximum solar cell efficiencies comparable to that of $CH_3NH_3PbI_3$. Unfortunately, until recently the main challenge was the development of effective and versatile synthesis methods to prepare such complex compounds[453]. As shown in this review, in the last years new efficient synthetic approaches for the colloidal synthesis of perovskite NCs were finally developed. These new strategies already delivered $Cs_2AgBiX_6$[118], $Cs_2AgInCl_6$[454] and Mn-doped $Cs_2AgInCl_6$[454] DP NCs. These materials represent only the tip of the iceberg, and the real potential of double perovskites is yet to be fully revealed. Also, it is likely that several elpasolite materials, which are predicted to be unstable in the bulk, could still be synthesized in the form of NCs, due to the non-negligible contribution of surface energy to the total energy balance in nanoscale materials. When this is coupled with the possibility to dope NCs with various ions, the potential for developing new materials for tailored optoelectronic applications is still enormous and remains vastly untapped to date.

Another critical research direction concerns the need for a deeper understanding of the role of the surface in the overall stability and photophysical properties of MHP NCs. A few studies have started to uncover the type of species binding to the surface of lead halide perovskite NCs. In this respect, the surface chemistry of standard $CsPbBr_3$ NCs (synthesized by OLA and OA) has been studied the most, and consensus has been reached about the presence of oleylammonium as a A-cation on surface of the NCs.[312, 314] Unfortunately, the dynamic nature of the bonding between the NCs surface and such capping ligands[54, 455], with alkylammonium ligands being much more mobile and susceptible than carboxylates to washing with polar solvents[124], leads to an important drawback: these materials are still facing major challenges of colloidal and structural instability that remain only partially addressed to date. For instance, as prepared LHPs NCs can even exhibit near-unity PLQY, but such optimal optical properties deteriorate drastically upon washing with polar solvents or during long-term storage. Promising approaches here consist in the use of zwitterionic ligands[145], or in the encapsulation of NCs in polymeric or inorganic matrixes. We believe that the exploration of alternative traditional ligands, such as phosphonic acids, thiols or sulfonic acids could be another exciting direction in order to achieve a more "static" surface passivation and/or to get a more complete picture of the surface stabilization of MHP NCs.[456] Among them, considering the strong chelating capability of organophosphonates toward metal cations, phosphonic acids represent promising surface ligands.[457-458] On the other hand, an extended and systematic study on the binding of different organic molecules on the surface of these NCs is still missing, with most published work focusing primarily on lead-based halide perovskites only. What is currently lacking is an extensive understanding on the type of surface termination, surface energy, and overall contribution to electronic band structure depending on what ligand molecules are passivating the surface of metal halide NCs. From a computational point of view, a more extended version of the recent work by ten Brinck and Infante on LHPs would be a good start[309].

Long term stability under harsh conditions (for example high temperatures, high humidity), needs to be carefully addressed, especially for applications in lighting applications, such as down-converting LEDs. Encapsulation of NCs in various inorganic matrixes has been actively pursued (for example, amorphous $CsPbBr_x$,[254] $TiO_2$[255] $SiO_2$[256] alumina,[355, 359, 362] and $PbSO_4$–oleate,[288] as discussed in this review). However it has to be demonstrated that such "shelling" entirely preclude oxygen, moisture and other chemicals from accessing the NCs inside the matrix over extended periods of time. This leads us to another open challenge: that of creating stable heterostructures between halide perovskite NCs and inorganic compounds. Encouraging results have been reported on stable epitaxial interfaces between lead halide perovskites and Pb chalcogenides[240, 459], and also growth of ZnS domains has been demonstrated, as discussed in this review.[250] Still, it would be highly desirable to prepare core/shell nanostructures with an epitaxial oxide shell.

Also, with the aim of improving and optimizing MHP NCs for optoelectronics a further in-depth understanding of their fundamental properties is essential. One of the trickiest parameters that still has to be determine is the exciton binding energy due to a both frequency- and temperature-dependent dielectric function. One strategy that could be now pursued, thanks to new synthesis tool developed so far, would be to finely tune the size of colloidal NCs from the smallest one accessible up to the bulk, Also, the observed fine structure and matching theory for cubic NCs will need to be verified, as do the existence and magnitude of the Rashba effect, for other compositions and crystal morphologies, such as NPLs or nanowires. Parameters such as Auger recombination rates, exciton vs. biexciton formation rates and gain profiles will need to be determined for uses in high-power LEDs and lasers. Multiple carrier generation obviously is of great interest for photovoltaic applications and could help to shatter the Shockley-Queisser limit.

## AUTHOR INFORMATION

### Corresponding Authors
* luca.detrizio@iit.it,
* liberato.manna@iit.it

### Present Addresses
†Javad Shamsi is now working as a research associate (post doc) at the Department of Physics, University of Cambridge; js2452@cam.ac.uk

## ACKNOWLEDGMENT

We acknowledge funding from the European Union under grant agreements n. 614897 (ERC Grant TRANS-NANO).